\documentclass[journal, onecolumn]{IEEEtran}

\usepackage{amsmath,amsfonts,amssymb,dsfont}
\usepackage{algorithmic}
\usepackage{algorithm}
\usepackage{array}
\usepackage[caption=false,font=normalsize,labelfont=sf,textfont=sf]{subfig}
\usepackage{textcomp}
\usepackage{stfloats}
\usepackage{url}
\usepackage{verbatim}
\usepackage{graphicx}
\usepackage{cite}
\usepackage{bbm}
\usepackage{xcolor}
\usepackage{hyperref}

\newcommand{\bR}{\mathbb{R}}
\newcommand{\bN}{\mathbb{N}}

\newcommand{\bP}{\mathbb{P}}
\newcommand{\bE}{\mathbb{E}}

\newcommand{\cI}{\mathcal{I}}

\newcommand{\cA}{\mathcal{A}}

\newcommand{\cH}{\mathcal{H}}

\newcommand{\cK}{\mathcal{K}}

\newcommand{\cJ}{\mathcal{J}}
\newcommand{\cR}{\mathcal{R}}
\newcommand{\cM}{\mathcal{M}}
\newcommand{\cE}{\mathcal{E}}

\newcommand{\cF}{\mathcal{F}}
\newcommand{\cV}{\mathcal{V}}

\newcommand{\cG}{\mathcal{G}}
\newcommand{\cW}{\mathcal{W}}

\newcommand{\Exp}{{\sf E}}
\newcommand{\Var}{{\sf Var}}

\newcommand{\Pro}{{\sf P}}

\newcommand{\sfE}{{\sf E}}

\newcommand{\sfP}{{\sf P}}
\newcommand{\sfF}{{\sf F}}
\newcommand{\sfG}{{\sf G}}
\newcommand{\sfH}{{\sf H}}

\newcommand{\rt}{\right}
\newcommand{\lt}{\left}

\newcommand{\e}{E}
\newcommand{\cU}{{\cal U}}

\def \med{\;\Big|\;}

\newtheorem{theorem}{Theorem}[section]

\newtheorem{proposition}{Proposition}
\newtheorem{corollary}{Corollary}[section]
\newtheorem{lemma}{Lemma}[section]
\newtheorem{remark}{Remark}[section]  
\newtheorem{example}{Example}[section]

\begin{document}
\allowdisplaybreaks


\title{\huge Round Robin Active Sequential Change Detection\\ for Dependent Multi-Channel Data}

\author{Anamitra Chaudhuri, Georgios Fellouris,\;~\IEEEmembership{Member,~IEEE,} and Ali Tajer,\;~\IEEEmembership{Senior Member,~IEEE}
\thanks{This work was supported in part by the U.S. National Science Foundation through grant NSF ATD 1737962.}

\thanks{The results in this paper were presented in part at the
2021 IEEE International Symposium on Information Theory (ISIT) \cite{ChFeAl21}. 
}
\thanks{Anamitra Chaudhuri is with the Department of Statistics, Texas A\&M University,
College Station, TX 77843 (e-mail: ac27@tamu.edu).}
\thanks{Georgios Fellouris is with the Department of Statistics, University of Illinois at Urbana Champaign, Champaign, IL 61820 (e-mail: fellouri@illinois.edu).}
\thanks{Ali Tajer is with the Department of Electrical, Computer, and Systems Engineering,
Rensselaer Polytechnic Institute, Troy, NY 12180 (e-mail: tajer@ecse.rpi.edu).}
}

\maketitle
\begin{abstract}
This paper considers the problem of sequentially detecting a change in the joint distribution of multiple data sources under a sampling constraint. Specifically, the channels or sources generate observations that are independent over time, but not necessarily independent at any given time instant. The sources follow an initial joint distribution, and at an unknown time instant, the joint distribution of an unknown subset of sources changes. Importantly, there is a hard constraint that only a fixed number of sources are allowed to be sampled at each time instant. The goal is to sequentially observe the sources according to the constraint, and stop sampling as quickly as possible after the change while controlling the false alarm rate below a user-specified level. The sources can be selected dynamically based on the already collected data, and thus, a policy for this problem consists of a joint sampling and change-detection rule. A non-randomized policy 
is studied, and an upper bound is established on its worst-case conditional expected detection delay with respect to both the change point and the observations from the affected sources before the change. 
It is shown that, in certain cases, this rule achieves first-order asymptotic optimality as the false alarm rate tends to zero,
simultaneously under every possible post-change distribution and among all schemes that satisfy the same sampling and false alarm constraints.
These general results are subsequently applied to the problems of (i) detecting a change in the marginal distributions of (not necessarily independent) information sources, and (ii) detecting a change in the covariance structure of  Gaussian information sources. 
\end{abstract}


\section{Introduction} \label{sec:introduction}

In statistical decision-making, growth in the data's complexity, dimension, and scale increases the data acquisition and computational costs, if not rendering them prohibitive altogether. Enforcing proper data-acquisition (sampling) constraints is a natural measure to contain such costs. For example, when the data of interest become available at multiple locations or sources, it may be practical and economically efficient to take observations from only a small fraction of these locations. There are many areas in science and engineering where these types of scenarios arise, e.g., multi-sensor networks, surveillance systems, cyber security, and power grids. Such sampling constraints have been considered in the sequential anomaly detection/identification problem \cite{CohenZhao2015,HuangCohenZhao2019,TsopFell20,Draga96,zigan66}, where the processes of interest are assumed to be statistically independent, and the goal is to identify the anomalous ones. Similar constraints are imposed in \cite{HeyTaj16} and \cite{HeyTaj17}, where only one source is sampled at each time instant, and observations from different sources are assumed to exhibit temporal dependence. 
Furthermore, such sampling constraints can be embedded into the  general framework of ``sequential design of experiments'' 
(see, e.g., \cite{Cher59}~and~\cite{NitAtiaVeer2013}).

A related problem to sequential testing is that of sequential change detection, where the goal is to detect as quickly as possible a change in the distribution of the underlying process \cite{OlympiaPoorBook,Tart14}. The problem of sequential change detection with controlled sensing, that is, in the presence of a set of actions that can influence the distribution of the observations, has been considered in   \cite{zhang-mei2020banditQCD_journal} and 
\cite{veeravalli2023quickest}. The concrete setup where there are multiple independent sources of observations, the change affects the distributions of an unknown subset of them, and it is possible to sample only a fixed number of sources at each time instant has been considered in  \cite{XuMeiMous21, XuMeiPostUncert, XuMeiPostUncert_SeqAn, XuMei2streams, LiMeiShi15, HeyTaj17_2}. In the first four references, the sampled processes are statistically independent,  whereas in the latter, the change can affect the dependence structure of a subset of these processes. 
Some other relevant works in the sequential literature involving inference regarding the observed processes' dependence structure are \cite{ChauFell20, chaudhuri2022joint, Xie2020, LaiZhang01}.

In this paper, we assume that there are $K$ sources generating data that are independent over time but not necessarily across sources.  That is,  
the samples from two different sources, generated at the same time, are not necessarily independent.  
We assume that the samples initially follow a completely specified joint distribution until an unknown time instant at which the joint distribution of an unknown subset of sources changes. The post-change regime is not necessarily completely specified, but a finite set of post-change alternatives is postulated.   The objective is to sequentially detect this change as quickly as possible while controlling the false alarm rate, 
under the constraint that only $m$ sources can be sampled at each time instant, where $m$ is a user-specified integer smaller than $K$.

This problem can be embedded into the general framework of sequential change detection problems with controlled sensing, where the action at each time instant is the selection of the sampled sources. In this context,  an asymptotically optimal procedure can be obtained by employing a sufficiently frequent random exploration of the action space and using a sufficiently large memory of past observations \cite{veeravalli2023quickest}. Our focus in the present paper is on a computationally simple procedure that does not employ any randomization and does not require memory of any past observations.  According to it, a family of  size-$m$ subsets of sources is specified, and its members, to which we refer as units, are sampled in a round-robin manner. 
For each unit, a cumulative sum (CUSUM)-like statistic, which enjoys a recursive structure, is computed, and
if its value exceeds a pre-specified threshold, then 
an alarm indicating that the change has occurred is raised;  if its value becomes smaller or equal to 0, the next unit is sampled;  otherwise, the same unit continues being sampled.


The motivating question for this paper is the following: under what conditions, if any, is this computationally simple procedure also statistically efficient? To answer this and quantify this test's detection performance, we adopt a version of Lorden's criterion \cite{Lorden1971}, in which the worst-case conditional expected detection delay is considered with respect to the change-point and the data until the change only from the  \textit{affected} sources. Then, under certain conditions on the pre-and post-change distributions, we establish a non-asymptotic upper bound for the above policy with respect to this delay metric. Subsequently, we show that,  in certain cases,  this policy is first-order asymptotically
optimal,  as the false alarm rate goes to $0$,  within the class of all procedures that satisfy the user-specified sampling and false alarm constraints.  In particular, this occurs if (i)~the post-change distribution is completely specified for every unit, (ii)~either  a single unit is affected by the change or the signal-to-noise ratio is the same for  all affected units, and (iii)~the family of units includes the size-$m$ subset of sources that is the affected the most by the change.



We specialize this general asymptotic theory to two concrete setups. The first one is the problem of detecting a change in the marginal distributions of multiple, not necessarily independent
data sources.
In fact,  the policy that we study in this work is inspired by, and also generalizes, the ones considered in \cite{Draga96}, and more recently in 
\cite{XuMeiMous21, XuMeiPostUncert_SeqAn}. Specifically, it reduces to the policy presented in  \cite{XuMeiMous21}
when, in particular,  
(i)~it is possible to sample \emph{only one of them} at each time instant $(m=1)$, and (ii)~there is a completely specified post-change distribution for each source. In \cite{XuMeiMous21}, this scheme was shown to be second-order asymptotically optimal, i.e., to achieve up to a constant term the optimal worst-case delay that would be achievable were all sources sampled at all times, \textit{when only one source is affected by the change}.  On the contrary, we do not compare the policy's performance with that of the optimal policy in the full-sampling case, which is not attainable in general. Instead, we compare it 
with the optimal one in the family of policies that satisfy the same 
sampling constraints.  Thus,  we show that  the rule in 
\cite{XuMeiMous21} is \textit{first-order}
asymptotically optimal not only when a single source is affected by the change but also \textit{when any possible number of data sources experience the change} as long as the signal-to-noise ratio is the same in all of them. 


The formulation in the paper is not limited to the case that the change affects the marginal distributions of the sources.  In fact, one of the motivating problems for this work is the detection of a change in the dependence structure. Thus, the second concrete setup to which we apply the general theory is that of detecting a change in the correlation structure of multiple Gaussian information sources. 
The theoretical results in this setup are also illustrated in certain simulation studies, which provide further insights.

The remainder of our paper is organized as follows.
In Section~\ref{sec:formulate}, we formulate the problem. In Section~\ref{sec:asymptotic optimality}, we establish a general universal asymptotic lower bound on the worst-case conditional expected detection delay. In Section \ref{sec: proposed procedure}, we introduce and analyze the policy that we study in the present work. 
Next, we apply the previous results to the problems of (i)~detecting a change in the marginal distributions of multiple, not necessarily independent, sources in Section~\ref{ex:mean_change}, and (ii)~detecting a change in the correlation structure of Gaussian information sources in Section~\ref{sec:change_corr}. In Section \ref{sec:sim_study}, we present the results of three simulation studies, and in Section~\ref{sec:conclusion}, we provide concluding discussions. The proofs and the technical details are relegated to the appendices.

We collect some notations that are used throughout the paper. We denote the set of natural numbers by $\bN$, i.e., $\bN := \{1, 2, \dots\}$, and for any $n \in \bN$ we set $[n] := \{1, \dots, n\}$.
We denote the set of real numbers by $\bR$ and for any $a,b \in \bR$, we set 
 $a \vee b \equiv \max\{a, b\}$ and $a \wedge  b \equiv \min\{a, b\}$.   The cardinality of a finite set $A$ is represented by $|A|$. For a square matrix $M$, we denote its determinant by $\det(M)$.  For any $p \in \bN$, we denote by $N_p(\mu, \Sigma)$ the $p$-variate Gaussian distribution with mean vector $\mu$ and covariance matrix $\Sigma$, and omit the subscript when $p = 1$. We denote by $\mathds{1}$ the indicator function, and for any $k \in \bN$, $\boldsymbol{1}_k$ denotes the $k$-dimensional vector of all ones. Finally, for any sequences of positive numbers $(x_n)$ and $(y_n)$, we write   $x_n \sim y_n$ for  $\lim_n  (x_n/y_n) = 1$ and  $x_n \lesssim  y_n$ for  $\limsup_n (x_n/y_n) \leq 1.$

\section{Problem Formulation} \label{sec:formulate}

\subsection{Multi-channel change-point model}
We consider $K$ data sources or channels that generate sequences of random elements sequentially and independently over time. For each $n \in \bN$, we denote the random element whose value is generated by source $k \in [K]$ at time $n$ by $X^k_n$, and we set
$X_n := (X^1_n, \dots, X^K_n)$. 
We refer to the 
distribution of $X_n$ as the \emph{global} distribution at time $n$,   and we assume that it changes at an unknown, deterministic time $\nu \in \{0,1, \ldots,\}$. We denote the pre-change global distribution by  $\sfF$ and the post-change global distribution by $\sfG$. Thus, we have
\begin{align}
\label{globalproblem}
X_n \sim \begin{cases}
\sfF,\ &\text{for}\ n \leq \nu\\
\sfG, \ &\text{for}\ n > \nu
\end{cases}\ .
\end{align}
We assume that while the pre-change distribution, $\sfF$, is completely specified, this is not the case for the post-change distribution. To emphasize this, we define $\cG$ as the family that encompasses all plausible post-change global distributions. 
We emphasize that the change does not necessarily affect all sources, and the subset of sources whose joint distribution changes is \textit{unknown}.  To formalize this, for any $\e\subseteq [K]$  we denote the set of samples from the sources in $\e$ at  time $n \in \bN$ by $X_n^E$, i.e.,
$X^\e_n:=\{X^{k}_n: k \in \e\}$. Furthermore,
we denote the   distributions of $X_n^\e$ under $\sfF$  and  $\sfG \in \cG$ by  $\sfF^\e$  and $\sfG^\e$, respectively. We  refer to the  former  as 
the \emph{local} distribution of $\e$ before the change and to the latter as the \emph{local} distribution of $\e$ after the change under $\sfG$. Therefore,  when the global post-change distribution is  $\sfG \in \cG$,
a subset of sources  $\e \subseteq [K]$ experiences the change if and only if \textit{its local distributions under $\sfF$ and $\sfG$ are different}, that is, 
\begin{align}
\label{localproblem1_1}
\text{if} & \quad \sfG^\e = \sfF^\e, \quad \text{then} \quad X_n^\e \sim \sfF^\e, \ \ \text{for every}\ n \in \bN\ ,\\
\label{localproblem1_2} \text{if} & \quad \sfG^\e \neq \sfF^\e, \quad \text{then} \quad
X_n^\e \sim \begin{cases}
\sfF^\e\ &\text{for}\ n \leq \nu\\
\sfG^\e \ &\text{for}\ n > \nu
\end{cases}\ .
\end{align}
 Finally,  for any  $\e\subseteq [K]$  we denote the family of all  plausible  post-change local distributions of $\e$ by  $\cG^\e$, i.e., 
\begin{align}\label{eq:G^E}
\cG^\e := \{\sfG^\e : \sfG \in \cG, \; \sfG^E \neq \sfF^E \}\ . 
\end{align}

\subsection{Sequential change detection with dynamic sampling control}
The problem we consider in this work is the detection of the change as quickly as possible under the hard constraint that  \emph{only $m$ of the $K$ sources can be sampled at any time instant}. That is, at each time $n \in \bN$ until stopping,  only $m$ coordinates of $X_n$ are observed. These coordinates, nevertheless, are selected dynamically, based on the already collected observations. Thus, at each time instant we have two intertwined tasks: (i)~to decide whether to terminate the sampling process declaring that the change has occurred (\emph{detection task}) or to continue sampling, and (ii)~when the decision is to continue sampling, to also select the sources to be sampled next (\emph{sampling task}).

To formalize these two tasks, we denote the family of subsets of $[K]$ with cardinality $m$ by $\cK_m$, i.e.,
\begin{align}
\cK_m &:= \{A \subseteq [K] :\;  |A| = m\}\ .
\end{align}
Subsequently, we define a \emph{sampling rule}  as a sequence  $S:= \{S_n :\;  n \in \bN\}$, where each   $S_{n}$ is a $\cK_m$-valued random variable that represents the set of the $m$ sources that we select to sample at time $n \in\bN$, and is a function of the collected data up to time $n-1$, i.e.,  $\{X_t^{S_t}:\; t \in [n-1]\}$.  We emphasize that  a sampling rule $S$ induces the  filtration of the collected observations, which can be  defined recursively as follows:
\begin{align}
\cF_{n}^S := \sigma\lt( \cF_{n-1}^S, X_n^{S_n}\rt), \quad  n \in \bN\ ,
\end{align}
where  $\cF_0^S$  is 
a $\sigma$-algebra 
independent of 
$\{X_n:n \in \bN\}$. The problem, subsequently, is to specify a sampling rule, $S$, and an $\{\cF^S_n\}$-measurable stopping time, $T$, at which it is declared that a change has occurred. That is,  $T$ must be a positive, integer-valued  random variable such that \ $\{T = n\}$$ \in \cF^S_n$ \ for every  $n \in \bN$.  
 We refer to such a pair $(S, T)$  as a \textit{joint sequential sampling rule and change-detection policy}, and we denote by $\Delta$ the family of all such policies.

\subsection{Change detection criteria}
We next introduce the metrics that we adopt in order to quantify the false alarm rate and the speed of detection of a policy. To this end,  we denote the underlying probability measure when there is no change by $\bP_{\infty}$, i.e., 
\begin{align}\bP_{\infty} = \sfF \times \sfF \times \dots\ ,
\end{align}
and denote the underlying probability measure  when the change happens at time $\nu$  and the  global post-change distribution is $\sfG$ by $\bP_{\nu}^{\sfG}$, i.e., 
\begin{align} 
\bP_{\nu}^{\sfG} &= \underset{\nu}{\underbrace{\sfF \times  \ldots \times \sfF}} \times \sfG  \times \sfG \times \ldots \ . 
\end{align}
Accordingly, we denote by $\bE_{\nu}^{\sfG}$ and $\bE_\infty$  the expectations under  $\bP_{\nu}^{\sfG}$ and  $\bP_{\infty}$, respectively. To control the false alarm rate, we focus on policies whose expected time until raising an alarm 
when there is no change is at least $\gamma$, where $\gamma>1$ is a user-specified level. Thus, such policies belong to 
\begin{align}
\Delta(\gamma) := \{(S, T) \in \Delta : \; \bE_{\infty}[T] \geq \gamma\}\ . 
\end{align}
To describe the delay in detecting the change,  we adopt a modified version of  Lorden's criterion \cite{Lorden1971}, as in \cite{XuMeiMous21}.  Specifically, for each 
 $\sfG \in \cG$  we denote by $\cA(\sfG)$
 the family of subsets of $[K]$ of size $m$ whose local distributions are affected when the  post-change  global distribution is $\sfG$, as described in~\eqref{globalproblem}, i.e., 
\begin{align}
\label{eq:A}
\cA(\sfG) := \{
\e \in \cK_m :\;  \sfG^\e \neq \sfF^\e\} \ .
\end{align}
Without loss of generality, we assume that this set is non-empty, i.e.,  $|\cA(\sfG)|\geq 1$ for every $\sfG \in \cG$.
Accordingly, for a policy $(S, T)$ we denote by  $D^{\sfG}(S, T)$  the worst-case  (with respect to the change-point) conditional expected detection delay given the worst possible history of observations up to the change-point \textit{from the sources whose distributions are affected by the change} when the post-change global distribution is $\sfG$.   Specifically, 
\begin{equation}\label{Det-Del}
D^{\sfG}(S,T) := \sup_{\nu \geq 0}\ \text{esssup}\ \bE^{\sfG}_\nu\left[T - \nu \, | \, \cF^S_\nu(\sfG), T > \nu \right]\, ,
\end{equation}
where  $\cF_\nu^S(\sfG)$ is the $\sigma$-algebra  generated by the sources in $\cA(\sfG)$, defined in \eqref{eq:A}, up to time $\nu$ when the sources are sampled according to the sampling rule $S$,  i.e.,
\begin{align}
\cF_\nu^S(\sfG) := \sigma\lt( S_t, X_t^{S_t} :  t \in [\nu], \;\; S_t \in \cA(\sfG)\rt)\ .
\end{align}


\subsection{Distributional assumptions}
Throughout the paper, we assume that for every   $\sfG \in \cG$ and $\e \in \cA(\sfG)$  
the local distributions $\sfG^\e$ and $\sfF^\e$ are mutually absolutely continuous, and we denote 
the  Kullback-Leibler (KL) divergence between $\sfG^\e$ and $\sfF^\e$ by $\cI^\e_0(\sfG)$, i.e.,  
\begin{align}\label{eq:defn_IG}
\cI^\e_0(\sfG) := \int \log \lt(\frac{d\sfG^\e }{d\sfF^\e}\rt)d\sfG^\e\ . 
\end{align}
We  refer to $\cI^\e_0(\sfG)$ as the \emph{information number} corresponding to $\e$ under $\sfG$, as it measures 
how much the local distribution of $\e$ gets affected when the global distribution changes from $\sfF$ to $\sfG$.

\section{Universal Asymptotic Lower Bound }\label{sec:asymptotic optimality}
In this  section, we  fix an arbitrary  $\sfG \in \cG$ and state an asymptotic, as $\gamma \to \infty$,   lower bound  on
\begin{equation} \label{infim}\inf_{(S, T) \in \Delta(\gamma)} D^{\sfG}(S,T)\ ,\end{equation}
which is the optimal,  among all policies in  $\Delta(\gamma)$, worst-case conditional expected detection delay when the global post-change distribution is $\sfG$. This lower bound is inversely proportional to the information number that corresponds to the subset in $\cA(\sfG)$, defined in \eqref{eq:A}, whose local distribution is altered the \text{most} under the change. The proof combines ideas and techniques from 
\cite{Cher59} and 
 \cite{Lai98}. As in the former,  we need to make the following second-moment assumption:
\begin{align}
\label{eq:assm1}
\int \left( \log \frac{d\sfG^E}{d\sfF^E} \right)^2 d\sfG^E< \infty \quad \forall \; E \in \cA(\sfG)\ .\\ \notag
\end{align}

\begin{theorem}[Universal asymptotic lower bound]\label{thm:lb}
If \eqref{eq:assm1} holds, then  as $\gamma \to \infty$ we have
\begin{align}\label{eq:lower_bound}
\inf_{(S, T) \in \Delta(\gamma)} D^{\sfG}(S,T) \geq \frac{\log \gamma}{ \max_{\e \in \cA(\sfG)} \cI^\e_0(\sfG)}(1 + o(1))\ .
\end{align}
\end{theorem}

\begin{IEEEproof} 
See Appendix \ref{pf:thm:lb}. \\
\end{IEEEproof}


\section{Round Robin Active Sequential Detection Policy} 
\label{sec: proposed procedure}
In this section,  we introduce the policy $(\tilde{S}, \tilde{T})$, and refer it as \textit{Round Robin CUSUM} due to the nature of its sampling rule. 
Under certain assumptions it
achieves the asymptotic lower bound of the previous section up to a multiplicative constant and, in certain cases, with equality. For this purpose, we first fix an arbitrary, non-empty subset of $\cK_m$, which we denote by $\cU$. We refer to the members of $\cU$ as \textit{units}.   For each  unit $\e \in \cU$, we assume that $\cG^\e$ defined in \eqref{eq:G^E}, i.e., the family of all  plausible post-change  local  distributions of $\e$, 
is non-empty and finite. We denote the mixture over all its members by $\sfH^\e$, i.e.,
\begin{align}\label{eq:P_mix}
\sfH^\e := \frac{1}{|\cG^\e|}\sum_{\sfP \in \cG^\e}\sfP\ , 
\end{align}
and for every $n \in \bN$ we denote the log-likelihood ratio of $\sfH^\e$ versus $\sfF^\e$ based on $X_n^\e$ by  $\xi_n^\e$ , i.e.,
\begin{align}\label{eq:eta_mix}
\xi_n^\e := &\log \frac{d\sfH^\e}{d\sfF^\e} (X_n^\e)\ .
\end{align}

\subsection{Description of the policy}
First,  we fix an arbitrary permutation of the units, $\e_1, \ldots, \e_{|\cU|}$, and a constant $A>0$. 
The first unit 
in the above permutation is sampled first, i.e.,  $\tilde{S}_1 = \e_1$. Then, for each time $n \in \bN$ up to stopping,  the following CUSUM-like statistic is computed,
\begin{align}
Y_n = \max\left\{Y_{n-1}, 0\right\} + \xi_n^{\tilde{S}_n}\ , \quad \text{where}  \qquad Y_0 = 0\ .
\end{align}
Based on the value of $Y_n$ we distinguish the following cases: 
\begin{enumerate}
\item If $Y_n \in (0, A)$, then the  same unit continues being sampled at the next time instant, i.e.,  
$
\tilde{S}_{n+1} = \tilde{S}_n$. 
\item If $Y_n \leq 0$, then the  next unit in the chosen permutation is  sampled at the next time instant,  i.e., 
\begin{align}
\tilde{S}_{n+1} = 
\begin{cases}
\e_{d+1}  &\text{if} \; \;  \tilde{S}_n=\e_{d} \;\; \text{and} \; \; d < |\cU|\\
\e_1  &\text{if}  \; \; \tilde{S}_n=\e_{|\cU|}
\end{cases}\ .
\end{align}
\item If $Y_n \geq A$, then an alarm is raised and sampling is terminated.
\end{enumerate}
Thus, the  stopping time of this procedure is
\begin{align}
\tilde{T} :=  \inf\left\{n \in\bN :\; Y_n \geq A\right\}\ .
\end{align}


\begin{algorithm}[H]
\small 
\caption{Round Robin CUSUM Policy $(\tilde{S}, \tilde{T})$ }\label{alg:proposed}
\begin{algorithmic}[1]
\STATE {\text{Fix an arbitrary permutation}}  $\left(\e_1, \ldots, \e_{|\cU|}\right)$\ 
\STATE {\text{Set}} \; $A > 0$\ 
\STATE {\text{Set}} \;  $Y \gets 0$\ 
\STATE {\text{Set}} \;  $\xi \gets 0$\ 
\STATE {\text{Set}} \;  $d \gets 1$\ 
\STATE {\text{Set}} \;  $n \gets 0$\ 

\STATE \hspace{0.5cm}$\textbf{while } \; Y < A \; \textbf{ do}$ 
\STATE \hspace{1cm}$ n \gets n + 1$
\STATE \hspace{1cm}$\tilde{S} \gets E_d$\ 
\STATE \hspace{1cm}$ \xi \gets \log\frac{d\sfH^E }{d\sfF^E}(X_n^{\tilde{S}})$
\STATE \hspace{1cm}$ Y \gets \max\left\{Y, 0\right\} + \xi$

\STATE \hspace{1cm}$ \textbf{if } \; Y < 0 \; \textbf{ then}$
\STATE \hspace{1.5cm} $d \gets (d+1) \; \; {\rm mod} \; |\cU|$
\STATE \hspace{1cm}$ \textbf{end if }$

\STATE \hspace{0.5cm}\textbf{return} \; $\tilde{T} = n$
\end{algorithmic}
\end{algorithm}

\subsection{Performance Analysis}
We start analyzing the performance of $(\tilde{S}, \tilde{T})$ by first establishing its false alarm control. \\
\begin{theorem}[False alarm control]\label{thm:fa_extn} 
For every $\gamma>1$ we have   $(\tilde{S}, \tilde{T}) \in \Delta(\gamma)$  
when $A = \log\gamma$.
\end{theorem}

\begin{IEEEproof}
See  Appendix \ref{app:extn_ub}.\\
\end{IEEEproof}

Next, we establish an asymptotic, as $A \to \infty$,   upper bound on the worst-case conditional expected detection delay of  $(\tilde{S}, \tilde{T})$ under an arbitrary post-change global distribution $\sfG \in \cG$. For this, at  least one unit must be  affected by the change, i.e., 
\begin{align}\label{assum:pos_I_mix_1}
\cA(\sfG) \cap \cU  \neq \emptyset\ .
\end{align}
Furthermore, for every unit $E$, the statistic $\xi_n^E$, defined in \eqref{eq:eta_mix}, needs to have negative and finite drift when $n \leq \nu$, i.e., before the change. That is,  \begin{align}
\cJ^\e_\infty &:= \Exp_\infty[ \xi_1^\e]= \int \log \left( \frac{d\sfF^\e}{d\sfH^\e} \right) d\sfF^E \in (0, \infty) \ ,  \qquad  \text{for every} \quad \e \in \cU\ .
\label{assum:pos_I_mix_1_b}
\end{align}

Finally,  for every \textit{affected} unit $E$, $\xi_n^E$  needs to have positive and finite drift when $n >\nu$, i.e.,  after the change. That is,
\begin{align}\label{assum:pos_I_mix_1_a}
\cJ^\e_0(\sfG) &:= 
\Exp_0^{\sfG}[ \xi_1^\e]=
\int \log \left( \frac{d\sfH^\e}{d\sfF^\e} \right) d\sfG^E \in (0, \infty) \ , \qquad  \text{for every} \quad \e \in \cA(\sfG) \cap \cU\ . 
\end{align}
The following proposition and its corollary provide sufficient conditions for the validity of  \eqref{assum:pos_I_mix_1_b}-\eqref{assum:pos_I_mix_1_a}.

\begin{proposition}\label{prop:on_assum_mix_0}
Assumptions \eqref{assum:pos_I_mix_1_b}-\eqref{assum:pos_I_mix_1_a} hold if the following conditions are satisfied:
\begin{itemize}
\item[(i)] 
for every $ E \in \cU$, \ $\sfH^E \neq \sfF^E $ and $\cJ_\infty^E<\infty$; and 
\item[(ii)] for   every $E\in \cA(\sfG) \cap \cU$, the expectation of $\xi_1^\e$ is finite and the same under any  $\sfP \in \cG^\e$.
\end{itemize}
\end{proposition}

\begin{IEEEproof}
When condition (i) is satisfied, assumption \eqref{assum:pos_I_mix_1_b} holds directly by the definition of $\cJ^\e_\infty$  in \eqref{assum:pos_I_mix_1_b}, and the positivity of Kullback-Leibler divergence when the two distributions are different.  Next,
 for every $\e \in \cA(\sfG) \cap \cU$ we have
\begin{align}
\cJ^\e_0(\sfG)
& = \int\log \left( \frac{d\sfH^\e}{d\sfF^\e} \right) \, d\sfG^\e \\
\label{eq:J_0_a} &= \frac{1}{|\cG^\e|} \sum_{\sfP \in \cG^\e } \int\log \left(\frac{d\sfH^\e}{d\sfF^\e} \right) \, d\sfP\\
 &= \int\log \left(\frac{d\sfH^\e}{d\sfF^\e} \right) \, d  \left( \frac{1}{|\cG^\e|} \sum_{\sfP \in \cG^\e } \sfP \right) \\
&= \int \log \left( \frac{d\sfH^\e}{d\sfF^\e} \right) \, d\sfH^\e > 0\ .
\end{align}
The second equality holds because $\sfG \in \cG$, and by condition (ii), it follows that all terms of the sum are equal. The third one holds by the additivity of the integral, and the last one holds due to the definition of $\sfH^E$ in   \eqref{eq:P_mix}. The inequality, i.e., the positivity of $\cJ^\e_0(\sfG)$, holds by the positivity of Kullback-Leibler divergence when the two distributions are different and condition (i),  which states that $\sfH^\e \neq \sfF^\e$. Finally, the finiteness of $\cJ^\e_0(\sfG)$ follows from the finiteness of the expectations in condition (ii). \\
\end{IEEEproof}

\begin{corollary}
\label{rem:on_assum_mix_0}
Assumptions \eqref{assum:pos_I_mix_1_b}-\eqref{assum:pos_I_mix_1_a} hold if the following conditions are satisfied:
\begin{itemize}
\item[(i)]  for every $ E \in \cU$, \ $\cG^\e$ is a singleton,
and 
$\cJ_\infty^E<\infty$,
\item[(ii)] for   every $E\in \cA(\sfG) \cap \cU$, \
$\cI_0^E(\sfG)< \infty$.
\end{itemize}
\end{corollary}

\begin{IEEEproof}
When condition (i) is satisfied,  by the definitions of $\cG^\e$ in \eqref{eq:G^E} and $\sfH^\e$ in \eqref{eq:P_mix} we have $\sfH^\e \neq \sfF^\e$ for every $\e \in \cU$.  This implies that condition (i) in Proposition \ref{prop:on_assum_mix_0} holds. Furthermore, for every $\e \in \cA(\sfG) \cap \cU$, since $\cG^\e$ is a singleton, we clearly have $\cG^\e = \{\sfG^\e\}$.
This, according to \eqref{eq:P_mix}, implies that $\sfH^\e = \sfG^\e$, and hence, under condition (ii),
\begin{align}\label{eq:I=J}
\cJ^\e_0(\sfG) = \int \log \left( \frac{d\sfH^\e}{d\sfF^\e} \right) d\sfG^\e = \int \log \left(\frac{d\sfG^\e}{d\sfF^\e} \right) d\sfG^\e = \cI^\e_0(\sfG) < \infty\ .
\end{align}
Therefore, condition (ii) in Proposition \ref{prop:on_assum_mix_0} holds. The proof is complete.\\
\end{IEEEproof}

We next state  the asymptotic  upper bound on the worst-case conditional expected  detection delay of 
 $(\tilde{S}, \tilde{T})$ as $A \to \infty$. \\

\begin{theorem}[Asymptotic upper bound on detection delay] \label{thm:ub_extn}
If  \eqref{assum:pos_I_mix_1}-\eqref{assum:pos_I_mix_1_a} hold, then,  as $A \to \infty$, we have 
\begin{align}\label{eq:ub_general}
D^{\sfG}(\tilde{S}, \tilde{T}) \leq  
\max_{\e \in \cA(\sfG) \cap \cU} \; \frac{A}{\cJ^\e_0(\sfG)} + o(A)
\ .
\end{align} 
 \end{theorem}

\begin{IEEEproof}
See  Appendix \ref{pf:thm:ub_extn}.\\
\end{IEEEproof}

In what follows, for simplicity, we set $A=\log \gamma$. Theorem \ref{thm:fa_extn} states that this choice ensures $(\tilde{S}, \tilde{T}) \in \Delta(\gamma)$.

\subsection{Asymptotic optimality}
As before, we fix an arbitrary post-change global distribution $\sfG \in \cG$. We define the  \textit{asymptotic
relative efficiency}  of $(\tilde{S}, \tilde{T})$ as
\begin{align}\label{eq:ARE}
{\sf ARE}^{\sfG}(\tilde{S}, \tilde{T}) := \limsup_{\gamma \to \infty} \; \frac{D^{\sfG}(\tilde{S}, \tilde{T})}{\inf \{D^{\sfG}(S, T) : (S, T) \in \Delta(\gamma)\}} \ , 
\end{align}
and we say that   
$(\tilde{S}, \tilde{T})$ is \textit{asymptotically optimal under $\sfG$} if,  as $\gamma \to \infty$, we have 
\begin{align}
D^{\sfG}(\tilde{S}, \tilde{T}) \sim \inf \{D^{\sfG}(S, T) : (S, T) \in \Delta(\gamma)\}\ .
\end{align}
Based on Theorem \ref{thm:ub_extn}, and the universal asymptotic lower bound in Theorem \ref{thm:lb}, we next obtain an upper bound on the asymptotic relative efficiency of 
$(\tilde{S}, \tilde{T})$, and establish sufficient conditions for its asymptotic optimality. \\

\begin{theorem}
\label{cor:ARE_bd} 
Suppose \eqref{eq:assm1} holds. 
\begin{itemize}
\item[(i)]
If \eqref{assum:pos_I_mix_1}-\eqref{assum:pos_I_mix_1_a}
hold, then 
\begin{align}\label{eq:ARE_bd}
{\sf ARE}^{\sfG}(\tilde{S}, \tilde{T}) 
\leq \frac{\max \{\cI^\e_0(\sfG): \e \in \cA(\sfG)\} }{\min\{\cJ^\e_0(\sfG): \e \in \cA(\sfG)\cap \cU\}}\ .
\end{align}
\item[(ii)] If, also,  $\cG^\e$ is a singleton  for every $\e \in \cA(\sfG) \cap \cU$,  and
\begin{align}
\label{cond_opt}
\max_{\e \in \cA(\sfG)} \; \cI^\e_0(\sfG) = \min_{\e \in \cA(\sfG)\cap \cU} \; \cI^\e_0(\sfG)\ ,
\end{align}
 then $(\tilde{S}, \tilde{T})$ is  asymptotically optimal under $\sfG$.
 \end{itemize}
\end{theorem}

\begin{IEEEproof}
(i) This follows directly from 
the asymptotic upper bound \eqref{eq:ub_general} in Theorem~\ref{thm:ub_extn}, and the asymptotic lower bound~\eqref{eq:lower_bound} in Theorem \ref{thm:lb}. (ii) 
If   $\cG^\e$ is a singleton  for every $\e \in \cA(\sfG) \cap \cU$, then  $
\cJ^\e_0(\sfG) =\cI^\e_0(\sfG)$
for every $\e \in \cA(\sfG) \cap \cU$ (recall \eqref{eq:I=J}), and the  upper bound  in \eqref{eq:ARE_bd} is equal to 1 if and only if \eqref{cond_opt} holds. \\
\end{IEEEproof}

\begin{remark}\label{rem:foranyU}
Given that there is at least one unit that is affected by the change, i.e., \eqref{assum:pos_I_mix_1} holds, 
condition \eqref{cond_opt} clearly holds when $\cI^\e_0(\sfG)$ is the same (and finite) for every $\e \in \cA(\sfG)$. This is, trivially, the case  when the local distribution changes in only one element of $\cK_m$, i.e., $|\cA(\sfG)| = 1$.
\\
\end{remark}

\begin{corollary}
\label{corr:homogU}
Suppose there is at least one unit affected by the change, i.e., 
\eqref{assum:pos_I_mix_1} holds, and for every unit there are unique pre-change and post-change local distributions $\sfF^*$ and $\sfG^*$ respectively, i.e., 
\begin{align}\label{eq:homogU}
\begin{split}
\sfF^\e &:= \sfF^*\ , \qquad  \text{for every} \quad \e \in \cU\ ,\\
\sfG^\e &:= \sfG^* \ ,  \qquad  \text{for every} \quad \e \in \cA(\sfG) \cap \cU\ ,
\end{split}
\end{align}
such that 
\begin{align}
\label{eq:assm_FG^*}
\int \log \left( \frac{d\sfG^*}{d\sfF^*} \right) d\sfG^* &< \infty \   \qquad  \text{and} \qquad  \int \log \left( \frac{d\sfF^*}{d\sfG^*} \right) d\sfF^* < \infty\ .
\end{align}
Then, conditions 
\eqref{assum:pos_I_mix_1_b}-\eqref{assum:pos_I_mix_1_a} are satisfied. Additionally, if \eqref{eq:assm1} also holds, and either $\cU=\cK_m$ or, more generally,
\begin{align}\label{eq:max=I}
\max_{\e \in \cA(\sfG)} \cI^\e_0(\sfG) = \int \log \left( \frac{d\sfG^*}{d\sfF^*} \right) d\sfG^*\ , 
\end{align}
then
$(\tilde{S}, \tilde{T})$ is  asymptotically optimal under $\sfG$.
\end{corollary}

\begin{IEEEproof}
For every $\e \in \cU$ we have 
 $\cG^\e = \{\sfG^*\}$. Furthermore, when \eqref{eq:assm_FG^*} holds, we have 
\begin{align}
\cJ^\e_\infty &= \int \log \left( \frac{d\sfF^*}{d\sfG^*} \right) d\sfF^* < \infty  \qquad  \text{for every} \quad \e \in \cU\ ,  \\
\cI^\e_0(\sfG) &= \int \log \left( \frac{d\sfG^*}{d\sfF^*} \right) d\sfG^* < \infty \qquad  \text{for every} \quad \e \in \cA(\sfG) \cap \cU\ . 
\end{align}
Thus, following Corollary \ref{rem:on_assum_mix_0},  conditions \eqref{assum:pos_I_mix_1_b}-\eqref{assum:pos_I_mix_1_a} are satisfied. Furthermore,  for every $\e \in \cA(\sfG) \cap \cU$  we have $\cG^\e = \{\sfG^*\}$, i.e.,  $\cG^\e$ is a singleton, and
\begin{align}
\cI^\e_0(\sfG) =\int \log \left( \frac{d\sfG^*}{d\sfF^*} \right) d\sfG^*\ .
\end{align}
Thus, condition \eqref{cond_opt} holds when \eqref{eq:max=I} holds, which is clearly always the case when $\cU = \cK_m$.
\end{IEEEproof}



\subsection{A non-asymptotic upper bound}
Under slightly stronger moment assumptions,  we can establish an explicit, non-asymptotic upper bound for the worst-case detection delay of the Round Robin CUSUM policy. 
To be specific, for every unit $\e \in \cU$ we introduce the first descending and ascending ladder variables of the random walk with increments $\{\xi_t^E: t \in \bN\}$ as follows. 
\begin{align}\label{eq:des_asc}
\zeta^\e_{-} := \inf\left\{n \geq 1 : \sum_{t = 1}^{n} \xi_t^\e < 0\right\} \ \qquad \text{and} \qquad \
\zeta^\e_+ := \inf\left\{n \geq 1 : \sum_{t = 1}^{n} \xi_t^\e > 0\right\}\ .
\end{align}

When $\e \notin \cA(\sfG)$, we denote by  $q^\e_+$ 
the probability under $\sfF$ that
this random walk 
never exceeds  $0$, i.e.,
\begin{align}
q^\e_+ := \sfF(\zeta^\e_+ = \infty) \ .
\end{align} 

When $\e \in \cA(\sfG)$, we denote  by $q^\e_-(\sfG)
$ the probability under $\sfG$ that 
this random walk never falls below 0, i.e., 
\begin{align}
q^\e_-(\sfG) := \sfG(\zeta^\e_- = \infty)\ .
\end{align}

We note that under assumptions \eqref{assum:pos_I_mix_1_b}-\eqref{assum:pos_I_mix_1_a}, both these probabilities are strictly positive (see, e.g.,\cite[Corollary 8.39]{BookSeig}, and  Lemma \ref{lem:mu/p} in Appendix \ref{app_subsec:implems}). \\

\begin{proposition}
\label{rem:ub_extn_nas}
Suppose  \eqref{assum:pos_I_mix_1}
 -\eqref{assum:pos_I_mix_1_a} hold. If we have 
\begin{align}\label{eq:assm_2nd_mom_extn}
 \cW^\e_0(\sfG) &:= \int \lt(\xi_1^E - \cJ^\e_0(\sfG)\rt)^2d\sfG
 < \infty  \qquad  \text{for every} \quad \e \in \cA(\sfG) \cap \cU\ ,
\end{align}
then 
\eqref{eq:ub_general} holds
for every large enough $A > 0$, with $o(A)$ replaced by 
\begin{align}\label{o(A)_replaced_extn}
\begin{split}
& \frac{1}{1 - \prod_{\e \in  \cA(\sfG) \cap \cU} (1 - q^\e_-(\sfG))} \, \sum_{\e \in \cU \setminus \cA(\sfG)} \frac{1}{q^\e_+} \; + \; \max_{\e \in \cA(\sfG) \cap \cU} \frac{1}{q^\e_-(\sfG)}\lt(1 + \frac{\cW^\e_0(\sfG)}{\cJ^\e_0(\sfG)^2}\rt) + \; C\ ,
\end{split}
\end{align}
where $C > 0$ is a finite constant independent of $A$.
\end{proposition}
\begin{IEEEproof}
See Appendix \ref{pf:rem:ub_extn_nas}.    \\
\end{IEEEproof}

\begin{remark}\label{rem:diffU}
A further upper bound on  \eqref{o(A)_replaced_extn} takes the form
\begin{equation}
\label{o(A)_replaced_extn2}
\text{constant} \; + \; \frac{1}{p_+} \cdot \frac{|\cU \setminus \cA(\sfG)|}{\left(1 - \left(1 - p_{-}\right)^{|\cA(\sfG) \cap \cU|}\right)}\ , 
\end{equation}
where ``constant'' refers to a term that does not depend on $\gamma$, and 
\begin{align}
p_- := \min_{\e \in \cA(\sfG) \cap \cU} q^\e_-(\sfG) \  \qquad \text{and} \qquad \ p_+ := \min_{\e \in \cU} q_+^\e \ . 
\end{align}
\end{remark}

\begin{remark}\label{rem:alt_U}
As implied by Corollary \ref{corr:homogU}, and we will see in some examples that follow, $(\tilde{S}, \tilde{T})$ may  be asymptotically optimal  even with a family of units,
$\cU$, that is smaller than $\cK_m$. 
The upper bounds presented in Proposition
\ref{rem:ub_extn_nas} as well as in Remark \ref{rem:diffU},
in which the second-order term grows with the number of units, suggest that the smaller family of units may not only be more computationally efficient but also preferable in terms of its detection performance. 
\end{remark}



\section{Change in the  Marginal Distributions of the Sources}\label{ex:mean_change}

In this section,  we illustrate the model, algorithm, and theoretical results of the previous sections in the problem of detecting a change in the marginal distributions of the sources.  Specifically, we assume that at an unknown time $\nu$  the global distribution changes from $\sfF$ to some $\sfG$ so that the marginal distribution of at least one source changes. 
Since it is possible to detect this change even when sampling one source at a time, we focus on the case that $m=1$. Furthermore, we do not assume any prior information regarding the subset of sources that are affected by the change. Thus,  the family of units, $\cU$, has to be selected as the family of all singletons of $[K]$, i.e.,   $\cK_1= \{ \{k\}: k \in [K]\}$. Note also that we do not assume independence among the sources,  in fact, the dependence structure may also change after $\nu$.  To simplify the notation, we use $k$ instead of 
$\{k\}$ as superscript in the local distributions. Hence, we have 
\begin{align}\label{globalproblem_Gaussian_mean_ext}
X^k_n \sim \left\{
\begin{array}{lll}
\sfF^k\  &\qquad &\text{for}\ n \leq \nu, \quad \; k \in [K] \\
\sfF^k\  & \;\;\qquad  &\text{for}\ n > \nu, \quad 
\{k\} \notin \cA(\sfG)\\
\sfG^k\ & \;\;\qquad &\text{for}\ n > \nu, \quad \{k\} \in \cA(\sfG) \\
\end{array}\right. \ .
\end{align}
Furthermore,  for every $k \in [K]$ and  $n \in \bN$ we set 
\begin{align}\label{eq:Hk}
\sfH^k = \frac{1}{|\cG^k|} \sum_{\sfP \in \cG^k } \sfP\ \  \qquad \text{and} \qquad \ \xi_n^k := &\log \frac{d\sfH^k}{d\sfF^k} (X_n^k)\ ,
\end{align}
where $\cG^k$ is the family of all possible post-change  local  distributions of source $k$.    Then,  the second-moment assumption \eqref{eq:assm1} is equivalent to 
\begin{align}\label{assum:wrtFG2}
\int \lt(\log \left(\frac{d\sfG^k}{d\sfF^k} \right) \rt)^2d\sfG^k < \infty \ , \qquad  \text{for every} \quad \{k\} \in \cA(\sfG)\ , 
\end{align}
and assumptions \eqref{assum:pos_I_mix_1_b}-\eqref{assum:pos_I_mix_1_a} are equivalent to 
\begin{align}\label{assum:wrtFG_a}
\cJ^k_\infty &= \int \log \left( \frac{d\sfF^k}{d\sfH^k} \right) d\sfF^k \in (0, \infty) \ ,  \qquad  \text{for every} \quad k \in [K]\ ,  \\
\cJ^k_0(\sfG) &= \int \log \left( \frac{d\sfH^k}{d\sfF^k} \right) d\sfG^k \in (0, \infty) \ , \qquad  \text{for every} \quad \{k\} \in \cA(\sfG)\ .
\label{assum:wrtFG_b}
\end{align}

The following proposition specializes Theorem \ref{cor:ARE_bd} to
the setting of this section. \\

\begin{proposition}\label{prop:ARE_marg}
Suppose that  \eqref{assum:wrtFG2} holds 
for every $k \in [K]$.
\begin{itemize}
\item[(i)] If   $\sfH^k \neq \sfF^k$ and $\cJ^k_\infty < \infty$ for every $k \in [K]$, 
and  the expectation of $\xi_1^k$ is finite and has the same value for  every $\sfP \in \cG^k$ and $\{k\} \in \cA(\sfG)$, 
then  
\begin{align}\label{eq:AREbd_FG}
 {\sf ARE}^{\sfG}(\tilde{S}, \tilde{T}) &\leq \frac{\max \{\cI^k_0(\sfG): \{k\} \in \cA(\sfG)\} }{\min \{ \cJ^k_0(\sfG): 
 \{k\} \in \cA(\sfG) \} }\ .
\end{align}
\item[(ii)] If, also, there is exactly one possible post-change marginal distribution  for every affected source, i.e., $\cG^k$ is a singleton for every $k \in [K]$, and $\cI^k_0(\sfG)$ has the same value for every $ \{k\} \in \cA(\sfG)$, 
then 
$(\tilde{S}, \tilde{T})$ is asymptotically optimal  under $\sfG$.
\end{itemize}
\end{proposition}

\begin{IEEEproof}
(i) Due to \eqref{assum:wrtFG2}, assumption \eqref{eq:assm1} is satisfied, and under the conditions stated above, assumptions \eqref{assum:wrtFG_a}-\eqref{assum:wrtFG_b}, or equivalently, \eqref{assum:pos_I_mix_1_b}-\eqref{assum:pos_I_mix_1_a} are also satisfied by  Proposition \ref{prop:on_assum_mix_0}. 
Furthermore, since $\cU = \cK_1$, we have $\cA(\sfG) \cap \cU = \cA(\sfG)$, and thus, \eqref{eq:AREbd_FG} follows from \eqref{eq:ARE_bd}. (ii) It suffices to check the conditions of  Corollary \ref{rem:on_assum_mix_0}. Indeed, condition (i) in Corollary \ref{rem:on_assum_mix_0} is satisfied when $\cG^k$ is a singleton for every $k \in [K]$. 
Moreover, when the quantities $\{\cI^k_0(\sfG) : \{k\} \in \cA(\sfG)\}$ are identical, by  Remark \ref{rem:foranyU} it follows that condition (ii) in Corollary \ref{rem:on_assum_mix_0}  is also satisfied. 

\end{IEEEproof}

\begin{remark}
The setup of Proposition  \ref{prop:ARE_marg}(ii)  coincides with the one in \cite{ XuMeiMous21}  with the differences that we allow   (i)  the data sources to be dependent and
(ii) any  number of data sources to be affected by the change. In
\cite{XuMeiMous21}, a second-order asymptotic optimality property is established when only a single
data source can be affected by the change. Proposition \ref{prop:ARE_marg} shows that the same scheme  is asymptotically optimal also  when an arbitrary, unknown number of 
data sources can be affected by the change, as long as the magnitude of the change is the same in all affected sources. This is, in particular,  the case  in the context of the following proposition,
where there is a common pre-change and a common post-change marginal distribution for all sources.\\
\end{remark}

\begin{proposition} 
If there are $\sfF^*$ and $\sfG^*$ such that
\begin{align}
    \sfF^k &= \sfF^*, \quad \text{for every} \; k \in [K]\ ,  
    \\
\sfG^k &= \sfG^*, \quad \text{for every} \; \{k\} \in \cA(\sfG), \
\sfG \in \cG\ , 
    \end{align}
and \eqref{eq:assm_FG^*} and \eqref{assum:wrtFG2} hold, 
then $(\tilde{S}, \tilde{T})$
is asymptotically optimal under every $\sfG \in \cG$. 
\end{proposition}

\begin{IEEEproof}
It suffices to check the conditions of  Corollary \ref{corr:homogU}.
  Conditions \eqref{assum:pos_I_mix_1} and \eqref{eq:homogU} are satisfied for every $\sfG \in \cG$ when $\cU = \cK_1$.  
Again, \eqref{eq:max=I} holds since $\cU = \cK_1$, and  condition \eqref{eq:assm1} is satisfied  due to \eqref{assum:wrtFG2}.
\end{IEEEproof}
Next, we illustrate the above results with two concrete examples that involve a change in the mean of a Gaussian sequence. In the first one, the post-change local distribution of each affected source is completely specified, whereas in the second the mean of each source may either increase or decrease.   \\

\begin{example}
Suppose that   for every $k \in [K]$ we have 
$\sfF^k = N(0, 1)$ and 
$\cG^k = \{N(\mu_k, 1)\}$, 
where $\mu_k \in \bR$.
Then, condition  \eqref{assum:wrtFG2}
clearly holds, and conditions  
\eqref{assum:wrtFG_a}-\eqref{assum:wrtFG_b} are satisfied with 
\begin{align}
\cJ^k_\infty(\sfG)=\cJ^k_0(\sfG) = \cI^k_0(\sfG) = \frac{1}{2} \, \mu_k^2\ .
\end{align}
Therefore, following Proposition \ref{prop:ARE_marg} we have
\begin{align}
{\sf ARE}^{\sfG}(\tilde{S}, \tilde{T}) \leq 
\frac{\max_{\{k\} \in \cA(\sfG)} \; \mu_k^2}{\min_{\{k\} \in \cA(\sfG)} \; \mu_k^2}\ ,
\end{align}
and 
$(\tilde{S}, \tilde{T})$ is   asymptotically optimal if $|\mu_k|$ is the same for every $\{k\} \in \cA(\sfG)$,  
i.e.,  the absolute magnitude of the post-change mean is the same for every affected source. Clearly, this is always the case when $\cA(\sfG)$ is a singleton, i.e., only a single source is affected by the change, or when $|\mu_k|$ is the same for every $k \in [K]$. \\

\end{example}

\begin{example}
Suppose that for every  $k \in [K]$ we have $\sfF^k = N(0, 1)$ and 
$\cG^k = \{N(- \delta_k, 1), \;  N( \delta_k, 1)\}
$ for some $\delta_k>0$. 
This clearly implies that $\sfH^k \neq \sfF^k$  for every  $k \in [K]$, and  that \eqref{assum:wrtFG2} holds.
Furthermore,  for every $k \in [K]$ we have 
\begin{align}
\xi_1^k & = \log \; \bigg\{ \frac{1}{2} \, \exp\lt(+\frac{\delta_k}{2}\lt(2  X_1^k - \delta_k\rt)\rt)+ \frac{1}{2} \, \exp\lt(-\frac{\delta_k}{2}\lt(2 X_1^k + \delta_k\rt)\rt) \bigg\}\ .
\end{align}
Since $\xi^k_1$ is symmetric with respect to $+\delta_k$ and $-\delta_k$,  it is identically distributed under $N(- \delta_k, 1)$ and $N(\delta_k, 1)$.   By  Jensen's inequality, it follows that the expectations of $\xi_1^k$ under $\sfF^k$ and any  $\sfP$ in $\cG^k$ are finite. Therefore, following Proposition \ref{prop:on_assum_mix_0}, assumptions  \eqref{assum:wrtFG_a}-\eqref{assum:wrtFG_b} hold. 


\end{example}

\section{Change in Correlation}\label{sec:change_corr}

In this section, we illustrate the models, algorithms, and theoretical results of the previous sections
in the problem of detecting a change in the correlation structure of sources that follow
a multivariate Gaussian distribution. For simplicity, we assume that 
the sources are initially independent, each with distribution $N(0,1)$. After  $\nu$, 
the marginal distributions still remain the same, i.e., $N(0,1)$, while an unknown subset of sources become dependent.  For each pair of sources that is affected by the change, we assume that the correlation coefficient takes a value in a finite set $\cR\subset(0,1)$.
Hence, we have 
\begin{align}\label{globalproblem_Gaussian_covariance}
X_n \sim \left\{
\begin{array}{lll}
N_K( \boldsymbol{0},  \, I) \ ,  &\text{for}\ n \leq \nu \\
N_K( \boldsymbol{0},  R)\ ,    &\text{for}\ n > \nu 
\end{array}\right. \ ,
\end{align}
where $R\in\mathbb{R}^{K\times K}$ is a correlation matrix such that  
\begin{align}\label{eq:R}
\begin{split}
\; &R_{ii}= 1 \, \quad \forall \; i \in [K]\ , \\
\; &R_{ij} \in \cR \cup \{0\} \quad \forall \; i, j \in [K], \; i < j\ , \;  \\ 
&R \neq I\ .
\end{split}
\end{align}
While the global pre-change distribution is fixed to  
$\sfF \equiv N_K( \boldsymbol{0},  I)$,  for the family of post-change distributions,  $\cG$, we  will consider different setups depending on the specification of  $\cR$ and  the 
 available topological information regarding where correlations are induced.  In all of them,  it is clear that it is necessary to sample at least $2$ sources at each time, i.e., $m$ must be at least $2$. First, we will  consider the case that we can sample only two sources per time instant, i.e., 
 $m = 2$, and  then the general case  where  $m \geq 2$. In all cases, it is readily verified that condition \eqref{eq:assm1} is satisfied.

\subsection{Sampling two sources at a time ($m=2$) }\label{ex:dep_change}
In this setting, $\cK_m$ is the set of all pairs of sources, i.e., 
$\{\{i, j\} : 1 \leq i < j \leq K\}$, and for any post-change global distribution  $\sfG  \in \cG$, the set  $\cA(\sfG)$, defined in \eqref{eq:A},  is the set of all pairs of sources that become correlated after the change under  $\sfG$, i.e., 
\begin{align}
\cA(\sfG) = \{\{i, j\} \in \cK_2 : R_{ij} \neq 0\}\ . 
\end{align}
We start with the case where every affected pair of sources becomes positively correlated after the change with a specified correlation $\rho \in (0, 1)$. We consider two cases regarding the prior topological  information under which the policy  $(\tilde{S}, \tilde{T})$ achieves 
asymptotic optimality.

\begin{proposition} 
\label{prop:homo_corr}
Suppose that  
$\cR = \{\rho\}$, where $\rho \in (0, 1)$.

\begin{itemize}
\item[(i)] If there is no prior information regarding the correlated pairs of sources, i.e., 
\begin{align}
\cG = \{N_K( \boldsymbol{0}, R)\ : \;  R \; \; \text{satisfies}~\eqref{eq:R} \}\ ,
\end{align}
then the policy $(\tilde{S}, \tilde{T})$, with 
$\cU=\cK_2$,
is asymptotically optimal.

\item[(ii)] If it is a priori known that there exists at least one pair of \textit{consecutive} sources, i.e., 
\begin{align}\label{eq:sub_fam}
\cG=
\lt\{N_K(\boldsymbol{0}, R)\ : \; R  \;\;  \text{satisfies}~\eqref{eq:R}    \;\;  \text{and}  \;\;  R_{i(i+1)}=\rho
\; \; \text{for some}\;\;   i\in[K-1]\rt\} \ ,
\end{align}
then the policy $(\tilde{S}, \tilde{T})$, with
$\cU$ either $\cK_2$ or  $\{\{i, i+1\} : i \in [K-1]\}$,  
is asymptotically optimal.
\end{itemize}
\end{proposition}

\begin{IEEEproof}
In both (i) and (ii), all conditions of Corollary \ref{corr:homogU} 
are satisfied with 
\begin{align}
\sfF^* \equiv  N_2(\boldsymbol{0}, I) \  \qquad \text{and} \ \qquad
\sfG^* \equiv N_2(\boldsymbol{0}, R_+)\ , \quad \mbox{where} \quad R_+ := \begin{bmatrix}1 &\rho\\\rho &1 \end{bmatrix} \ .
\end{align}
\end{IEEEproof}

\begin{remark}\label{rem:rem_to_homo_corr}
The previous proposition provides an example of a setup where, due to the presence of prior topological information,  asymptotic optimality is achieved with a family of units, $\cU$, that is smaller than the largest possible, $\cK_2$. Specifically,  its size is only $(K-1)$ compared to 
$K(K-1)/2$, i.e, the size of $\cK_2$ (recall  Remark  \ref{rem:alt_U}).\\
\end{remark}
Next, we consider the case where every affected pair of sources becomes either positively or negatively correlated after the change.

\begin{proposition}
Suppose $\cR = \{+\rho, -\rho\}$, where $\rho \in (0,1)$. If  there is no prior information regarding the correlated pairs of sources, i.e., 
\begin{align}
\cG = \lt\{N_K(\boldsymbol{0},R)\ : \;  R \; \; \text{satisfies} ~\eqref{eq:R} \rt\},
\end{align}
 then conditions  \eqref{assum:pos_I_mix_1_b}-\eqref{assum:pos_I_mix_1_a} hold for every $\sfG \in \cG$ with $\cU = \cK_2$.
\end{proposition}

\begin{IEEEproof}
It suffices to verify the conditions of Proposition \ref{prop:on_assum_mix_0}.  
For any pair of  sources  $\e \in \cK_2$ that is  affected by the change, the covariance matrix changes from the identity to either $R_+$ or $R_{-}$, where 
\begin{align}
R_{-} := \begin{bmatrix}1 &-\rho\\ -\rho &1 \end{bmatrix} \qquad \text{and}
\qquad R_{+} := \begin{bmatrix}1 &+\rho\\ +\rho &1 \end{bmatrix}\ .
\end{align}
Therefore, we have 
\begin{align}\sfF^\e = N_2( \boldsymbol{0} , I)  \qquad \text{and} \qquad \cG^\e = \{N_2( \boldsymbol{0} , R_+), \, N_2( \boldsymbol{0} , R_-)\} \ . \end{align}
This yields 
\begin{align}
\begin{split}
\xi_1^\e & = \log \; \bigg\{\exp\lt(-\frac{1}{2}(X_1^\e)^T(R_{-}^{-1} - I_2) \, X_1^\e \rt)\;+\; \exp\lt(-\frac{1}{2}(X_1^\e)^T(R_{+}^{-1} - I_2)\,  X_1^\e \rt) \bigg\}  -\log\lt(2\sqrt{1-\rho^2}\rt)\ . 
\end{split}
\end{align}
Since $\xi_1^\e$ is symmetric with respect to $+\rho$ and $-\rho$, it is identically distributed under $N_2(\boldsymbol{0} ,R_+)$ and $N_2(\boldsymbol{0} , R_-)$. Thus,  by Jensen's inequality, 
the 
expectations of $\xi_1^\e$ under $\sfF^E$ and any  $\sfP$ in $\cG^E$ are  finite.

\end{IEEEproof}

\subsection{Sampling $m$ sources, where $m\geq 2$} \label{ex:dep_change_m}

In this case, 
for any $\sfG  \in \cG$,  $\cA(\sfG)$ is  the set of all subsets of sources of size $m$ that contain at least one pair of sources that is  correlated after the change when the  post-change global distribution is $\sfG$, i.e., 
\begin{align}
\cA(\sfG) = \{ \{i_1, &, \dots, i_m\} \in \cK_m : \ \text{there exist} \ i \neq j, \;\; i, j \in \{i_1, i_2, \dots, i_m\} \ \text{such that} \ R_{ij} \neq 0 \}\ .
\end{align}

The following proposition describes a setup where the policy
$(\tilde{S}, \tilde{T})$ achieves asymptotic optimality. \\

\begin{proposition} 
Suppose that the total number of sources is a multiple of $m$, i.e., $K = Cm$ for some $C \geq 2$, and $\cR = \{\rho\}$ for some $\rho\in (0,1)$. Also, suppose that after the change 
at least one of the \emph{disjoint} $m$-tuples of consecutive sources in
\begin{align}
\cM_k:= \{(k-1)m+1, (k-1)m+2, \dots, km - 1, km\}\ , \quad k \in [C]\ ,
\end{align}
 follows an equi-correlated multivariate Gaussian distribution with common 
 positive correlation $\rho \in (0, 1)$ and is independent of the rest of the sources, that is,
\begin{align}
\cG = \Bigg\{N_K(\boldsymbol{0}, R)\ : \  &\mbox{$R$ satisfies~\eqref{eq:R}  \; and  }\\
&\exists \; A \subseteq [C] \quad \mbox{such that} \quad R_{ij} = 
\begin{cases}
\rho &\text{if}\ \{i,j\}\in\cM_k \;\;\mbox{for some } k \in A\\
0 &\text{otherwise}
\end{cases} \quad \Bigg\}\ .
\end{align}
Then, the policy $(\tilde{S}, \tilde{T})$, with $\cU$ either $\cK_m$ or  
$\{\cM_k : k \in [C]\}$,  is asymptotically optimal.
\end{proposition}

\begin{IEEEproof}
 Under $\sfG$,  at least one element from $\{\cM_k, k \in [C]\}$ is affected by the change, and its local distribution changes from the standard $m$-variate Gaussian to the equi-correlated one with common positive correlation $\rho$. 
Thus, it suffices to verify the conditions of Corollary \ref{corr:homogU}  with
\begin{align}
\sfF^* \equiv N_m(\boldsymbol{0}, I) \qquad \text{and} \qquad \ \sfG^* \equiv N_m(\boldsymbol{0} , R_m)\ , 
\end{align}
where, for every $k \in \bN$, we set
\begin{align}
R_k := 
\begin{bmatrix}1 &\rho &\cdots &\rho\\ \rho &1 &\cdots &\rho\\ 
\vdots &\vdots &\ddots &\vdots\\ \rho &\rho &\cdots &1
\end{bmatrix}\ \in \bR^{k \times k}\ .
\end{align}
It can be readily verified that \eqref{eq:assm_FG^*} holds. Specifically, we have 
\begin{align}\label{eq:dep_change3}
\int \log \left( \frac{d\sfG^*}{d\sfF^*} \right) d\sfG^* = -\frac{1}{2}\log{\det(R_m)}\ . 
\end{align}
Now, for every $\e \in \cA(\sfG)$, we have
\begin{equation}\label{gauss_info}
\cI^\e_0(\sfG) = -\frac{1}{2}\log({\det(\Sigma^\e_\sfG)})\ ,
\end{equation}
where $\Sigma_\sfG^\e$ denotes the covariance matrix of $X_1^\e$ under $\sfG$. Clearly, the above quantity 
varies over $\e \in \cA(\sfG)$. Therefore, in order to prove  \eqref{eq:max=I}, it suffices to show that 
\begin{align}\label{eq:dep_change3}
\int \log \left( \frac{d\sfG^*}{d\sfF^*} \right) d\sfG^*  \geq \cI^\e_0(\sfG) \ \quad \text{for every} \quad \e \in \cA(\sfG)\ ,
\end{align}
or equivalently that \begin{align}\label{eq:dep_change3}
 -\frac{1}{2}\log{\det(R_m)} \geq  \frac{1}{2}\log({\det(\Sigma^\e_\sfG)}) \ \quad \text{for every} \quad \e \in \cA(\sfG)\ .
\end{align}
Indeed, for every $\e \in \cA(\sfG)$ the  covariance matrix $\Sigma^\e_\sfG$ takes the following block-matrix form:
\begin{align}
\Sigma^\e_\sfG = 
\begin{bmatrix}
R_s &0 &0\\ 
0 &R_t &0\\
0 &0 &I_{m - (s + t)}
\end{bmatrix},
\quad \text{for some} \quad s, t \in \bN \quad \text{such that} \quad s + t \leq m\ .
\end{align}
Hence, in order to prove \eqref{eq:dep_change3}, it is enough to show that for every $s,t \in \bN$ such that $s+t \leq m$ we have 
\begin{align}
-\frac{1}{2}\log({{\rm det}(R_m)}) \geq -\frac{1}{2}\log({{\rm det}(R_s)\; {\rm det}(R_t)})\ ,
\end{align}
or, equivalently, 
$\det(R_m) \leq \det(R_s)\; \det(R_t)$. 
This inequality  follows by the explicit expression for the determinant of the autocorrelation matrix of order $k$, i.e., 
\begin{align}
\det(R_k) = (1 - \rho)^{k-1}(1 + (k-1)\rho)\ , \quad k \in \bN\ .
\end{align}
\end{IEEEproof}
\section{Simulation Studies}\label{sec:sim_study}

\subsection{Description of the model and settings}
In this section, we present the results of various simulation studies. In all of them,   we apply the policy 
$(\tilde{S}, \tilde{T})$, introduced in  Section \ref{sec: proposed procedure},   
  to the problem of detecting a change  in the correlation structure of
  $K = 10$ Gaussian information sources (Section \ref{sec:change_corr}).
 After the change, every affected pair of sources becomes positively correlated with correlation coefficient $\rho \in (0, 1)$, i.e., $\cR = \{\rho\}$, but there is no prior information regarding the post-change correlation structure. That is,  it is
not known a priori which or even how many sources become correlated. 
We denote the unknown number of affected sources by $s$, 
and without loss of generality, we assume that the sources in $\{K-s+1, K-s+2, \dots, K\}$ are the ones that become correlated to each other after the change.  
In each simulation study, we vary $s$ in  $\{2, \dots, K\}$, and in this way the true number of correlated pairs 
varies in $\left\{{s \choose 2} : 2 \leq s \leq K\right\}$. Given the above, we consider three simulation setups.
\begin{enumerate}
\item In the first one, we set $\rho=0.7$,  we are allowed to sample only $2$ sources at each time, i.e., $m = 2$, and we consider two cases for  $\gamma$, $10^2$ and  $10^5$.

\item  In the second, we set $\rho = 0.7$,  $\gamma = 10^2$, and we 
consider two cases for the number of sources we are allowed to sample at each time, $m=2$ and $m=3$.

\item In the third study, we consider the same setup as in the second with the only difference that we set $\rho = 0.95$.
\end{enumerate}

In all studies, for the implementation of the policy  $(\tilde{S}, \tilde{T})$ the threshold is 
$A=\log \gamma$, the family of units is 
$\cU=\cK_m$, and the 
permutation of  elements of $\cK_m$ for $m=2$ is
\begin{align}\{1, 2\}, \{1, 3\}, \dots, \{1, K\}, \{2, 3\}, \dots, \{2, K\}, \dots, \{K-1, K\}\ ,
\end{align}
and for $m=3$ it is 
\begin{align}
\{1, 2, 3\}, \{1, 2, 4\}, \dots, \{1, K-1, K\}, \{2, 3, 4\}, \dots, \{2, K-1, K\}, \dots, \{K-2, K-1, K\}\ .
\end{align}
For all studies, we consider the case that the change happens 
at time $\nu = 0$,
and we focus on the worst-case expected detection delay, in the sense that the unaffected (less informative) subsets, either pairs or triplets,  of sources are the ones sampled first,  
and the affected (more informative) ones are sampled later. Given these settings,  we estimate the expected detection delay of $(\tilde{S}, \tilde{T})$ using $4,000$ Monte Carlo replications for each  $s \in \{2, \dots, K\}$. In all cases, the Monte Carlo standard error in the estimation of every expectation did not exceed $5\%$  of the corresponding estimate.

In the first study, the goal is to investigate the sharpness of the non-asymptotic upper bound in Proposition \ref{rem:ub_extn_nas}.  In the other two, the goal is to explore the impact of sampling 2 or 3 sources at a time on the performance of  $(\tilde{S}, \tilde{T})$. Expectedly, sampling 3 sources at a time improves the first-order asymptotic performance of 
$(\tilde{S}, \tilde{T})$, but there is a considerable increase in the second-order term.  Furthermore, by Proposition \ref{prop:homo_corr}(i) it follows that $(\tilde{S}, \tilde{T})$  is asymptotically optimal when $m = 2$, but this is not the case when $m = 3$,
as there are multiple possible local post-change distributions for any triplet of sources.  

\begin{figure}[H]
\centering
\includegraphics[width=0.4\linewidth]{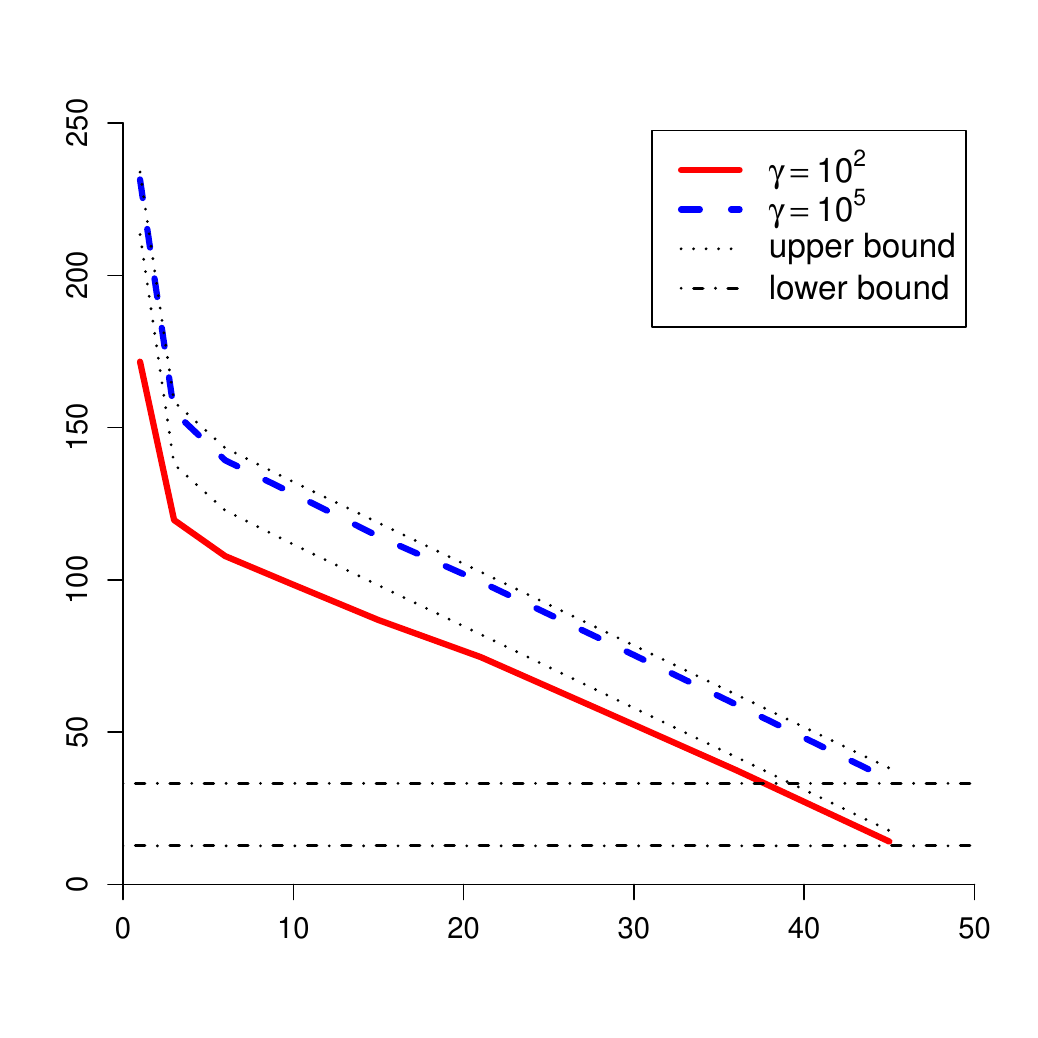}  
\caption{Expected detection delay of the rules under the first study.}
\label{fig:study1}
\end{figure}

We present the results for the first study in Figure~\ref{fig:study1}, for the  second  in Figure~\ref{fig:study2}, and for the third one  in Figure~\ref{fig:study3}. In all figures, 
the horizontal axis corresponds to  the true number of correlated pairs and  the vertical axis  to  the expected detection delay of the implemented procedures. 

\subsection{Simulation results}
From Figure \ref{fig:study1},
we observe that the non-asymptotic upper bound established in  Proposition \ref{rem:ub_extn_nas}  is very close to the expected detection delay,
especially for the larger value of $\gamma$. Furthermore, we can see that when the true number of correlated pairs is large enough, the detection delay decreases linearly with it. This can be predicted by the form of upper bound described in Remark \ref{rem:diffU}. 
Indeed, the numerator in the second term of this upper bound takes the form ${K \choose 2} - |\cA(\sfG)|$, whereas the denominator changes very slowly and behaves like a constant as the true number of correlated pairs $|\cA(\sfG)|$ increases to ${K \choose 2}$. Hence, 
this upper bound overall exhibits an almost linear trend with a negative slope. Finally, we observe that  the performance  of the Round Robin CUSUM rule essentially agrees with the first-order approximation of the corresponding lower bound (Theorem \ref{thm:lb}) in both cases for $\gamma$ when all sources become correlated after the change. 
\begin{figure}[H]
\centering
\makebox[\linewidth]{
\subfloat[]{\label{fig:study2} 
\includegraphics[width=0.4\linewidth]{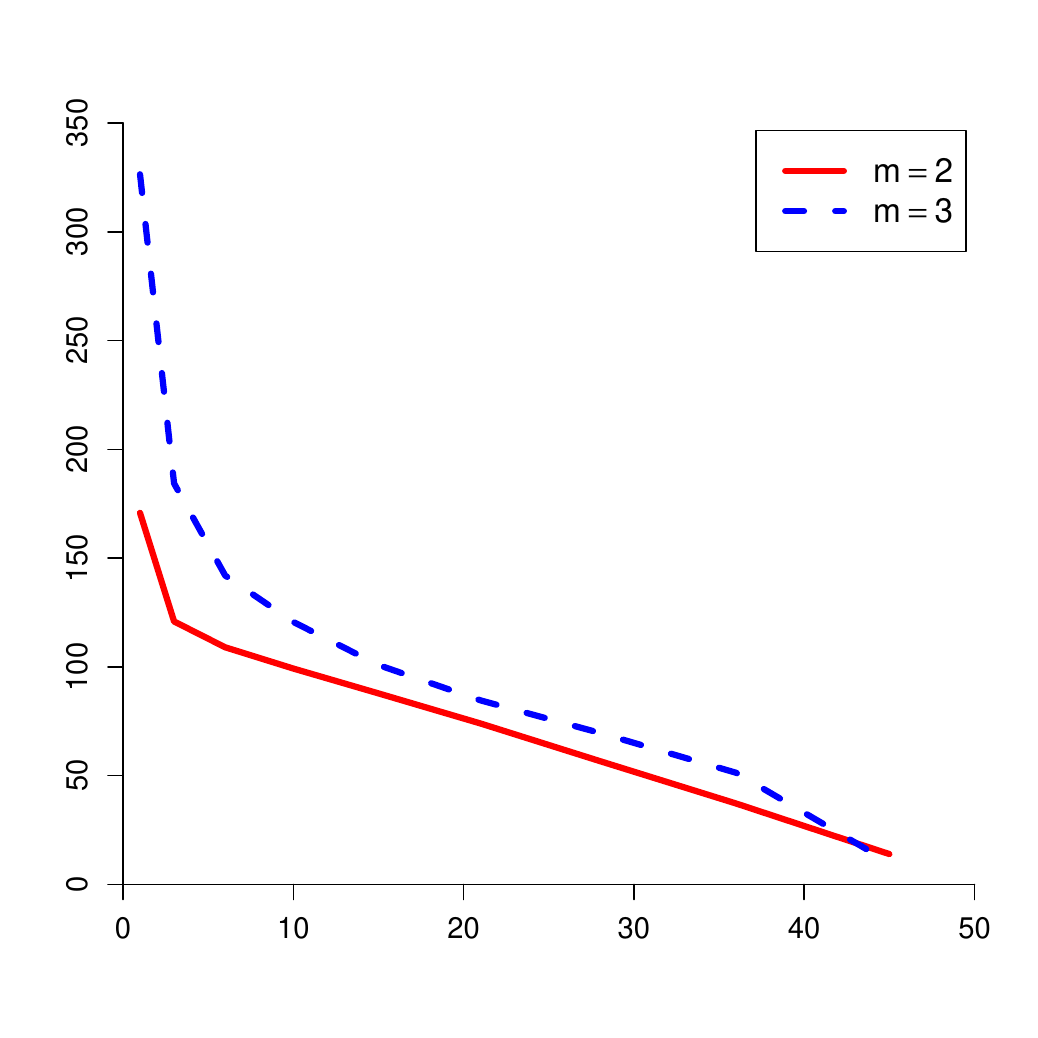}  
}
\subfloat[]{\label{fig:study3}  
\includegraphics[width=0.4\linewidth]{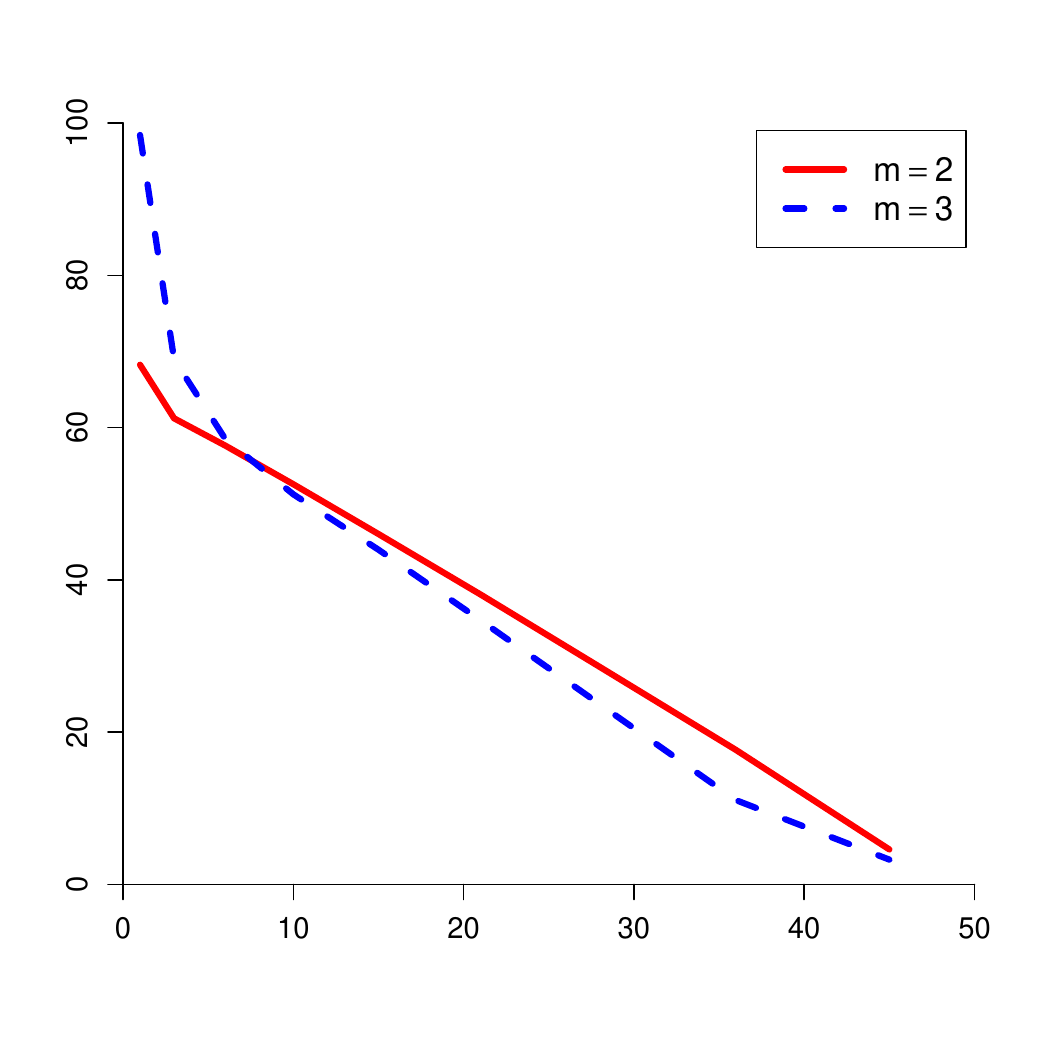}  
}
}
\caption{Expected detection delay of the rules under the second study (a) and third  study (b). 
}
\label{fig:allstudies}
\end{figure}

 From Figure \ref{fig:study2}, we observe that 
 when $\rho = 0.7$, the detection delay when sampling $3$ sources at each time is worse than that when  sampling $2$  sources at a time.  However, as shown in Figure \ref{fig:study3},
 the situation is mostly reversed when $\rho= 0.95$. 
 This phenomenon could be explained by the  possible trade-off between the strength of correlation and the advantageousness of asymptotic optimality, and this leads us to a potential direction for future research.

\section{Concluding Remarks and Future Directions}\label{sec:conclusion}
In this paper, we have considered an environment of multiple information sources or channels generating data streams in discrete time. The generated data are independent over time, but the observations from different sources can be statistically dependent. At an unknown time, the underlying joint distribution undergoes a change. The objective has been to detect the change with minimal delay, subject to a sampling constraint, when at each time, it is allowed to sample only a pre-fixed number of sources, of our choice.  In the class of all sequential procedures that satisfy the sampling constraint and control the false alarm rate below a certain pre-specified level, we have established a universal asymptotic lower bound for 
a modified version of Lorden's criterion \cite{Lorden1971}, as the false alarm rate vanishes. 
Subsequently, we have considered a simple, computationally efficient, and easily implementable sequential procedure that belongs to the above-mentioned class and achieves asymptotic optimality under certain assumptions regarding the underlying change-point model. Furthermore, under more general conditions, we have shown that the asymptotic relative efficiency of the proposed rule remains bounded. In contrast with most previous studies in the literature, the results in the present paper have not assumed independence among the data sources and have not relied on any prior information regarding how the sources are affected by the change. Thus, we have applied the procedures and the results to the problems of detecting a change in (i) the marginal distributions of multiple, not necessarily independent, data sources and (ii) the correlation structure of multiple Gaussian data sources. In the context of the first problem,  we have generalized the results in \cite{XuMeiMous21, XuMeiPostUncert, XuMeiPostUncert_SeqAn}  by relaxing the assumption that exactly one source is affected by the change, and establishing an asymptotic optimality property that holds for any possible number of sources affected by the change, as long as the signal to noise ratio is the same in all affected sources.

There are several potential open questions for further research, such as a more precise description of the optimal worst-case detection delay,  
and the consideration of more general setups that may involve temporal dependence and/or in which the pre-change and post-change distributions belong to parametric families.
Other directions of interest include considering  alternative error and delay metrics, studying an asymptotic regime where the number of sources also goes to infinity as the false alarm rate vanishes, modeling the data streams using spatial models, e.g., a  Markov random field, where the special underlying dependence could be used to obtain more efficient rules in practice. Finally, it would be of interest to design procedures that provide  guarantees for the correct isolation of the sources that are affected by the change under such sampling constraints.

\appendices

\renewcommand\thesection{\Alph{section}}

\section{Some Useful Lemmas for Lower Bound Analysis}

In this section, we assume that assumption~\eqref{eq:assm1} holds and present some lemmas that are critical in establishing the universal asymptotic lower bound in Theorem \ref{thm:lb}. We denote by  $\lambda_n^\e(\sfG)$  the LLR of $\sfG$ versus $\sfF$ based on the samples from the sources in $\e$ at time $n$, i.e., 
\begin{align}
\lambda_n^\e(\sfG) :=  
\log\frac{d\sfG^\e }{d\sfF^\e}\left(X_n^\e\right)\ .
\end{align}
Therefore, from the definition of $\cA(\sfG)$, it is noted that
\begin{align}
\lambda_n^\e(\sfG) = 0 \ , \quad \text{if} \quad \e \notin \cA(\sfG)\ .
\end{align}
Note that, for every $\sfG \in \cG$ and $\e \in \cA(\sfG)$,
the LLRs $\{\lambda_n^\e(\sfG) : n \in \bN\}$ are identically distributed under $\sfG$, with mean $\cI^\e_0(\sfG)$ and variance 
\begin{align}\label{eq:bdd2moment}
\cV^\e_0(\sfG) := \sfE_{\sfG}\lt[\lt(\lambda_1^\e(\sfG) - \cI^\e_0(\sfG)\rt)^2\rt] < \infty\ .
\end{align}
The finiteness in~\eqref{eq:bdd2moment} follows from \eqref{eq:assm1}. 

\begin{lemma}\label{lem:use_of_indep}
Suppose $\sfG \in \cG$, $\nu \geq 0$, and $S$ is any sampling rule. 
Then for every $n \geq \nu + 1$ we have
\begin{flalign}
&\text{(i)} \qquad \qquad  \bE_\nu^{\sfG}\lt[\lt(\lambda_{n}^{S_{n}}(\sfG) - \cI^{S_{n}}_0(\sfG)\rt) \;\Bigg|\; \cF_{n-1}^S\rt] = \; 0\ ,&&\\
&\text{(ii)} \qquad \quad \; \; \, \bE_\nu^{\sfG}\lt[\lt(\lambda_{n}^{S_{n}}(\sfG) - \cI^{S_{n}}_0(\sfG)\rt)^2 \;\Bigg|\; \cF_{n-1}^S \rt] = \; \cV^{S_n}_0(\sfG) \, \mathbbm{1}\{S_n \in \cA(\sfG)\}\ .
\end{flalign}
\end{lemma}

\begin{IEEEproof}
By assumption, $X_{n}^{S_{n}}$  is conditionally independent of  $\cF_{n-1}^S$ given $S_n$, i.e.,
\begin{align}\label{eq:markov_type}
X_{n}^{S_{n}} \med  \cF_{n-1}^S  \; \overset{d}{=} \; X_{n}^{S_{n}}\ . 
\end{align}
Thus, for every $n \geq \nu + 1$ 
\begin{align}
\bE_\nu^{\sfG} \lt[\lt(\lambda_{n}^{S_{n}}(\sfG) - \cI^{S_{n}}_0(\sfG)\rt) \;\Bigg|\; \cF_{n-1}^S \rt] 
&= \sum_{E}  \bE_\nu^{\sfG}\lt[\lambda_{n}^{E}(\sfG) - \cI^{E}_0(\sfG) \rt] \, \mathbbm{1}\{S_n=E\} \\
&= 0\ ,
\end{align} 
and 
\begin{align}
\bE_\nu^{\sfG}\lt[\lt(\lambda_{n}^{S_{n}}(\sfG) - \cI^{S_{n}}_0(\sfG)\rt)^2 \;\Bigg|\; \cF_{n-1}^S \rt]\
&= \sum_{E}  \bE_\nu^{\sfG}\lt[\lt(\lambda_{n}^{E}(\sfG) - \cI^{E}_0(\sfG)\rt)^2 \rt] \, \mathbbm{1}\{S_n=E\} \\
&= \sum_{E}   \cV^\e_0(\sfG) \, \mathbbm{1}\{S_n=E\} \\
&=  \cV^{S_n}_0(\sfG) \, \mathbbm{1}\{S_n \in \cA(\sfG)\}\ ,
\end{align}
where the first equality is due to \eqref{eq:markov_type} and the second equality follows by noting that for any $n \geq \nu + 1$ and $\e \in \cA(\sfG)$ we have
\begin{align}
&X_{n}^{S_{n}} \med  S_{n} = \e \sim \sfG^\e\ ,
 \end{align}
which implies
\begin{align}
&\bE_\nu^{\sfG}\lt[\lt(\lambda_{n}^{S_{n}}(\sfG) - \cI^{S_{n}}_0(\sfG)\rt)^2 \med  S_{n} = \e\rt]
= \bE_\nu^{\sfG}\lt[\lt(\lambda_{n}^\e(\sfG) - \cI^{\e}_0(\sfG)\rt)^2\rt] = \cV^\e_0(\sfG)\ .
 \end{align}
Also, for any $\e \notin \cA(\sfG)$ 
 \begin{align}
&X_{n}^{S_{n}} \med  S_{n} = \e \sim \sfF^\e\ ,
\end{align}
which implies 
\begin{align}
&\lt[\lt(\lambda_{n}^{S_{n}}(\sfG) - \cI^{S_{n}}_0(\sfG)\rt)^2 \med  S_{n} = \e\rt] \equiv 0\ .
 \end{align}
This completes the proof.
\end{IEEEproof}


\begin{lemma}\label{lem:Zmartingale}
Suppose $\sfG \in \cG$, $\nu \geq 0$, and $S$ is any sampling rule. For any $n \geq 0$, let
\begin{align}\label{eq:ZnAn}
Z_n := \log \frac{d\bP_0^{\sfG}}{d\bP_{\infty}}(\cF^S_n)\  \qquad \text{and} \qquad \
A_n := \sum_{t = 1}^{n} \cI^{S_t}_0(\sfG)\ ,
\end{align}
where $Z_0$ and $A_0 \equiv 0$.
Then under $\bP_\nu^{\sfG}$, the process 
\begin{align}
\{(Z_n- A_n) - (Z_\nu - A_\nu) : n > \nu\}
\end{align}
is a zero-mean martingale with respect to the filtration $\{\cF_n^S : n > \nu\}$.
\end{lemma}
\begin{IEEEproof}
Note that, for every $n \in \bN$, 
\begin{align}
Z_n - A_n = \sum_{t = 1}^n \lt(\lambda_t^{S_t}(\sfG) - \cI^{S_t}_0(\sfG)\rt)\ .
\end{align}
Consequently, for every $n > \nu$ we have
\begin{align}\label{eq:znznu}
(Z_n- A_n) - (Z_\nu - A_\nu) = \sum_{t = \nu + 1}^n \lt(\lambda_t^{S_t}(\sfG) - \cI^{S_t}_0(\sfG)\rt)\ .
\end{align}
Then, using \eqref{eq:znznu}, for every $n > \nu$, from \eqref{eq:assm1} we obtain
\begin{align}
\bE_\nu^{\sfG}\lt[|(Z_n - A_n) - (Z_\nu - A_\nu)|\rt] 
&\leq \sum_{t = \nu + 1}^n \bE_\nu^{\sfG}\lt[|\lambda_t^{S_t}(\sfG) - \cI^{S_t}_0(\sfG)|\rt]\\
&= \sum_{t = \nu + 1}^n \sum_{\e \in \cA(\sfG)} \bE_\nu^{\sfG}\lt[|\lambda_t^{\e}(\sfG) - \cI^{\e}_0(\sfG)| \mathbbm{1}\{S_t = \e\}\rt]\\
&\leq \sum_{t = \nu + 1}^n \sum_{\e \in \cA(\sfG)} \bE_\nu^{\sfG}\lt[|\lambda_t^{\e}(\sfG) - \cI^{\e}_0(\sfG)|\rt]\\
&= (n - \nu) \sum_{\e \in \cA(\sfG)} \bE_\nu^{\sfG}\lt[|\lambda_{\nu + 1}^{\e}(\sfG) - \cI^{\e}_0(\sfG)|\rt] < \infty\ ,
\end{align}
where the last inequality follows from \eqref{eq:bdd2moment}.
Furthermore, we have
\begin{align}
\bE_\nu^{\sfG}\lt[(Z_n - A_n) - (Z_\nu - A_\nu) \; |\; \cF_{n-1}^S\rt] &= \bE_\nu^{\sfG}\bigg[\sum_{t = \nu + 1}^{n-1} \lt(\lambda_t^{S_t}(\sfG) - \cI^{S_t}_0(\sfG)\rt)+ \lt(\lambda_n^{S_n}(\sfG) - \cI^{S_n}_0(\sfG)\rt) \;\Bigg|\; \cF_{n-1}^S\bigg]\\  \label{eq:Zn-1}
&= (Z_{n-1} - A_{n-1}) - (Z_\nu - A_\nu) + \bE_\nu^{\sfG}\lt[\lt(\lambda_{n}^{S_{n}}(\sfG) - \cI^{S_{n}}_0(\sfG)\rt) \;\Bigg|\; \cF_{n-1}^S\rt]\ .
\end{align}
Therefore, it suffices to show that the second term in \eqref{eq:Zn-1} is $0$, which follows from Lemma \ref{lem:use_of_indep}. 
\end{IEEEproof}

\begin{lemma}[Optional Stopping Theorem]\label{lem:trunc_opt_stop}
For some $\nu \geq 0$, let $\{M_n : n > \nu\}$ be a submartingale with respect to some filtration $\{\cG_n : n > \nu\}$, and $T$ be a stopping time such that $T \in \{\nu + 1, \dots, N\}$, where $N \in \bN$. Then
\begin{align}
    M_{T} \leq \sfE\lt[M_N|\cG_T\rt]\ ,
\end{align}
where $\sfE[\cdot]$ denotes the underlying expectation. 
\end{lemma}
\begin{IEEEproof}
This is a standard and fundamental result in probability theory. 
see, e.g., \cite[Theorem 2.3.1]{Tart14}.
\end{IEEEproof}
\begin{lemma}[Conditional Doob's Submartingale Inequality]\label{lem:cond_Doob}
For some $\nu \geq 0$, let $\{M_n : n > \nu\}$ be a submartingale with respect to some filtration $\{\cG_n : n > \nu\}$. Then for every $a > 0$
and $n \geq 1$ the following holds almost surely:
\begin{align}
\sfP\lt(\max_{\nu+1 \leq k \leq \nu + n} M_k \geq a \; \;\Bigg|\; \; \cG_{\nu}\rt) \leq \frac{\sfE\lt[M_{\nu + n}|\cG_{\nu}\rt]}{a}\ ,
\end{align}
where $\sfP$ and $\sfE[\cdot]$ are the underlying probability measure and its corresponding expectation, respectively. 
\end{lemma}
\begin{IEEEproof}
We fix $a>0$ and $n \geq 1$, and 
define the following stopping time
\begin{align}
    T := \inf\lt\{k \geq \nu + 1 : M_k \geq a\rt\} \wedge (\nu + n)\;,
\end{align}
which clearly satisfies $\nu + 1 \leq T \leq \nu + n$ and
\begin{align}\label{eq:cond_Doob}
\lt\{\max_{\nu+1 \leq k \leq \nu + n} M_k \geq a\rt\} = \lt\{M_T \geq a\rt\}\ .
\end{align}
Then,
\begin{align}
\sfP\lt(\max_{\nu+1 \leq k \leq \nu + n} M_k \geq a \;\;\Bigg|\; \;\cG_{\nu} \rt)
&= \sfP \lt( M_T \geq a  \, \;\Bigg|\; \, \cG_{\nu} \rt) \\
&\leq \frac{1}{a}\sfE\lt[M_T \mathbbm{1}\lt\{M_T \geq a\rt\}  \;\Bigg|\; \cG_{\nu} \rt]\\ \label{ineq:Mnupn}
&\leq \frac{1}{a}\sfE\lt[M_{\nu + n} \mathbbm{1}\lt\{M_T \geq a\rt\}  \;\Bigg|\; \cG_{\nu} \rt]\\
&\leq \frac{1}{a}\sfE\lt[M_{\nu + n}  \;\Bigg|\; \cG_{\nu} \rt]\ ,
\end{align}
where  the inequality in \eqref{ineq:Mnupn} holds, since for any $A \in \cG_{\nu}\subseteq \cG_{T}$ we have $A \cap \{M_T \geq a\} \in \cG_T$ and, thus, by Lemma \ref{lem:trunc_opt_stop} we have
\begin{align}
\sfE\lt[M_T \mathbbm{1}\lt\{A \cap \lt\{M_T \geq a\rt\}\rt\} \rt] \leq \sfE\lt[M_{\nu + n} \mathbbm{1}\lt\{A \cap \lt\{M_T \geq a\rt\}\rt\}\rt]\ .
\end{align}
The proof is complete.
\end{IEEEproof}

\begin{lemma}\label{lem:cond_variance}
Suppose $\sfG \in \cG$ and $(S, T)$ is any sequential change detection policy. Then for any $\nu, n \in \bN$ we have
\begin{align}
\begin{split}
\bE_{\nu}^{\sfG}\lt[\lt\{\sum_{t = \nu + 1}^{\nu + n} \lt(\lambda_t^{S_t}(\sfG) - \cI^{S_t}_0(\sfG)\rt)\rt\}^2 \;\Bigg|\; T > \nu\rt] 
\leq n \; \max_{\e \in \cA(\sfG)} \cV^\e_0(\sfG)\ . 
\end{split}
\end{align}
\end{lemma}
\begin{IEEEproof}
For any $\nu + 1 \leq t \leq \nu + n$, we have
\begin{align}
\label{eq:lam^2} \bE_{\nu}^{\sfG}\lt[\lt(\lambda_t^{S_t}(\sfG) - \cI^{S_t}_0(\sfG)\rt)^2 \; \;\Bigg|\; \; T > \nu\rt] 
&=\bE_{\nu}^{\sfG}\lt[\bE_{\nu}^{\sfG}\lt[\lt(\lambda_t^{S_t}(\sfG) - \cI^{S_t}_0(\sfG)\rt)^2 \;\Bigg|\; \cF^S_{t-1}\rt] \;\Bigg|\; \; T > \nu\rt]\\ \label{eq:lamtoV}
&= \bE_{\nu}^{\sfG}\lt[\cV^{S_t}_0(\sfG) \mathbbm{1}\{S_t \in \cA(\sfG)\} \;\Bigg|\; \; T > \nu\rt]\\
&\leq \bE_{\nu}^{\sfG}\lt[\mathbbm{1}\{S_t \in \cA(\sfG)\} \max_{\e \in \cA(\sfG)} \cV^\e_0(\sfG) \;\Bigg|\; \; T > \nu\rt]\\
&\leq \max_{\e \in \cA(\sfG)} \cV^\e_0(\sfG) \; \bP_{\nu}^{\sfG}\lt(S_t \in \cA(\sfG) \;\Bigg|\; \; T > \nu\rt)\ ,
\end{align}
where \eqref{eq:lam^2} follows by using the tower property of conditional expectation and the fact that, $\{T > \nu\} \in \cF^S_{\nu - 1} \subseteq \cF^S_{t - 1}$, 
and \eqref{eq:lamtoV} follows from Lemma \ref{lem:use_of_indep}(ii).
\end{IEEEproof}


\section{Proof of Universal Asymptotic Lower Bound}\label{pf:thm:lb}

\; Fix any arbitrary $\sfG \in \cG$, $\nu \geq 0$, $(S, T) \in \Delta(\gamma)$ and consider the sequences $\{Z_n, A_n : n \geq 0\}$ as defined in \eqref{eq:ZnAn}.
Then by using Lemma \ref{lem:Zmartingale}, under $\bP_\nu^\sfG$ and with respect to the filtration $\{\cF_n^S : n > \nu\}$, the process $\{(Z_n - A_n) - (Z_\nu - A_\nu) : n > \nu\}$ is a zero-mean martingale,
which further implies that the process
\begin{align}
\{\{(Z_n - A_n) - (Z_\nu - A_\nu)\}^2 : n > \nu\}\ ,
\end{align}
is a submartingale.
Note that, for any $m, n \in \bN$, such that $m < n$ we have 
\begin{align}\label{eq:ineq_for_An}
A_n - A_m \leq (n - m) \; \cI_*\ , \qquad \text{where} \quad \cI_* := \max_{\e \in \cA(\sfG)} \cI^\e_0(\sfG)\ .
\end{align}
Now, for an arbitrary $0 < \epsilon < 1$ and $\gamma>1$ we define 
\begin{align}
N_{\gamma, \epsilon} := \frac{\log \gamma}{\cI_*}(1 - \epsilon)\ .
\end{align}
and 
\begin{equation}\label{eta_gamma}
\log\eta_{\gamma} :=  (1-\epsilon^2)\log\gamma= (1+\epsilon) N_{\gamma, \epsilon}  \cI_*\ .
\end{equation}
Then by the definition of $D^{\sfG}(S,T)$ in  \eqref{Det-Del} and   Markov's inequality  it follows that, for any $\nu \geq 0$,  
\begin{align}\label{markov}
\begin{split}
D^{\sfG}(S,T) &\geq \bE_{\nu}^{\sfG}[T - \nu \,| \,  T > \nu] \\
&\geq N_{\gamma, \epsilon}  \, \bP_{\nu}^{\sfG}\left(T - \nu > N_{\gamma, \epsilon} \,| \,  T > \nu\right) \\
&= N_{\gamma, \epsilon}  \, (1- \bP_{\nu}^{\sfG}\left(T - \nu \leq  N_{\gamma, \epsilon} \,| \,  T > \nu\right)\\
&= N_{\gamma, \epsilon}  \, \left(1- p(\nu, T) - q(\nu, T) \right)\ ,
\end{split}
\end{align}
where we have defined
\begin{align}
p(\nu, T) & := \bP_{\nu}^{\sfG}\left(T - \nu \leq N_{\gamma, \epsilon}, \; e^{Z_T - Z_{\nu}} > \eta_{\gamma}| T > \nu\right)\ , \\
q(\nu, T)  &:= \bP_{\nu}^{\sfG}\left(T - \nu \leq N_{\gamma, \epsilon}, \; e^{Z_T - Z_{\nu}} \leq \eta_{\gamma}| T > \nu\right)\ .
\end{align}
First, we find an upper bound on $p(\nu, T)$ for any $\nu \geq 0$. To this end, note that
\begin{align}
&\lt\{T \leq \nu + N_{\gamma, \epsilon}, e^{Z_T - Z_{\nu}} > \eta_{\gamma}\rt\}\\
&\subseteq \lt\{\max_{\nu + 1 \leq k \leq \nu + N_{\gamma, \epsilon}} {Z_k - Z_{\nu}} > \log\eta_{\gamma}\rt\}\\
&\subseteq \lt\{\max_{\nu + 1 \leq k \leq \nu + N_{\gamma, \epsilon}} \big\{(Z_k - A_k) - (Z_{\nu} - A_{\nu})\big\} + \max_{\nu + 1 \leq k \leq \nu + N_{\gamma, \epsilon}} (A_k - A_\nu) > \log\eta_{\gamma}\rt\}\\
&\overset{\eqref{eta_gamma}}= \lt\{\max_{\nu + 1 \leq k \leq \nu + N_{\gamma, \epsilon}} \big\{(Z_k - A_k) - (Z_{\nu} - A_{\nu})\big\} + \max_{\nu + 1 \leq k \leq \nu + N_{\gamma, \epsilon}} (A_k - A_\nu) > N_{\gamma, \epsilon}\cI_* + \epsilon N_{\gamma, \epsilon}\cI_* \rt\}\\ \label{ineq:remove2max}
&\subseteq \lt\{\max_{\nu + 1 \leq k \leq \nu + N_{\gamma, \epsilon}} \big\{(Z_k - A_k) - (Z_{\nu} - A_{\nu})\big\} > \epsilon N_{\gamma, \epsilon}\cI_* \rt\}\\ \label{eq:pnu_events}
&\subseteq \lt\{\max_{\nu + 1 \leq k \leq \nu + N_{\gamma, \epsilon}} \big\{(Z_k - A_k) - (Z_{\nu} - A_{\nu})\big\}^2 > \epsilon^2 N_{\gamma, \epsilon}^2\cI_*^2 \rt\}\ .
\end{align}
The relationship in \eqref{ineq:remove2max} holds because for every $\nu + 1 \leq k \leq \nu + N_{\gamma, \epsilon}$, by using \eqref{eq:ineq_for_An} we have:
\begin{align}
A_k - A_\nu \leq (k - \nu) \cI_* \leq N_{\gamma, \epsilon} \cI_*\ .
\end{align}
Now, conditioning on the event $\{T > \nu\}$ and following \eqref{eq:pnu_events}, we have
\begin{align}
p(\nu, T) 
&= \bP_{\nu}^{\sfG}\left(T \leq \nu + N_{\gamma, \epsilon}, e^{Z_T - Z_{\nu}} > \eta_{\gamma}| T > \nu\right)\\
&\leq \bP_{\nu}^{\sfG}\bigg(\max_{\nu + 1 \leq k \leq \nu + N_{\gamma, \epsilon}} \big\{(Z_k - A_k) - (Z_{\nu} - A_{\nu})\big\}^2 > \epsilon^2 N_{\gamma, \epsilon}^2\cI_*^2 \;\Bigg|\; T > \nu\bigg)\\ \label{ineq:useconddoob}
&\leq \frac{\bE_\nu^\sfG\lt[\lt\{\lt(Z_{\nu + N_{\gamma, \epsilon}} - A_{\nu + N_{\gamma, \epsilon}}\rt) - \lt(Z_{\nu} - A_{\nu}\rt)\rt\}^2 \;\Bigg|\; T > \nu\rt]}{\epsilon^2 N_{\gamma, \epsilon}^2\cI_*^2}\\
&= \frac{\bE_{\nu}^{\sfG}\lt[\lt\{\sum_{t = \nu + 1}^{\nu + N_{\gamma, \epsilon}} \lt(\lambda_t^{S_t}(\sfG) - \cI^{S_t}_0(\sfG)\rt)\rt\}^2 \;\Bigg|\; T > \nu\rt]}{\epsilon^2 N_{\gamma, \epsilon}^2\cI_*^2}\\ \label{ineq:max_out}
&\leq \max_{\e \in \cA(\sfG)} \cV^\e_0(\sfG)  \; \frac{N_{\gamma, \epsilon}}{\epsilon^2 N_{\gamma, \epsilon}^2\cI_*^2}\\
&= \max_{\e \in \cA(\sfG)} \cV^\e_0(\sfG)\; \frac{1}{\epsilon^2 N_{\gamma, \epsilon} \cI_*^2}
\equiv \xi_{\epsilon}(\gamma)\ .
\end{align}
The inequality in \eqref{ineq:useconddoob} holds by using Lemma \ref{lem:cond_Doob} for the submartingale process,
$\{\{(Z_n - A_n) - (Z_\nu - A_\nu)\}^2 : n > \nu\}$, and the inequality in \eqref{ineq:max_out} follows from Lemma \ref{lem:cond_variance}.
Now, since for any given $\epsilon > 0$, $N_{\gamma, \epsilon} \to \infty$ as $\gamma \to \infty$, we have $\xi_{\epsilon}(\gamma) \to 0$ as $\gamma \to \infty$.

Next, we characterize an upper bound on $q(\nu, T)$. For this purpose, we follow the approach developed in the proof of Theorem 1 in \cite{Lai98}. 
Let $n_\gamma$ be the largest integer less than or equal to $\lt(\log\gamma\rt)^2$. Then, as shown in the proof of Theorem 1 in \cite{Lai98}, there exists $\nu_\gamma \geq 1$ such that
\begin{align}\label{eq:Lai'sresult}
\bP_{\infty}\lt(T \geq \nu_\gamma\rt) > 0 \ , \qquad \text{and} \qquad \bP_{\infty}\lt(T < \nu_\gamma + n_\gamma | T \geq \nu_\gamma\rt) \leq \frac{n_\gamma}{\gamma}\ .
\end{align}
Thus, for $\gamma$ large enough, 
\begin{align}
q(\nu_\gamma, T)&= \bP_{\nu_\gamma}^{\sfG}\left(T \leq \nu_\gamma + N_{\gamma, \epsilon}, e^{Z_T - Z_{\nu_\gamma}} \leq \eta_{\gamma}| T > \nu_\gamma\right)\\ \label{eq:nutoinfty}
&= \frac{\bP_{\nu_\gamma}^{\sfG}\left(\nu_\gamma < T \leq \nu_\gamma + N_{\gamma, \epsilon}, e^{Z_T - Z_{\nu_\gamma}} \leq \eta_{\gamma}\right)}{\bP_{\infty}(T > \nu_\gamma)}\\
&=\frac{\bE_{\infty}[e^{Z_T - Z_{\nu_\gamma}}\mathbbm{1}\{\nu_\gamma < T \leq \nu_\gamma + N_{\gamma, \epsilon}, e^{Z_T - Z_{\nu_\gamma}} \leq \eta_{\gamma}\}]}{\bP_{\infty}(T > \nu_\gamma)}\\
&\leq \frac{\eta_{\gamma} \, \bP_{\infty}\left(\nu_\gamma < T \leq \nu_\gamma + N_{\gamma, \epsilon}\right)}{\bP_{\infty}(T > \nu_\gamma)}\\
&= \eta_{\gamma} \, \bP_{\infty}\left(T \leq \nu_\gamma + N_{\gamma, \epsilon} | T > \nu_\gamma\right)\\ \label{ineq:largegamma}
&\leq \eta_{\gamma} \, \bP_{\infty}\left(T \leq \nu_\gamma + n_\gamma | T > \nu_\gamma\right)\\ \label{ineq:useofLai}
&\leq \eta_{\gamma} \, \frac{n_\gamma}{\gamma}\\
&\leq \gamma^{-\epsilon^2}\lt(\log\gamma\rt)^2 \\
& \equiv \xi'_{\epsilon}(\gamma)\ .
\end{align}
The equality in \eqref{eq:nutoinfty} follows from the fact that, $\bP_{\nu_\gamma}^{\sfG}(T > \nu_\gamma) = \bP_{\infty}(T > \nu_\gamma) $ since $\{T > \nu_\gamma\} \in \cF^S_{\nu_\gamma}$. The inequality in \eqref{ineq:largegamma} follows because for sufficiently large $\gamma$, $N_{\gamma, \epsilon} \leq n_\gamma$ for every $\epsilon > 0$, and the inequality in \eqref{ineq:useofLai} follows from \eqref{eq:Lai'sresult}.
It is important to observe that $\xi'_{\epsilon}(\gamma) \to 0$ as $\gamma \to \infty$ for any given $\epsilon > 0$.
Finally from \eqref{markov}, we have, for all large $\gamma$,
\begin{align}
D^{\sfG}(S, T) &\geq N_{\gamma, \epsilon}  \, \left(1- p(\nu_\gamma, T) - q(\nu_\gamma, T) \right)\\
&\geq \frac{\log \gamma}{\cI_*} (1 - \epsilon) \;  (1 - \xi_{\epsilon}(\gamma)- \xi'_{\epsilon}(\gamma))\ .
\end{align}
Now, we  let first $\gamma \to \infty$, and then $\epsilon \to 0$, and the proof is complete.

\section{Some Useful Lemmas for Performance Analysis}\label{app_subsec:implems}
In this appendix, 
we present some lemmas that are useful in proving 
Theorem \ref{thm:fa_extn} and Theorem \ref{thm:ub_extn}.
All random variables are defined on a probability space $(\Omega, \cF, \Pro)$, and we denote by $\sfE[\cdot]$  the corresponding expectation.

For the first two lemmas, namely Lemma \ref{lem:mu/p} and Lemma \ref{lem:first_cycle_worstcase}, we assume that $\{Z_n : n \in \bN\}$ is a sequence of independent and identically distributed (i.i.d.) random variables, 
and we set
\begin{align}
S_n := \sum_{t = 1}^{n} Z_t, \quad n \in \bN\ . 
\end{align}
Furthermore, we fix some  arbitrary $A > 0$, and for any  $x \in [0, A)$ we set $\pi(x) := \min\{\pi_-(x), \pi_+(x)\}$, where
\begin{align}
\pi_-(x) &:= \inf\lt\{n \geq 1 : S_n < -x\rt\}\ ,\\
\pi_+(x) &:= \inf\lt\{n \geq 1 : S_n > A-x\rt\}\ .
\end{align}
Furthermore, we assume that the first two moments of $Z_1$ exist, and we denote them as follows: 
\begin{align}
\cI := \sfE[Z_1] \ \qquad \text{and} \qquad \ \cV:= \sfE\lt[\lt(Z_1 - \cI\rt)^2\rt]\ .
\end{align}

\begin{lemma}\label{lem:mu/p}
If $0 < \cI < \infty$, then for any $x \in [0, A)$ we have, as $A \to \infty$,
\begin{align}\label{ineq:A-x_ub}
\frac{\sfE \lt[\pi(x)\rt]}{\sfP\lt(\pi_+(x) < \pi_-(x)\rt)} \leq
\frac{A-x}{\cI} + o(A)\ . 
\end{align}
If,  also, $\cV < \infty$, then the inequality \eqref{ineq:A-x_ub} holds for every $A > 0$  with the $o(A)$ term being replaced with
\begin{align}
\frac{1}{p}\lt(1 + \frac{\cV}{\cI^2}\rt), \qquad \text{where} \;\;\; p := \; \sfP\lt(\pi_-(0) = \infty\rt) > 0\ .
\end{align}
In particular, when $x = 0$, the upper bound  holds even without the factor $1/p$.
\end{lemma}
\begin{IEEEproof}
For any $x \in [0, A)$ we have
\begin{align}
 \cI \, \sfE\lt[\pi(x)\rt] 
&= \sfE\lt[S_{\pi(x)}\rt]
\label{eq:useofWald} \\
&= \sfE\lt[S_{\pi(x)}\mathbbm{1}\{\pi_+(x) < \pi_-(x)\}\rt] + \sfE\lt[S_{\pi(x)}\; \mathbbm{1}\{\pi_-(x) < \pi_+(x)\}\rt] \\ \label{ineq:part2<0}
&\leq  \sfE\lt[S_{\pi(x)}\mathbbm{1}\{\pi_+(x) < \pi_-(x)\}\rt] \\
&=  (A - x)  \; \sfP \lt( \pi_+(x) < \pi_-(x)\rt) +  \sfE\lt[ (S_{\pi(x)} - (A - x)) \; \mathbbm{1}\{\pi_+(x) < \pi_-(x)\}\rt]\\ \label{ineq:S+}
&\leq  (A - x)  \; \sfP \lt( \pi_+(x) < \pi_-(x)\rt) + o(A), 
\end{align}
where the equality in \eqref{eq:useofWald} follows from Wald's identity, the inequality in \eqref{ineq:part2<0} follows from the fact that
\begin{align}
S_{\pi(x)}\mathbbm{1}\{\pi_-(x) < \pi_+(x)\} = S_{\pi_-(x)}\mathbbm{1}\{\pi_-(x) < \pi_+(x)\} \leq 0\ ,
\end{align}
and the inequality in \eqref{ineq:S+} holds from \eqref{eq:useofWald} along with the fact, due to the elementary renewal theorem, that, as $(A-x) \to \infty$, 
\begin{align}\label{eq:o(A-x)}
\sfE\lt[S_{\pi(x)}\rt] = \cI \, \sfE\lt[\pi(x)\rt] \leq (A-x) + o(A-x)\ ,
\end{align}
and also, $o(A-x)/A \leq o(A-x)/(A-x) \to 0$. 
The inequality in \eqref{ineq:S+} further leads to
\begin{align}
\frac{ \cI \; \sfE\lt[\pi(x)\rt]}{\sfP\lt(\pi_+(x) < \pi_-(x)\rt)}
&\leq (A - x) + \frac{o(A)}{\sfP\lt(\pi_+(x) < \pi_-(x)\rt)}\\ \label{ineq:redtoQ12} 
&\leq (A - x) + \frac{o(A)}{\sfP\lt(\pi_-(x) = \infty\rt)}\\  \label{ineq:plus<mod} 
&\leq (A - x) + \frac{o(A)}{p}\ . 
\end{align}
The inequalities in \eqref{ineq:redtoQ12} and \eqref{ineq:plus<mod} follow by using    
the fact that 
\begin{align}\label{eq:pitop}
\sfP\lt(\pi_+(x) < \pi_-(x)\rt) \geq \sfP\lt(\pi_-(x) = \infty\rt) \geq \sfP\lt(\pi_-(0) = \infty\rt) = p\ ,     
\end{align}
where the second inequality in \eqref{eq:pitop} 
follows from the fact that,     
for every $x \geq 0$, $\pi_-(x) \geq \pi_-(0)$ as $\pi_-(x)$ is an increasing function with respect to $x$. 
Furthermore, the positivity of $p$ follows from the fact that, for large $A$,
\begin{align} \label{eq:positive_p}
p = \sfP\lt(\pi_-(0) = \infty\rt) = \frac{1}{\sfE\lt[\pi_+(A)\rt]} \geq \frac{1}{\sfE\lt[\pi_+(0)\rt]} \geq \lt(\frac{A + o(A)}{\cI}\rt)^{-1} > 0\ ,
\end{align} 
where, in \eqref{eq:positive_p}, the second equality follows from \cite[Corollary 8.39]{BookSeig}, the first inequality holds because $\pi_+(0) \geq \pi_+(A)$, as $\pi_+(x)$ is a decreasing function with respect to $x$, and the second inequality follows from what is shown next.
Note that the quantity 
\begin{align}
\frac{\sfE\lt[\pi(0)\rt]}{\sfP\lt(\pi_+(0) < \pi_-(0)\rt)}\ ,
\end{align}
represents the expected detection delay of a CUSUM process adapted to the sequence $\{S_n:n\in\bN\}$ \cite{Tart14}. 
Since for every $n \in \bN$ the CUSUM process at time $n$ can always be bounded below by $S_n$, the detection delay can be consequently bounded above by $\pi_+(0)$. Therefore, as $A \to \infty$,
\begin{align} 
\cI \, \frac{\sfE\lt[\pi(0)\rt]}{\sfP\lt(\pi_+(0) < \pi_-(0)\rt)}
&\leq \cI \, \sfE\lt[\pi_+(0)\rt]\\ \label{eq:exppi0}
&= \sfE\lt[S_{\pi_+(0)}\rt]\\ \label{ineq:excesspi0}
&\leq A + o(A)\ , 
\end{align}
where the equality in \eqref{eq:exppi0} follows from Wald's identity and the inequality in \eqref{ineq:excesspi0} follows similarly to \eqref{ineq:S+}
If, also, $\cV < \infty$, then by Theorem 1 in \cite{Lorden70} it follows that, for every $A > 0$, each of the above relations holds with the $o(A)$ term being replaced with 
\begin{align}
\frac{1}{\cI} \sfE\lt[\lt( {S_1^+}\rt)^2\rt] \leq \frac{\cV + \cI^2}{\cI}\ ,
\end{align}
where the inequality follows from the fact that $S_1^+ \leq |S_1|$.
This completes the proof.\\
\end{IEEEproof}

\begin{lemma}\label{lem:cond_decomp}
For any non-negative random variable $X$ and two  disjoint non-null events $H_1$ and $H_2$, we have
\begin{align}
    \sfE\lt[X \med H_1 \cup H_2\rt] \leq \sfE\lt[X \med H_1\rt] + \sfE\lt[X \med H_2\rt]\ .
\end{align}
\end{lemma}
\begin{IEEEproof}
    The proof follows from the below that 
    \begin{align}
        \sfE\lt[X \med H_1 \cup H_2\rt] &= \frac{\sfE\lt[X \mathbbm{1}\{H_1 \cup H_2\} \rt]}{\sfP\lt(H_1 \cup H_2\rt)}\\
        &=\frac{\sfE\lt[X \mathbbm{1}\{H_1\} \rt] + \sfE\lt[X \mathbbm{1}\{H_2\} \rt]}{\sfP(H_1) + \sfP(H_2)}\\
        &\leq \frac{\sfE\lt[X \mathbbm{1}\{H_1\} \rt]}{\sfP(H_1)} + \frac{\sfE\lt[X \mathbbm{1}\{H_2\} \rt]}{\sfP(H_2)} = \sfE\lt[X \med H_1\rt] + \sfE\lt[X \med H_2\rt]\ .
    \end{align}
\end{IEEEproof}

\begin{lemma}\label{lem:first_cycle_worstcase}
Let $W$ be a random variable independent of $\{Z_n : n \in \bN\}$ such that  
\begin{align}\label{eq:surv_prob}
\sfP(W > x) \leq e^{-x}, \quad x \geq 0\ . 
\end{align} 
If $-\infty < \cI < 0$, then for every $A > 0$ large enough, we have 
\begin{align}\label{ineq:piW_ub}
\sfE\lt[\pi(W) \med  W \in (0, A)\rt] \leq C\ ,
\end{align}
where $C < \infty$ is a constant independent of $A$.
\end{lemma}
\begin{IEEEproof}
Since $\{Z_n : n \in \bN\}$ is independent of $W$, for any $x \in (0, A)$ we have, as $x \to \infty$, and hence, $A \to \infty$,
\begin{align} \label{eq:uppbd_sprt}
\sfE\lt[\pi(W) \med  W = x\rt] \leq \sfE\lt[\pi_-(x)\rt] \leq -\frac{x}{\cI} + o(x)\ ,
\end{align}
where the second inequality follows from
the elementary renewal theorem (also see the derivation in the proof of Lemma \ref{lem:mu/p}). 
Fix any arbitrary $\epsilon > 0$. Then, there exists $M > 0$ such that for every $x \geq M$,
\begin{align}
    \frac{1}{x} \sfE\lt[\pi(W) \med  W = x\rt] \leq \cI_\epsilon := -\frac{1}{\cI} + \epsilon\ .
\end{align}
We assume that there exists $\kappa > 1$ such that $\sfP(W > M) > e^{-\kappa M}$. Otherwise, we have $\sfP(W \in (0, M]) = 1$, and thus, for every $A > M$,
\begin{align}
    \sfE\lt[\pi(W) \med  W \in (0, A)\rt] = \sfE\lt[\pi(W) \med  W \in (0, M]\rt] 
    \leq \sfE\lt[\pi_-(W)\mathbbm{1}\{W \in (0, M]\}\rt]  \leq \sfE\lt[\pi_-(M)\rt]\ ,
\end{align}
and the proof is complete.
Then, for every $A \geq \kappa M$, we have
\begin{align}
    &\sfE\lt[\pi(W) \med  W \in (0, A)\rt]\\ \label{eq:cond_dec_use}
    &\leq \sfE\lt[\pi(W) \med  W \in (0, M]\rt] + \sfE\lt[\pi(W) \med  W \in (M, A)\rt]\\
    &= \frac{\sfE\lt[\pi(W) \mathbbm{1}\{W \in (0, M]\}\rt]}{\sfP(W \in (0, M])} + \sfE\lt[\sfE\lt[\pi(W) \med W, W \in (M, A)\rt] \med W \in (M, A)\rt]\\ \label{eq:WinM}
    &\leq \frac{\sfE\lt[\pi_-(M)\rt]}{1 - e^{-M}} + \sfE\lt[\cI_\epsilon W \mathbbm{1}\{W \in (M, A) \med W \in (M, A)\rt]\ ,
\end{align}
where the inequality in \eqref{eq:cond_dec_use} follows from Lemma \ref{lem:cond_decomp} and the inequality in \eqref{eq:WinM} follows from \eqref{eq:surv_prob}.
Next, we bound the second term in \eqref{eq:WinM} as follows.
\begin{align}
    &\sfE\lt[\cI_\epsilon W \mathbbm{1}\{W \in (M, A) \med W \in (M, A)\rt]\\
    &= \cI_\epsilon \int_0^\infty \sfP\lt(W \mathbbm{1}\{W \in (M, A) > x \med W \in (M, A)\rt) dx\\
    &= \cI_\epsilon \int_0^A \frac{\sfP\lt(\{x \vee M\} < W < A\rt)}{\sfP\lt(M < W < A\rt)} dx\\
    &= \cI_\epsilon \lt\{\int_0^M \frac{\sfP\lt(M < W < A\rt)}{\sfP\lt(M < W < A\rt)}dx \; + \; \int_M^A \frac{\sfP\lt(x < W < A\rt)}{\sfP\lt(M < W < A\rt)}dx\rt\}\\
    &\leq \cI_\epsilon \lt\{\int_0^M dx \; + \; \int_M^A \frac{\sfP\lt(x < W\rt)}{\sfP\lt(M < W\rt) - \sfP\lt(A < W\rt)}dx\rt\}\\ \label{eq:WinA}
    &\leq \cI_\epsilon \lt\{M \; + \; \int_M^A \frac{e^{-x}}{\sfP\lt(M < W\rt) - e^{-A}}dx\rt\}\\
    &= \cI_\epsilon \lt\{M \; + \; \frac{e^{-M} - e^{-A}}{\sfP\lt(M < W\rt) - e^{-A}}dx\rt\}\\ \label{eq:AgeqkM}
    &\leq \cI_\epsilon \lt\{M \; + \; \frac{e^{-M}}{\sfP\lt(M < W\rt) - e^{-\kappa M}}\rt\}\ ,
\end{align}
where the inequality in \eqref{eq:WinA} follows from \eqref{eq:surv_prob} and the inequality in \eqref{eq:AgeqkM} follows from the fact that $A \geq \kappa M$. The proof is complete.
\end{IEEEproof}

The following lemma is a generalization of Wald's identity that we use repeatedly later. \\

\begin{lemma}\label{lem:cond_wald}
Let $\{Y_n : n \in \bN\}$ be a sequence of Bernoulli random variables, and  set 
\begin{align}
N := \inf\{n \geq 1 : Y_n = 1\}\ .
\end{align}
Furthermore, let
$W$ be a random variable 
so that $Z_1, Y_1, W$ are independent of $\{(Z_n, Y_n) : n \geq 2\}$, and set
\begin{align}
\mu_x \equiv \sfE\lt[Z_1 \, |\, W = x\rt]  \in  \bR\ \qquad \text{and}\  \qquad p_x \equiv \sfP(Y_1 = 1 \,|\, W = x)\ . 
\end{align} 
\begin{itemize}
\item[(i)]
If $\{(Z_n, Y_n) : n \geq 2\}$ are i.i.d. with 
\begin{align}
\sfE\lt[Z_n\rt] = \mu \in \bR \ \qquad \text{and} \qquad \ \sfP(Y_n = 1) = p >0\ ,
\end{align}
then 
\begin{align}
\sfE\lt[\sum_{t = 1}^N Z_t \;\Bigg|\; W = x\rt] = \mu_x + (1 - p_x)\lt(\frac{\mu}{p}\rt)\ .
\end{align}
\item[(ii)] If 
there exists an $L \in \bN$ such that, for every $i \in \{2, \dots, L + 1\}$,  $\{(Z_{jL + i}, Y_{jL + i}) : j \geq 0\}$ are i.i.d. with 
\begin{align}
\sfE\lt[Z_{i}\rt] = \mu_i \in \bR \ \qquad \text{and} \qquad \sfP(Y_{i} = 1) = p_i > 0\ ,
\end{align}
then 
\begin{align}
&\sfE\lt[\sum_{t = 1}^N Z_t \;\Bigg|\; W = x\rt] = p_x \times \frac{\mu_x}{p_x}  + {(1 - p_x)}  \times \sum_{i = 2}^{L+1}\; w_i \, \frac{\mu_i}{p_i}\ , 
\end{align}
where 
\begin{align} \label{w}
w_i \equiv \frac{ p_i \, \prod_{j = 2}^{i-1}\, (1-p_j)}{1 - \prod_{j = 2}^{L+1} \, (1 - p_j)}  \ , 1< i \leq L+1\ ,
\end{align}
\end{itemize}
and $\prod_{j}^i = 1$ whenever $j > i$.\\
\end{lemma}

\begin{IEEEproof}
We first observe that if  $\Exp[N | W = x]<\infty$, then
by Wald's identity we have 
\begin{align} \label{note}
\sfE\lt[\sum_{t = 1}^N Z_t \;\Bigg|\; W = x\rt] &= \sfE\lt[\sum_{t = 1}^N \Exp[ Z_t \, | \, W=x]  \;\Bigg|\; W = x\rt].
\end{align}

(i)  By \eqref{note} we have   
\begin{align}
\sfE\lt[\sum_{t = 1}^N Z_t \;\Bigg|\; W = x\rt] &= \mu_x + \mu \; \lt(\sfE[N | W = x] - 1\rt)\ .
\end{align}
To this point, it suffices to show that 
\begin{align}
\sfE\lt[N \; | \; W = x\rt] = 1 + \frac{1 - p_x}{p}\ .
\end{align}
Indeed, the conditional distribution of $N$ can be easily found as follows.
\begin{align}
\sfP\lt(N = k \, |\,  W = x\rt) =
\begin{cases}
p_x \quad &\text{if} \quad k = 1\ \\
(1 - p_x)(1 - p)^{k-2}p  \quad &\text{if} \quad k \geq 2\ 
\end{cases}\ ,
\end{align}
which further yields that
\begin{align}
\sfE\lt[N \,|\, W = x\rt] &= p_x + \sum_{k = 2}^\infty k (1 - p_x)(1 - p)^{k-2}p\\
&= p_x + (1 - p_x)\sum_{k = 1}^\infty (1 + k) (1 - p)^{k-1}p\\
&= p_x + (1 - p_x)\lt(\sum_{k = 1}^\infty (1 - p)^{k-1}p + \sum_{k = 1}^\infty k (1 - p)^{k-1}p\rt)\\
&= p_x + (1 - p_x)\lt(1 + \frac{1}{p}\rt)\\
&= 1 + \frac{1 - p_x}{p}\ .
\end{align}



(ii) By \eqref{note} we have  
\begin{align} 
\sfE\lt[\sum_{t = 1}^N Z_t \;\Bigg|\; W = x\rt] 
&=\sfE\lt[Z_1 | W = x\rt] + \sum_{t = 2}^\infty \sfE\lt[Z_t\mathbbm{1} \; \{N \geq t\} \;\Bigg|\; W = x\rt]\\ \label{eq:decomN>t} 
&= \mu_x + \sum_{t = 2}^\infty \sfE\lt[Z_t\rt] \; \sfP\lt(N \geq t \med  W = x\rt)\\ \label{eq:waldextn1}
&= \mu_x + \sum_{j = 0}^{\infty}\sum_{t = jL+2}^{(j+1)L+1} \sfE\lt[Z_t\rt] \sfP\lt(N \geq t \med  W = x\rt)\ ,
\end{align}
where the equality in \eqref{eq:decomN>t} follows from the fact that, since $N$ is a stopping time with respect to the filtrations generated by $\{W, Y_n : n \in \bN\}$, for every $t \geq 2$, we have
\begin{align}
\{N \geq t\} \in \sigma\lt(W, Y_k : k \in [t - 1]\rt)\ .
\end{align}
Thus, the event $\{N \geq t\}$ is independent of $(Z_t, Y_t)$, which is also independent of $W$.
To this end, also following \eqref{eq:waldextn1}, it is necessary to derive the expressions for the conditional tail probabilities 
\begin{align}
\sfP\lt(N \geq t \med  W = x\rt)\ , \qquad t \geq 2\ ,    
\end{align}
and we do it as follows. For every $t \geq 2$ in the form
$t = j_tL + i_t$,
for some $j_t \geq 0$ and $i_t \in \{2, \dots, L+1\}$, we have
\begin{align} \label{eq:defN}
\sfP\lt(N \geq t \med  W = x\rt) &= \sfP\lt(Y_k = 0 \;\; \text{for every} \;\; k = 1, \dots, t-1 \; | W = x\rt)\\ \label{eq:decomYs}
&= \sfP(Y_1 = 0 \,|\,  W= x) \prod_{k = 2}^{t-1}\sfP(Y_k = 0)\\ \label{eq:}
&= \sfP(Y_1 = 0 \, | \,  W= x) \prod_{j = 0}^{j_t - 1} \prod_{i = 2}^{L+1} \sfP(Y_{jL + i} = 0)
\times \prod_{i =2}^{i_t - 1}\sfP(Y_{j_tL+i} = 0)\\
&= (1 - p_x) \prod_{j = 0}^{j_t - 1} \prod_{i = 2}^{L+1} (1 - p_i) \times \prod_{i =2}^{i_t - 1} (1 - p_i)\\
&=(1 - p_x) \lt(\prod_{i = 2}^{L+1} (1 - p_i)\rt)^{j_t} \prod_{i =2}^{i_t - 1} (1 - p_i)\ ,
\end{align}
where the equality in \eqref{eq:defN} follows from the definition of $N$, and the equality in \eqref{eq:decomYs} holds since $\{Y_n : n \geq 2\}$ is independent of $W$. Thus, following \eqref{eq:waldextn1},
\begin{align}
& \sum_{j = 0}^{\infty}\sum_{t = jL+2}^{(j+1)L+1} \sfE\lt[Z_t\rt] \sfP\lt(N \geq t \med  W = x\rt)\\
&=  \sum_{j = 0}^{\infty}\sum_{i = 2}^{L+1} \sfE\lt[Z_{jL+i}\rt] \; \sfP\lt(N \geq jL+i \med  W = x\rt)\\
&= \sum_{j = 0}^{\infty}\sum_{i = 2}^{L+1} \mu_i (1- p_x) \lt(\prod_{k = 2}^{L+1} (1 - p_k)\rt)^{j} \prod_{k =2}^{i - 1} (1 - p_k)\\
&=  (1 - p_x) \sum_{j = 0}^{\infty} \lt(\prod_{k = 2}^{L+1} (1 - p_k)\rt)^{j} \sum_{i = 2}^{L+1} \mu_i  \prod_{k =2}^{i - 1} (1 - p_k)\\
&=  (1 - p_x)\lt\{\sum_{i = 2}^{L+1} \mu_i  \prod_{k =2}^{i - 1} (1 - p_k)\rt\}\sum_{j = 0}^{\infty} \lt(\prod_{k = 2}^{L+1} (1 - p_k)\rt)^{j}\\
&= (1 - p_x) 
\sum_{i = 2}^{L+1} \mu_i  \prod_{k =2}^{i - 1} (1 - p_k)  \frac{1}{1 - \prod_{k = 2}^{L+1} (1 - p_k)}\\
&= (1 - p_x)  \, \sum_{i = 2}^{L+1}\; w_i \, \frac{\mu_i}{p_i}\ ,
\end{align}
where $w_i, 2 \leq i \leq L+1$ is defined in \eqref{w}. 
\end{IEEEproof}

\section{Proof of False Alarm Control}
\label{app:extn_ub}
In this section, we prove
Theorem \ref{thm:fa_extn}.
For this, we need to introduce some additional notations, which are also used later in the proof of Theorem \ref{thm:ub_extn}. Specifically,
for every $\e \in \cU$, we  recall  from \eqref{eq:des_asc} the definition of the 
first descending 
ladder variable of the random walk with increments $\{\xi_t^E: t \in \bN\}$, 
\begin{align}\label{eq:ladder}
\zeta^\e_{-} := \inf\left\{n \geq 1 : \sum_{t = 1}^{n} \xi_t^\e < 0\right\}\ ,
\end{align}
for every $A>0$  we denote by $\zeta^\e_A$  the first time this random walk   crosses $A$ for the first time, i.e., 
\begin{align}\label{eq:crossA}
\zeta^\e_A := \inf\left\{n \geq 1 : \sum_{t = 1}^{n} \xi_t^\e > A\right\}\ , 
\end{align}
and we set  
\begin{align}
\zeta^\e := \min\{\zeta^\e_-, \zeta^\e_A\}\ \qquad \text{and} \qquad \ z^\e := \mathbbm{1}\{\zeta^\e_A < \zeta^\e_-\}\ .
\end{align}

\subsection{Proof of Theorem \ref{thm:fa_extn}}\label{pf:thm:fa_extn}

\begin{IEEEproof}
For every $n \in \bN$ we set 
\begin{align}
\tau_{n} &:=  \inf \{t > \tau_{n-1}: \tilde{S}_{t + 1} \neq \tilde{S}_{\tau_{n-1} + 1} \}\ , \quad \text{where} \quad \tau_0 \equiv 0\ ,\\
z_n &:= \mathbbm{1}\{ Y_{\tau_n} \geq A\}\ .
\end{align}
That is, $\tau_n$ is the time instant at which the sampling unit changes for the $n^{th}$ time, and $z_n$ is equal to 1 if the alarm is raised at time $\tau_n$ and  0 otherwise. Then, we have the 
following representation:
\begin{align}
\tilde{T} = \sum_{n=1}^{M} \zeta_n\ ,
\end{align}
where $M$ denotes the stage at which the process is terminated, i.e., 
\begin{align}
M := \inf\{n \in \bN: z_n = 1\}\ ,
\end{align}
and,  for every $n \in \bN$,  
$\zeta_n := \tau_n - \tau_{n-1}$.
Due to independence over time,  $\{(\zeta_n, z_n) : n \in \bN\}$ is a sequence of independent random vectors under $\bP_\infty$. Furthermore,  since for the sampling policy $\tilde{S}$ we consider a pre-specified permutation $\e_1, \e_2, \dots, \e_{|\cU|}$ of the elements in $\cU$ that remains unchanged over time, then, clearly,  for every $i \in [|\cU|]$, 
the sequence
\begin{align}
\{(\zeta_{j|\cU| + i}, z_{j|\cU| + i}) : j \geq 0\} 
\end{align}
 is i.i.d. under $\bP_\infty$ with  common expectation $(\bE_\infty[\zeta^{\e_i}], \ \bP_\infty(z^{\e_i} = 1))$.
Thus,
\begin{align}
\bE_\infty[\tilde{T}] &= \bE_\infty\lt[\sum_{n=1}^{M} \zeta_n\rt]\\ \label{eq:appl_mod_wald}
&= \sum_{i = 1}^{|\cU|}\bE_\infty[\zeta^{\e_i}] \; \frac{\prod_{j = 1}^{i-1}(1 - \bP_\infty(z^{\e_j} = 1))}{1 - \prod_{j = 1}^{|\cU|} (1 - \bP_\infty(z^{\e_j} = 1))}\\ \label{ineq:exp_stop1}
&\geq \sum_{i = 1}^{|\cU|}\frac{(1 - e^{-A})^{i-1}}{1 - (1 - e^{-A})^{|\cU|}} = e^A \, \frac{1 - (1 - e^{-A})^{|\cU|}}{1 - (1 - e^{-A})^{|\cU|}}\\
&= e^A\ ,
\end{align}
which proves that setting $A=\log \gamma$ guarantees the desired error control. The 
equality in \eqref{eq:appl_mod_wald} holds by an application of Lemma \ref{lem:cond_wald}(ii), whereas  the inequality in \eqref{ineq:exp_stop1} holds since $\zeta^{\e_i} \geq 1$ for every $i \in [|\cU|]$,
and also because 
by Wald's likelihood ratio identity we have, for every $\e \in \cU$,
\begin{align}
\bP_\infty(z^\e = 1) &= \bP_\infty(\zeta^\e_A < \zeta^\e_-) \\
&= \bP_\infty(\xi_{\zeta^\e}^\e > A) \leq e^{-A}\ . 
\end{align}
\end{IEEEproof}

\section{Proof of Asymptotic Upper Bound on Detection Delay 
}\label{pf:thm:ub_extn} In this appendix, we prove
Theorem \ref{thm:ub_extn}.
For this, we need to show that, for any $\nu \geq 0$, as $A \to \infty$ we have 
\begin{align}
\text{esssup}\ \bE^{\sfG}_\nu\left[\tilde{T} - \nu\med \cF^{\tilde{S}}_\nu(\sfG), \tilde{T} > \nu \right]
 \leq \max_{\e \in \cA(\sfG) \cap \cU} \; \frac{A}{\cJ^\e_0(\sfG)} + o(A)\ . 
\end{align}
Thus, throughout this appendix, we fix an arbitrary  $\nu \geq 0$. This proof is quite lengthy, and requires introducing some additional notation.
 
 \subsection{Stages}
 First, we introduce the  sequence of random times, \textit{starting from  $\nu$}, at which the sampled unit changes, i.e., 
\begin{align}
\tau_{n} &:=  \inf \{t > \tau_{n-1}: \tilde{S}_{t + 1} \neq \tilde{S}_{\tau_{n-1} + 1} \}\ , \quad n \in \bN\ , \\
 \tau_0 &:= \nu\ .
\end{align}
For every $n \in \bN$,
\begin{itemize}
\item  we refer to  the time interval 
$[\tau_{n-1} - \tau_{n}]$ as the $n^{\rm th}$ ``stage'' of the sampling procedure, 
\item we denote  by $\zeta_n$ the duration of the $n^{th}$ stage, i.e., 
\begin{align}
\zeta_n := \tau_n - \tau_{n-1}\ , 
\end{align}

\item we denote by $z_n$ the indicator that is equal to 1 if and only if  the process is terminated at the end of the $n^{\rm th}$ stage, i.e.,   
\begin{align}
z_n := \mathbbm{1}\{ Y_{\tau_n} \geq A\}\ ,
\end{align}
\item we denote by $w_n$ the  quantity that is equal to 1 if the unit that is sampled at the $n^{th}$ stage is actually affected by the change  and  $0$ otherwise, i.e., 
\begin{align}
w_n := \mathbbm{1}\{S_{\tau_n} \in \cA(\sfG) \}\ .
\end{align}
\end{itemize}
We stress that since the permutation of the units is pre-specified, the sequence $\{w_n : n \in \bN\}$ is deterministic. Furthermore, we denote by 
\begin{itemize}
\item $\{(\zeta'_n, z'_n) : n \in \bN\}$  the subsequence of $\{(\zeta_n, z_n) : n \in \bN\}$  that corresponds to $w_n = 1$, 

\item  $\{(\zeta''_n, z''_n) : n \in \bN\}$ the subsequence of $\{(\zeta_n, z_n) : n \in \bN\}$  that corresponds to  $w_n = 0$,

\item $M_\sfG$ the total number of stages until 
the process is terminated by a stage at which the sampled unit is actually affected by the change, i.e.,
\begin{align}
M_\sfG := \inf\{n \in \bN : z_n = 1, \quad w_n = 1\}\ ,
\end{align}
\item $M'_\sfG$ the number of stages until the above-mentioned termination 
at which the sampled units are affected by the change, i.e., 
\begin{align}
M_\sfG' := \sum_{n=1}^{M_\sfG} \mathbbm{1}\{w_n=1\}
= \inf \{n \in \bN : z'_n = 1\}\ .
\end{align}
\end{itemize}
We have the following representation that, given $Y_\nu$, 
\begin{align}\label{eq:delta1_extn}
\zeta_1 &= \inf\left\{n \geq 1 : Y_\nu + \sum_{t = 1}^{n} \xi_{\nu + t}^{\tilde{S}_\nu} \notin (0, A)\right\}\ ,\\
\zeta_n &= \inf\left\{n \geq 1 : \sum_{t = 1}^{n} \xi_{\tau_{n-1} + t}^{\tilde{S}_{\tau_{n-1}}} \notin (0, A)\right\}\ , \qquad \forall \;\; n \geq 2\ . \label{eq:delta2_extn}
\end{align}
Thus, due to the independence of the observations over time, under $\bP_\nu^\sfG$,
$\{(\zeta_n, z_n) : n \geq 2\}$  are independent as well as independent of $Y_\nu$ and $(\zeta_1, z_1)$. 

 \subsection{Cycles}
For every $k \in \bN$, \begin{itemize}
 \item 
 we refer to the stages from $\{(k-1)\, |\cU| + 2\}$ to $\{k\, |\cU| + 1\}$ as the $k^{\rm th}$ ``cycle'' of the sampling process.
To be precise, the $1^{\rm st}$ cycle starts just after the end of $1^{\rm st}$ stage, and afterwards, every time the process completes $|\cU|$ stages, we define its duration as a cycle. 
\item we denote by  $\sigma_k$ the total duration of only  those stages  in the  $k^{\rm th}$ cycle at which the units that are not affected by the change are sampled, i.e., 
\begin{align}
\sigma_k=\sum_{n=2}^{|\cU|+1} \zeta_{(k-1)|\cU|+n} \, \mathbbm{1}\{w_{(k-1)|\cU|+n} =0\}\ , \quad k \in \mathbb{N}\ . 
\end{align}

\item  we denote by  $N_\sfG$ the number of cycles the sampling process has undergone before its termination, i.e., \begin{align}
N_\sfG := 
\inf \left\{k \in \bN : ((k-1)|\cU| + 2) \leq M_\sfG \leq (k|\cU| + 1) \right\}\ .
\end{align} 
\end{itemize}

\subsection{An upper bound on the worst-case detection delay}

We next upper bound the worst-case detection delay of $\tilde{T}$ using the previous quantities. \\

\begin{lemma}\label{lem:decom_3parts}
On the event $\{\tilde{T}>\nu\}$ we have 
\begin{align}\label{eq:decom_3parts}
\tilde{T} - \nu \leq  \zeta_1 \cdot \mathbbm{1} \{\tilde{S}_\nu \notin \cA(\sfG)\} + \sum_{n=1}^{M'_\sfG} \zeta'_n + \sum_{m=1}^{N_\sfG} \sigma_m\ ,
\end{align}
and, consequently, 
\begin{align}
&\text{esssup}\ \bE^{\sfG}_\nu\left[\tilde{T} - \nu\med \cF^{\tilde{S}}_\nu(\sfG), \tilde{T} > \nu \right]\\  \label{eq:esssup_extn}
&\leq \max \Bigg\{\max_{\e \in \cA(\sfG) \cap \cU} \; \bE^{\sfG}_\nu\left[ \sum_{n = 1}^{M'_\sfG} \zeta'_n + \sum_{m=1}^{N_\sfG} \sigma_m \, \med \, \tilde{S}_\nu = \e,\,  Y_\nu, \,  Y_\nu \in (0,A)  \right]\ ,\\
&\qquad \qquad \,  \max_{\e \in \cU \setminus \cA(\sfG)} \; \bE^{\sfG}_\nu\left[\zeta_1+  \sum_{n = 1}^{M'_\sfG} \zeta'_n + \sum_{m=1}^{N_\sfG} \sigma_m  \, \med \, \tilde{S}_\nu = E, \, Y_\nu \in (0,A) \right]  \Bigg\}\ .
\end{align}
\end{lemma}

\begin{IEEEproof}
By the above definitions it is clear that 
\begin{align}
\tilde{T} - \nu \leq  \sum_{n = 1}^{M_\sfG} \zeta_n\ ,
\end{align} 
and
\begin{align} &\sum_{n=1}^{M_\sfG} \zeta_n \mathbbm{1}\{w_n=1\} =
\sum_{n=1}^{M'_\sfG} \zeta'_n\ .
\end{align}
Furthermore, we have
\begin{align}\label{eq:T2ineq1_extn}
\sum_{n=1}^{M_\sfG} \zeta_n \mathbbm{1}\{w_n=0\} = \zeta_1 \mathbbm{1}\{w_1=0\} + \sum_{n=2}^{M_\sfG} \zeta_n \mathbbm{1}\{w_n=0\} \leq \zeta_1 \mathbbm{1} \{\tilde{S}_\nu \notin \cA(\sfG)\} + \sum_{m=1}^{N_\sfG} \sigma_m\ ,
\end{align}
where the inequality holds since by definition, cycles are counted from the $2^{\rm nd}$ stage, and 
when $\tilde{S}_\nu \notin \cA(\sfG)$, i.e., at the first stage, sampling is done from a unit which is not affected by the change, we have $w_1 = 0$. 
This completes the first part. 

Now, we have
\begin{align} \label{eq:TQ0}
 \bE^{\sfG}_\nu\left[\tilde{T} -\nu \med \cF^{\tilde{S}}_\nu(\sfG), \tilde{T} > \nu \right]
&= \bE^{\sfG}_\nu\left[\tilde{T}-\nu \med \tilde{S}_\nu \in \cA(\sfG), \, \cF^{\tilde{S}}_\nu(\sfG), \, \tilde{T} > \nu \right]\\
&= \bE^{\sfG}_\nu\left[\tilde{T} -\nu \med \tilde{S}_\nu \in \cA(\sfG), \, \cF^{\tilde{S}}_\nu(\sfG),\,\tilde{T} > \nu \right]\mathbbm{1}\{\tilde{S}_\nu \in \cA(\sfG)\}\\
&\quad + \bE^{\sfG}_\nu\left[\tilde{T}-\nu  \med \tilde{S}_\nu \notin \cA(\sfG), \, \cF^{\tilde{S}}_\nu(\sfG),\, \tilde{T} > \nu \right]\mathbbm{1}\{\tilde{S}_\nu \notin \cA(\sfG)\}\\ \label{eq:TQ2}
&= \bE^{\sfG}_\nu\left[\tilde{T}-\nu  \med \tilde{S}_\nu \in \cA(\sfG), \, Y_\nu,\,Y_\nu \in (0,A)  \right] \mathbbm{1}\{\tilde{S}_\nu \in \cA(\sfG)\}\\ \label{eq:TQ3}
&\quad + \bE^{\sfG}_\nu\left[\tilde{T}-\nu \med \tilde{S}_\nu \notin \cA(\sfG), \,  Y_\nu \in (0,A)   \right]\mathbbm{1}\{\tilde{S}_\nu \notin \cA(\sfG)\}\ ,
\end{align}
where the equality in \eqref{eq:TQ0} follows from the fact that $\{\tilde{S}_\nu \in \cA(\sfG)\} \in \cF_\nu^{\tilde{S}}(\sfG)$. The equality in \eqref{eq:TQ2} holds because given $\tilde{S}_\nu \in \cA(\sfG)$ and $\tilde{T} > \nu$, the detection delay $\tilde{T} - \nu$ depends on the history $\cF_\nu^{\tilde{S}}(\sfG)$ only through $Y_\nu$, i.e., the value of the statistic at time $\nu$. However, given $\tilde{S}_\nu \notin \cA(\sfG)$,  $Y_\nu$ is not observed from the history $\cF_\nu^{\tilde{S}}(\sfG)$, hence, the detection delay is independent of $\cF_\nu^{\tilde{S}}(\sfG)$. The result now follows from the first part and the following identities:
\begin{align}
\{\tilde{S}_\nu \in \cA(\sfG)\} = \bigcup_{\e \in \cA(\sfG) \cap \cU} \; \{\tilde{S}_\nu = \e\} \ \qquad \text{and} \ \qquad 
\{\tilde{S}_\nu \notin \cA(\sfG)\} = \bigcup_{\e \in \cU \setminus \cA(\sfG)} \; \{\tilde{S}_\nu = \e\}\ .
\end{align}

\end{IEEEproof}
In view of Lemma \ref{lem:decom_3parts}, we need to upper bound the conditional expectation of each of the quantities in the right-hand side of \eqref{eq:decom_3parts}. This is done in Lemmas \ref{lem:M'_G_ij}, \ref{lem:zeta_1_notinA}, \ref{lem:M'_G_notinA} and \ref{lem:bd_cycle}. For some of these lemmas, we need to introduce some additional  notations and state some supporting lemmas.

\subsection{Random walk quantities}
Recall the stopping times introduced in \eqref{eq:ladder} and \eqref{eq:crossA}.
Now, for every $\e \in \cU$,
\begin{itemize}
\item we denote by  $q^\e_+$ 
the probability that, while sampling from $\e$, $\xi_n^E$  never exceeds  $0$ under $\sfF$, i.e.,
\begin{align}
q^\e_+ := \sfF(\zeta^\e_+ = \infty)\ , 
\end{align} 
\item when $\e \in \cA(\sfG)$, we denote  by $q^\e_-(\sfG)  
$ the probability that, while sampling from $\e$, the random walk never falls below 0 under $\sfG$, i.e., 
\begin{align}
q^\e_-(\sfG) := \sfG(\zeta^\e_- = \infty)\ .
\end{align}
[We note that under assumptions \eqref{assum:pos_I_mix_1_a}-\eqref{assum:pos_I_mix_1_b}, both these probabilities are  strictly positive (see, e.g.,\cite[Corollary 8.39]{BookSeig}, and also Lemma \ref{lem:mu/p} in Appendix \ref{app_subsec:implems}).]
\item when $\e \in \cA(\sfG)$, we denote by  $\mu^\e_A(\sfG)$ the expectation of $\zeta^\e$, and by $q^\e_A(\sfG)$  the probability that, while sampling from $\e$, the corresponding LLR process crosses $A$ for the first time before falling below $0$ under $\sfG$, i.e., 
\begin{align}
\mu^\e_A(\sfG) := \sfE_{\sfG}[\zeta^\e] \qquad \ \text{and} \ \qquad q^\e_A(\sfG) := \sfE_{\sfG}[z^\e] = \sfG(\zeta^\e_A < \zeta^\e_-)\ .
\end{align}
\end{itemize}

\subsection{Cycle quantities}
Recall that we have fixed an arbitrary permutation, $(\e_1, \dots, \e_{|\cU|})$, of the elements of $\cU$ to start the sampling process $\tilde{S}$. Next, for any $\sfG \in \cG$, order the indices in this particular permutation, which correspond to the elements of $\cA(\sfG)$, and denote them as follows.
\begin{align}\label{eq:permut}
1 \leq i_1(\sfG) < i_2(\sfG) < \dots < i_{|\cA(\sfG) \cap \cU|}(\sfG) \leq |\cU|\ .
\end{align}
It is important to note that the 
above ordering depends on what permutation we determine at the beginning of sampling.
Subsequently, we can also represent $\cA(\sfG)$ as 
\begin{align}
\cA(\sfG) \cap \cU = \{\e_{i_j(\sfG)} : j \in [|\cA(\sfG) \cap \cU|]\}\ .
\end{align}
Next, for any $j \in [|\cA(\sfG) \cap \cU|]$, 
\begin{itemize}
\item  when $\e$ is  the $j^{\rm th}$ element of $\cA(\sfG) \cap \cU$ in the  the above ordering, i.e.,
$\e=\e_{i_j(\sfG)} \in \cA(\sfG) \cap \cU$, we set 
\begin{align}
\tilde{\zeta}^j := \zeta^\e\ &,\quad \quad \ \tilde{z}^j := z^\e\ ,\\
\tilde{q}^j_{-}(\sfG) := q^\e_-(\sfG)\ &,\quad \quad \ \tilde{q}^j_{A}(\sfG) := q^\e_A(\sfG)\ ,\quad \quad \ \tilde{\mu}_A^j(\sfG) := \mu^\e_A(\sfG)\ ,\\
\tilde{\cJ}^j_0(\sfG) := \cJ^\e_0 (\sfG)\ &,\quad \quad \tilde{\cW}^j_0(\sfG) := \cW^\e_0(\sfG)\ ,
\end{align}

\item  for any $l \in \bN$, we denote by $\kappa(j, l)$ the $l^{\rm th}$ element of the following periodic sequence, which starts from $j$ and has period $|\cA(\sfG) \cap \cU|$,  
\begin{align}
\lt\{j, j+1, \dots, |\cA(\sfG) \cap \cU|, 1, 2, \dots, j-1, j, j+1, \dots, |\cA(\sfG) \cap \cU|, 1, 2, \dots\rt\}\ ,
\end{align}
i.e., more formally,
\begin{align}
\label{eq:k(j, l)}
\kappa(j, l) := (j + l-1)\; \; {\rm mod} \; |\cA(\sfG) \cap \cU| \; \in \; [|\cA(\sfG) \cap \cU|]\ ,
\end{align}

\item we define the \textit{$j$-cycle} to be the total duration in which sampling is performed only from the elements of $\cA(\sfG) \cap \cU$ starting with the element corresponding to the $j^{\rm th}$ index in \eqref{eq:permut}, i.e., $\e_{i_j(\sfG)}$, until the process stops or returns to start sampling again from $\e_{i_j(\sfG)}$,
\item  we introduce the quantity 
$\tilde{\cJ}^j_*(\sfG)$, which is a weighted harmonic mean of the 
information numbers corresponding to the elements of $\cA(\sfG) \cap \cU$, i.e. 
\begin{align}
\frac{1}{\tilde{\cJ}^j_*(\sfG)} := \sum_{l = 1}^{|\cA(\sfG) \cap \cU|} w^j_l \, \frac{1}{\tilde{\cJ}^{\kappa(j, l)}_0(\sfG)}\ ,
\end{align}
where the weight corresponding to any element is  the conditional probability that during the $j$-cycle we stop while sampling from that particular element given that we stop in the $j$-cycle, i.e.,  
\begin{align}\label{eq:qtt_w}
    w^j_l= \frac{{\tilde{q}_A^{\kappa(j, l)}(\sfG) \; \prod_{s = 1}^{l-1} (1 - \tilde{q}_A^{\kappa(j, s)}(\sfG))}}{1 - \prod_{s = 1}^{|\cA(\sfG) \cap \cU|} (1 - \tilde{q}_A^{\kappa(j, s)}(\sfG))}\ , \ \qquad 
\text{for every} \quad l \in [|\cA(\sfG) \cap \cU|]\ .
\end{align}
Since the inverse of the information number of any unit in $\cA(\sfG) \cap \cU$ represents the delay or difficulty of detection when sampling only from that element after the change,  $\tilde{\cJ}^j_*(\sfG)$ represents the overall difficulty of detection averaged over all elements in $\cA(\sfG) \cap \cU$ when $\e_{i_j(\sfG)}$ is the first element in $\cA(\sfG) \cap \cU$ that we sample from immediately after the change. Hence, $\tilde{\cJ}^j_*(\sfG)$ can be interpreted as the information number corresponding to the $j$-cycle.
\end{itemize}

\subsection{Some useful lemmas}

Next, in the following lemmas, we establish upper bounds of the conditional expectation of each of the quantities in the right-hand side of \eqref{eq:decom_3parts}, and combine them in the end to obtain the desired result. \\

\begin{lemma}\label{lem:k(j,l)}
Fix $\sfG \in \cG$ and
recall the definition of $\kappa(j, l)$ in \eqref{eq:k(j, l)} for any $j \in [|\cA(\sfG) \cap \cU|]$ and $l \in \bN$.
Then
\begin{align}
\kappa(j, l + 1) = \kappa(\kappa(j, 2), l)\ .
\end{align}
\end{lemma}
\begin{IEEEproof}
We have
\begin{align}
\begin{split}
\kappa(\kappa(j, 2), l)
&= \kappa((j + 1) \; {\rm mod} \;|\cA(\sfG) \cap \cU|, \ l)\\
&= ((j + 1) \; {\rm mod} \;|\cA(\sfG) \cap \cU| + l - 1)\; {\rm mod}\;  |\cA(\sfG) \cap \cU|\\
&= ((j + 1) \; {\rm mod} \;|\cA(\sfG) \cap \cU|) \; {\rm mod} \; |\cA(\sfG) \cap \cU| +  (l - 1) \; {\rm mod} \; |\cA(\sfG) \cap \cU|\\
&= (j + 1) \; {\rm mod} \;|\cA(\sfG) \cap \cU| +  (l - 1) \; {\rm mod}\;  |\cA(\sfG) \cap \cU|\\
& = (j + l) \; {\rm mod} \; |\cA(\sfG) \cap \cU|\\
&= (j + (l + 1) - 1) \; {\rm mod} \;|\cA(\sfG) \cap \cU| = \kappa(j, l + 1)\ ,
\end{split}
\end{align}
which completes the proof.
\end{IEEEproof}

\begin{lemma}\label{lem:M'_G_ij}
For any $j \in [|\cA(\sfG) \cap \cU|]$, 
\begin{align}
\bE^{\sfG}_\nu \left[\sum_{n=1}^{M'_\sfG} \zeta'_n  \, \med \, \tilde{S}_\nu = \e_{i_j(\sfG)}, Y_\nu, Y_\nu \in (0,A) \right] 
   &\leq \lt(\frac{A}{\tilde{\cJ}_0^{j}(\sfG)} \bigvee \frac{A}{\tilde{\cJ}_*^{\kappa(j, 2)}(\sfG)}\rt) + o(A)\ .
\end{align}

\end{lemma}

\begin{IEEEproof}
 First, fix any arbitrary $j \in [|\cA(\sfG) \cap \cU|]$ and consider conditioning on
\begin{align}
\{\tilde{S}_\nu = \e_{i_j(\sfG)}, Y_\nu, \tilde{T} > \nu\}\ .
\end{align}
Now, since $\tilde{S}_\nu = \e_{i_j(\sfG)}$, clearly, at the first stage sampling is done from a unit that belongs to $\cA(\sfG)$, i.e., $w_1 = 1$ or $\zeta'_1 = \zeta_1$. 
Thus, we define the following conditional expectations
\begin{align}
\mu^j_{Y_\nu} &:= \bE_\nu^\sfG[\zeta'_1 \, | \, \tilde{S}_\nu = \e_{i_j(\sfG)}, Y_\nu, Y_\nu \in (0, A)]\ ,\\ \label{eq:p_yj}
p^j_{Y_\nu} &:= \bP_\nu^{\sfG}(z_1' = 1 | \tilde{S}_\nu = \e_{i_j(\sfG)}, Y_\nu, Y_\nu \in (0, A))\ .
\end{align}
Furthermore, under $\bP_\nu^\sfG$, $\{(\zeta'_n, z'_n) : n \geq 2\}$ is independent of $\{\tilde{S}_\nu = \e_{i_j(\sfG)}\}$, $Y_\nu$ and $(\zeta'_1, z_1)$, and for every $l \in \{2, \dots, |\cA(\sfG) \cap \cU| + 1\}$,
the following sequence is i.i.d.:
\begin{align}
\{(\zeta'_{k|\cA(\sfG) \cap \cU| + l}, z'_{k|\cA(\sfG) \cap \cU| + l}) : \; k \geq 0\}\ , 
\end{align}
with the common expectation $(\tilde{\mu}_A^{\kappa(j, l)}(\sfG), \ \tilde{q}_A^{\kappa(j, l)}(\sfG))$.
Also, note that, in this case, $M'_\sfG \geq 1$ is a stopping time with respect to the filtration generated by $\{Y_\nu, Y_\nu \in (0, A), z'_n : n \in \bN\}$.
Then,
\begin{align}
&\bE_\nu^\sfG\lt[\sum_{n=1}^{M'_\sfG} \zeta'_n \;\Bigg|\; \tilde{S}_\nu = \e_{i_j(\sfG)}, Y_\nu, Y_\nu \in (0, A)\rt]\\ \label{eq:use_cond_lemma}
&= p^j_{Y_\nu} \times \frac{\mu_{Y_\nu}^j}{p^j_{Y_\nu}} + (1 - p^j_{Y_\nu})
\times \sum_{l = 2}^{|\cA(\sfG) \cap \cU|+1}w_{l-1}^{\kappa(j, 2)} \; \frac{\tilde{\mu}_A^{\kappa(j, l)}(\sfG)}{\tilde{q}_A^{\kappa(j, l)}(\sfG)}\\ 
&= p^j_{Y_\nu} \times \frac{\mu_{Y_\nu}^j}{p^j_{Y_\nu}} + (1 - p^j_{Y_\nu})
\times \sum_{l = 1}^{|\cA(\sfG) \cap \cU|}w_{l}^{\kappa(j, 2)} \; \frac{\tilde{\mu}_A^{\kappa(j, l+1)}(\sfG)}{\tilde{q}_A^{\kappa(j, l+1)}(\sfG)}\\ \label{eq:use_lem_k}
&= p^j_{Y_\nu} \times \frac{\mu_{Y_\nu}^j}{p^j_{Y_\nu}} + (1 - p^j_{Y_\nu})
\times \sum_{l = 1}^{|\cA(\sfG) \cap \cU|}w_{l}^{\kappa(j, 2)} \; \frac{\tilde{\mu}_A^{\kappa(\kappa(j, 2), l)}(\sfG)}{\tilde{q}_A^{\kappa(\kappa(j, 2), l)}(\sfG)}\\ \label{eq:use_lem_mu/p}
&\leq p^j_{Y_\nu}\lt(\frac{A - Y_\nu}{\tilde{\cJ}_0^{j}(\sfG)}\mathbbm{1}\{Y_\nu \in (0, A)\} + o(A)\rt) + (1 - p^j_{Y_\nu})\sum_{l = 1}^{|\cA(\sfG) \cap \cU|}w_{l}^{\kappa(j, 2)}\lt( \frac{A}{\tilde{\cJ}^{\kappa(\kappa(j, 2), l)}_0(\sfG)} + o(A)\rt)\\
&\leq p^j_{Y_\nu}\lt(\frac{A}{\tilde{\cJ}_0^{j}(\sfG)}\rt) + (1 - p_{Y_\nu})\lt(\frac{A}{\tilde{\cJ}_*^{\kappa(j, 2)}(\sfG)}\rt) + o(A)\\
&\leq \lt(\frac{A}{\tilde{\cJ}_0^{j}(\sfG)} \bigvee \frac{A}{\tilde{\cJ}_*^{\kappa(j, 2)}(\sfG)}\rt) + o(A)\ ,
\end{align}
where the constants $\{w_l^j:\; j, l \in [|\cA(\sfG) \cap \cU|]\}$ are defined in \eqref{eq:qtt_w}.
In the above, the equality in \eqref{eq:use_cond_lemma} follows by a direct application of Lemma \ref{lem:cond_wald}(ii), the one in \eqref{eq:use_lem_k} follows by using Lemma \ref{lem:k(j,l)}, and the inequality in \eqref{eq:use_lem_mu/p} follows by a direct application of Lemma \eqref{lem:mu/p}. 
The proof is complete.

\end{IEEEproof}

\begin{lemma}\label{lem:zeta_1_notinA}
For every $\e \in \cU \setminus \cA(\sfG)$ and $A > 0$ large enough, we have
\begin{align}
\label{eq:ranYnu} \bE_\nu^{\sfG}\lt[\zeta_1 \med  \tilde{S}_\nu = \e, Y_\nu \in (0, A) \rt]
&\leq  C\ ,
\end{align}
    where $C > 0$ is a finite constant independent of $A$.
\end{lemma}

\begin{IEEEproof}
We have
\begin{align} 
\bE_\nu^{\sfG}\lt[\zeta_1 \med  \tilde{S}_\nu = \e, \, Y_\nu \in (0, A) \rt] \label{eq:del''indep}
&= \bE_\nu^{\sfG}\lt[\zeta''_1 \med  \tilde{S}_\nu = \e, Y_\nu \in (0, A)\rt]\\ \label{eq:directLemmaB2}
&\leq C\ ,
\end{align}
where the equality in \eqref{eq:del''indep} follows from the facts that, given $\tilde{S}_\nu = \e$, we have $\zeta_1 = \zeta''_1$, 
and the one in \eqref{eq:directLemmaB2} follows by a direct application of Lemma \ref{lem:first_cycle_worstcase} since the unit sampled in $\zeta''_1$, under $\bP_\nu^\sfG$, follows the local distribution $\sfF^\e$.\\
\end{IEEEproof}

\begin{lemma}\label{lem:M'_G_notinA}
For  any $\e \in \cU \setminus \cA(\sfG)$,
\begin{align}
\bE^{\sfG}_\nu\left[\sum_{n = 1}^{M'_\sfG} \zeta'_n  \med \tilde{S}_\nu = \e, \tilde{T} > \nu \right] \leq \max_{j \in [|\cA(\sfG) \cap \cU|]} \; \frac{A}{\tilde{\cJ}_*^j (\sfG)} + o(A)\ .
\end{align}
\end{lemma}

\begin{IEEEproof}
Suppose $j \in [|\cA(\sfG) \cap \cU|]$ is such that $\e_{i_j(\sfG)}$ is the immediate next unit after $\e$, from where we are supposed to sample according to the pre-fixed permutation of the units.    
Then, $\{\zeta'_n : n \in \bN\}$ is a sequence of positive and independent random variables such that, under $\bP_\nu^\sfG$, for every $l \in \{1, \dots, |\cA(\sfG) \cap \cU|\}$,
the following sequence is i.i.d.:
\begin{align}
\{(\zeta'_{k|\cA(\sfG) \cap \cU| + l}, z'_{k|\cA(\sfG) \cap \cU| + l}) : \; k \geq 0\}\ , 
\end{align}
with the common expectation $(\tilde{\mu}_A^{\kappa(j, l)}(\sfG), \ \tilde{q}_A^{\kappa(j, l)}(\sfG))$,
and is independent of $\{\tilde{T} > \nu\}$. As a consequence, $M'_\sfG$ is a stopping time with respect to the filtration generated by $\{\zeta'_n : n \in \bN\}$, and
\begin{align} \label{eq:applyLemC1toM}
\bE_\nu^{\sfG}\lt[\sum_{n = 1}^{M'_\sfG} \zeta'_n \;\Bigg|\; \tilde{S}_\nu = \e\rt]
&= \sum_{l = 1}^{|\cA(\sfG) \cap \cU|}
w_l^j \; 
\frac{\tilde{\mu}_A^{\kappa(j, l)}(\sfG)}{\tilde{q}_A^{\kappa(j, l)}(\sfG)} \\ \label{ineq:AI_ineq}
&\leq \sum_{l = 1}^{|\cA(\sfG) \cap \cU|}
w_l^j \; 
\lt(\frac{A}{\tilde{\cJ}_0^{\kappa(j, l)}(\sfG)} + o(A) 
\rt)\\
&\leq \frac{A}{\tilde{\cJ}_*^j (\sfG)} + o(A)\\ 
&\leq \max_{j \in [|\cA(\sfG) \cap \cU|]} \; \frac{A}{\tilde{\cJ}_*^j (\sfG)} + o(A)\ , 
\end{align}
where the equality in \eqref{eq:applyLemC1toM} follows by using Lemma \ref{lem:cond_wald}(ii), and the inequality in \eqref{ineq:AI_ineq} follows by a direct application of Lemma \ref{lem:mu/p}.\\ 
\end{IEEEproof}

\begin{lemma}\label{lem:bd_cycle}
For  any $\e \in \cU \setminus \cA(\sfG)$ and  $j \in [|\cA(\sfG) \cap \cU|]$, we have 
\begin{align}   
\bE^{\sfG}_\nu\left[ \sum_{m=1}^{N_\sfG} \sigma_m \,  \med \, \tilde{S}_\nu = \e_{i_j(\sfG)}, Y_\nu, Y_\nu \in (0, A)\right] 
  \; \leq \; \bE_\nu^\sfG\lt[\sum_{m=1}^{N_\sfG} \sigma_m \;\Bigg|\; \tilde{S}_\nu = \e, Y_\nu \in (0, A)\rt] \leq 
  \; \text{constant}\ , 
\end{align}
where 
\begin{align}
\text{constant} \; =
\frac{1}{1 - \prod_{j = 1}^{|\cA(\sfG) \cap \cU|} (1 - \tilde{q}_{-}^j(\sfG))} \sum_{\e \in \cU \setminus \cA(\sfG)} \frac{1}{q^\e_{+}}\ .
\end{align}

\end{lemma}

\begin{IEEEproof} Since $\zeta_1 = \zeta''_1$, we have $M_{\sfG} \geq 1$ and thus, $N_\sfG \geq 1$ is a Geometric random variable with parameter, say $q_A$. 
Therefore, we have
\begin{align}
\label{ineq:T2mindelNQ} \bE_\nu^\sfG\lt[\sum_{m=1}^{N_\sfG} \sigma_m \;\Bigg|\; \tilde{S}_\nu = \e, Y_\nu \in (0, A)\rt] \; = \; \bE_\nu^\sfG\lt[\sum_{m=1}^{N_\sfG} \sigma_m \;\Bigg|\; \tilde{S}_\nu = \e \rt]
\; = \; \frac{\bE_\nu^\sfG[\sigma_1]}{q_A}\ ,
\end{align}
where the first equality in \eqref{ineq:T2mindelNQ} follows from the fact that the distributions of the cycles are independent of $\{Y_\nu \in (0, A)\}$, and the second one follows by an application of Wald's identity, where $q_A$ 
is the probability that the process is terminated in the first cycle, i.e.,  
\begin{align}
q_A := \bP_\nu^\sfG( z_n=1 \;\;  \text{for some} \;\; 2 \leq n \leq |\cU| + 1 \;\; \text{with} \;\; w_n =1 )\ .
\end{align}

Since we stop only while sampling a unit in $\cA(\sfG)$, when $\tilde{S}_\nu \in \cA(\sfG)$, we have
\begin{align}
    N_\sfG = 0 \quad \text{if and only if} \quad z'_1 = 1\ .
\end{align}
Otherwise, since the cycles are i.i.d., the probability of termination of the process at any of the cycles is $q_A$. Therefore, under $\bP_\nu^\sfG$, the conditional distribution of $N_\sfG$ is
\begin{align}
&\bP_\nu^\sfG\lt(N_\sfG = k \med  \tilde{S}_\nu = \e_{i_j(\sfG)}, Y_\nu, Y_\nu \in (0, A)\rt)
=\begin{cases}
p^j_{Y_\nu} \qquad &\text{if} \quad k = 0\\
(1 - p^j_{Y_\nu})(1 - q_A)^{k-1}q_A  \qquad &\text{if} \quad k \geq 1
\end{cases}\ ,
\end{align}
where $p^j_{Y_\nu}$ is defined in \eqref{eq:p_yj}.
Therefore, we have
\begin{align}
\bE_\nu^\sfG\lt[\sum_{m=1}^{N_\sfG} \sigma_m \;\Bigg|\; \tilde{S}_\nu = \e_{i_j(\sfG)}, Y_\nu, Y_\nu \in (0, A)\rt]
&= (1 - p^j_{Y_\nu})\, \frac{\bE_\nu^\sfG[\sigma_1]}{q_A} \leq \frac{\bE_\nu^\sfG[\sigma_1]}{q_A}\ ,
\label{ineq:comb_qndel}
\end{align}
where the equality in \eqref{ineq:comb_qndel} follows from Lemma \ref{lem:cond_wald}(ii).  Finally, since every unit in $\cU \setminus \cA(\sfG)$ has exactly one corresponding stage that appears in every cycle, we have
\begin{align}\label{eq:exp_cycle_extn}
\bE_\nu^{\sfG}[\sigma_1] = \sum_{\e \in \cU \setminus \cA(\sfG)} \sfE_{\sfF}[\zeta^\e]\ .
\end{align}
 Furthermore, for any $\e \in \cU \setminus \cA(\sfG)$, we have
\begin{align}\label{ineqdel''_extn}
\sfE_{\sfF}[\zeta^\e] \leq \sfE_{\sfF}[\zeta^\e_-] = \frac{1}{\sfF(\zeta^\e_+ = \infty)} = \frac{1}{q^\e_+}\ ,\end{align}
where the first equality follows from  \cite[Corollary 8.39]{BookSeig}. \\
Furthermore,  we have
\begin{align}
1-q_A  &= \bP_\nu^\sfG( z_n = 0 \;\;  \text{for every} \;\; 2 \leq n \leq |\cU| + 1 \;\; \text{with} \;\; w_n =1 ) \label{eq:indepzn'}   \\
&=   \prod_{\substack{2 \leq n \leq |\cU| + 1,\\ w_n=1}} \sfG(z_n = 0)\\ \label{eq:distn.m-tuple} 
&=   \prod_{j = 1}^{|\cA(\sfG) \cap \cU|} \sfG(\tilde{z}^j = 0) = \prod_{j = 1}^{|\cA(\sfG) \cap \cU|} (1 - \tilde{q}_A^j(\sfG))\\ \label{ineqN_extn}
&\leq  \prod_{j = 1}^{|\cA(\sfG) \cap \cU|} (1 - \tilde{q}_{-}^j(\sfG))\ ,
\end{align}
where the equality in \eqref{eq:indepzn'} follows from the fact that, $\{z'_n : n \geq 2\}$ is a sequence of independent random variables under $\bP_\nu^{\sfG}$, the one in \eqref{eq:distn.m-tuple} holds because every unit in $\cA(\sfG) \cap \cU$ has exactly one corresponding stage that appears in every cycle, which implies that
\begin{align} 
\prod_{\substack{2 \leq n \leq |\cU| + 1,\\ w_n=1}} \; \mathbbm{1}\{z_n = 0\} \; \overset{d}{=} \; \prod_{j = 1}^{|\cA(\sfG)|} \; \mathbbm{1}\{\tilde{z}^j = 0\}\ .
\end{align}
The inequality in \eqref{ineqN_extn} holds because, by the definitions of $\tilde{q}_A^j(\sfG)$ and $\tilde{q}^j_{-}(\sfG)$, for $\e = \e_{i_j(\sfG)}$ we have 
\begin{align}\label{eq:lbp'_extn}
\tilde{q}_A^j(\sfG) = \sfG(\zeta^\e_- > \zeta^\e_A) \geq \sfG(\zeta^\e_- = \infty) = \tilde{q}^j_{-}(\sfG)\ .
\end{align} 
The proof is complete.\\
\end{IEEEproof}

Finally, we combine every result in this appendix to prove Theorem \ref{thm:ub_extn}.

\subsection{Proof of Theorem \ref{thm:ub_extn}}

\begin{IEEEproof}
Following Lemma \ref{lem:decom_3parts}, Lemma \ref{lem:M'_G_ij}, Lemma \ref{lem:zeta_1_notinA}, Lemma \ref{lem:M'_G_notinA} and Lemma \ref{lem:bd_cycle}, we have 
\begin{align} &\text{esssup}\ \bE^{\sfG}_\nu\left[\tilde{T} - \nu\med \cF^{\tilde{S}}_\nu(\sfG), \tilde{T} > \nu \right] \\
&\leq \max_{j \in [|\cA(\sfG) \cap \cU|]} \lt(\frac{A}{\tilde{\cJ}_0^{j}(\sfG)} \bigvee \frac{A}{\tilde{\cJ}_*^{\kappa(j, 2)}(\sfG)} \bigvee \frac{A}{\tilde{\cJ}_*^j (\sfG)}\rt) + o(A) + \; \text{constant} \\
&= \max_{\e \in \cA(\sfG) \cap \cU} \lt(\frac{A}{\cJ_0^\e(\sfG)}\rt) + o(A) + \; \text{constant}\ .
\end{align}
The last equality holds due to the following. 
For every $j \in [|\cA(\sfG) \cap \cU|]$, note that, 
\begin{align}
\min_{\e \in \cA(\sfG) \cap \cU} \cJ^\e_0(\sfG) = \min_{l \in [|\cA(\sfG) \cap \cU|]} \tilde{\cJ}_0^l(\sfG) \leq \tilde{\cJ}^j_*(\sfG)\ ,
\end{align}
where the inequality follows by the definition of $\tilde{\cJ}^j_*(\sfG)$. Consequently, we have
\begin{align}
\min_{\e \in \cA(\sfG) \cap \cU} \cJ^\e_0(\sfG) = \min_{l \in [|\cA(\sfG) \cap \cU|]} (\tilde{\cJ}_0^l(\sfG) \wedge \tilde{\cJ}^{\kappa(l, 2)}_*(\sfG) \wedge \tilde{\cJ}^l_*(\sfG))\ .
\end{align}
\end{IEEEproof}

\subsection{Proof of Proposition \ref{rem:ub_extn_nas}}\label{pf:rem:ub_extn_nas}


\begin{IEEEproof}
By using Lemma \ref{lem:mu/p} 
if in the proof of Theorem \ref{thm:ub_extn} we appropriately  replace 
the $o(A)$ terms 
with the following constant
\begin{align}
\max_{\e \in \cA(\sfG) \cap \cU} \; \frac{1}{q^\e_-(\sfG)}\lt(1 + \frac{\cW^\e_0(\sfG)}{\cJ^\e_0(\sfG)^2}\rt) \ ,  
\end{align}
then the non-asymptotic upper bound is obtained.
\end{IEEEproof}

\section{An Improvement of The Non-asymptotic Upper Bound}\label{pf:rem:ub_extn_nas_omega}

In this appendix, we further improve the non-asymptotic upper bound established in Proposition \ref{rem:ub_extn_nas} under an additional distributional assumption. In the following, first, we present some lemmas that are useful later.

\subsection{Some useful lemmas}

\begin{lemma}\label{lem:f1(x)}
Suppose $A > 0$, \; $\cI_1 \geq \cI_2 > 0$ and $\omega < 0$. Let
\begin{align}
    f(x) := \frac{A - x}{\cI_1} + Ae^{\omega x}\lt(\frac{1}{\cI_2} - \frac{1}{\cI_1}\rt)\ , \quad \text{for} \;\; x \in (0, A)\ .
\end{align}
Then 
\begin{align}
\sup_{x \in (0, A)} f(x) = \frac{A}{\cI_2}\ .
\end{align}
\end{lemma}
\begin{IEEEproof}
We have
\begin{align}
f'(x) = - \frac{1}{\cI_1} + A\omega e^{\omega x}\lt(\frac{1}{\cI_2} - \frac{1}{\cI_1}\rt)\ .
\end{align}
If $\cI_1 \geq \cI_2$, or equivalently, $\lt(\frac{1}{\cI_2} - \frac{1}{\cI_1}\rt) \geq 0$, then $f'(x) < 0$ for every $x \in (0, A)$, which implies that $f$ is decreasing, and
\begin{align}
\sup_{x \in (0, A)} f(x) = \lim_{x \to 0} f(x) = \frac{A}{\cI_2}\ .
\end{align}
\end{IEEEproof}

\begin{lemma}\label{lem:f2(x)}
Suppose $A > 0$, \; $\cI_2 > \cI_1 > 0$ and $\upsilon > 0 > \omega$. Let
\begin{align}
    f(x) := \frac{A - x}{\cI_1} + \frac{A(e^{\omega x} - e^{\omega A})}{\upsilon}\lt(\frac{1}{\cI_2} - \frac{1}{\cI_1}\rt)\ , \quad \text{for} \;\; x \in (0, A)\ .
\end{align}
Then 
\begin{align}
\sup_{x \in (0, A)} f(x) = \frac{A +\frac{1}{\omega}\lt(\log (-A\omega) + \log \lt(1 - \frac{\cI_1}{\cI_2}\rt) - \log \upsilon + 1\rt)}{\cI_1} - \frac{Ae^{\omega A}}{\upsilon}\lt(\frac{1}{\cI_2} - \frac{1}{\cI_1}\rt)\ .
\end{align}
\end{lemma}
\begin{IEEEproof}
We have
\begin{align}
f'(x) = - \frac{1}{\cI_1} + \frac{A\omega e^{\omega x}}{\upsilon}\lt(\frac{1}{\cI_2} - \frac{1}{\cI_1}\rt)\ .
\end{align}
Since $\cI_2 > \cI_1$, $f'(x) = 0$ has the solution
\begin{align}
x^* = -\frac{1}{\omega}\lt(\log(-A\omega) + \log \lt(1 - \frac{\cI_1}{\cI_2}\rt) - \log\upsilon\rt)\ ,
\end{align}
which leads to
\begin{align}
\sup_{x \in (0, A)} f(x) = f(x^*) = \frac{A +\frac{1}{\omega}\lt(\log (-A\omega) + \log \lt(1 - \frac{\cI_1}{\cI_2}\rt) - \log \upsilon + 1\rt)}{\cI_1} - \frac{Ae^{\omega A}}{\upsilon}\lt(\frac{1}{\cI_2} - \frac{1}{\cI_1}\rt)\ .
\end{align}
\end{IEEEproof}

\subsection{Improvement of the non-asymptotic upper bound}
\label{rem:ub_extn_nas_omega}

\begin{proposition}
If,  in addition to \eqref{eq:assm_2nd_mom_extn},  we further assume that for every $\sfG \in \cG$ and $\e \in \cA(\sfG) \cap \cU$ the following equation has only one positive real solution,
\begin{align}\label{assum:root_cond}
\int \lt(\frac{d\sfH^\e}{d\sfF^\e}\rt)^{-t}d\sfG = 1\ ,
\end{align}
then the non-asymptotic upper bound in Proposition \ref{rem:ub_extn_nas} can be further improved as follows. 
To be specific, when \eqref{eq:assm_2nd_mom_extn} and \eqref{assum:root_cond} hold, then 
there exist functions $\{\Phi_j : j \in [|\cA(\sfG) \cap \cU|]\}$ such that for every $A > 0$, \eqref{eq:ub_general} holds with the $o(A)$ term being replaced with \eqref{o(A)_replaced_extn}, and the first term with
\begin{align}
\max_{j \in [|\cA(\sfG) \cap \cU|]} \lt(\frac{A}{\tilde{\cJ}^j_*(\sfG)} \bigvee  \frac{A + \Phi_j(A)}{ \tilde{\cJ}_*^{\kappa(j, 2)}(\sfG)} \bigvee  \frac{A + \Phi_j(A)}{ \tilde{\cJ}_0^j(\sfG)}\rt)\ ,
\end{align}
where, for every $j \in [|\cA(\sfG) \cap \cU|]$, \ $\Phi_j(\cdot)$ are deterministic functions such that \ $\Phi_j(A)/A$ goes to $0$ as $A \to \infty$. 
\end{proposition}

\begin{IEEEproof}
We provide an alternative upper bound for the quantity under consideration in Lemma \ref{lem:M'_G_ij}, which will lead us to the desired result.
We start with \eqref{eq:use_lem_k} and proceed as follows.
\begin{align}\label{ineq:obj_fn_g}
\mu_{Y_\nu}^j + (1 - p^j_{Y_\nu}) 
\times \sum_{l = 1}^{|\cA(\sfG) \cap \cU|}w_{l}^{\kappa(j, 2)} \; \frac{\tilde{\mu}_A^{\kappa(\kappa(j, 2), l)}(\sfG)}{\tilde{q}_A^{\kappa(\kappa(j, 2), l)}(\sfG)} \;
\leq \; \max_{x \in [0, A)} \; g(x)\ ,
\end{align}
where the objective function in \eqref{ineq:obj_fn_g} is defined by
\begin{align}
g(x) := \mu_x^j + (1 - p^j_x)
\times \sum_{l = 1}^{|\cA(\sfG) \cap \cU|}w_{l}^{\kappa(j, 2)} \; \frac{\tilde{\mu}_A^{\kappa(\kappa(j, 2), l)}(\sfG)}{\tilde{q}_A^{\kappa(\kappa(j, 2), l)}(\sfG)}\ , \qquad \text{for} \; x \in [0, A)\ .
\end{align}
Leveraging the following relationship that,
\begin{align}\label{eq:max_of_g}
\max_{x \in [0,A)} g(x) = g(0) \bigvee \max_{x \in (0,A)} g(x)\ ,
\end{align}
it suffices to provide upper bounds for both quantities on the right-hand side. First, for any $x \in (0, A)$, we upper bound $g(x)$ as follows.
\begin{align}
g(x)
&= \mu_x^j + (1 - p^j_x)
\times \sum_{l = 1}^{|\cA(\sfG) \cap \cU|}w_{l}^{\kappa(j, 2)} \; \frac{\tilde{\mu}_A^{\kappa(\kappa(j, 2), l)}(\sfG)}{\tilde{q}_A^{\kappa(\kappa(j, 2), l)}(\sfG)}\\ \label{eq:dir_lemC2}
&\leq\mu_x^j + (1 - p^j_x)
\times \sum_{l = 1}^{|\cA(\sfG)|} w_{l}^{\kappa(j, 2)} \; \lt(\frac{A}{\tilde{\cJ}_0^{\kappa(\kappa(j, 2), l)}(\sfG)} + \; \text{constant}
\rt)\\
&\leq\mu_x^j + (1 - p^j_x)\;\frac{A}{\tilde{\cJ}_*^{\kappa(j, 2)}(\sfG)} + \; \text{constant} 
\\ \label{ineq:obj_fn}
&\leq \max_{x \in (0,A)} \; \lt(\mu^j_x + (1 - p^j_x)\;\frac{A}{\tilde{\cJ}_*^{\kappa(j, 2)}(\sfG)} \rt)
+ \; \text{constant} \ , 
\end{align}
where 
the inequality in \eqref{eq:dir_lemC2} 
follows from Lemma \ref{lem:mu/p}. 
Now, we use Wald's bounds to solve the maximization problem in \eqref{ineq:obj_fn}, where the objective function is denoted by
\begin{align}
h(x) := \lt(\mu^j_x + (1 - p^j_x)\;\frac{A}{\tilde{\cJ}_*^{\kappa(j, 2)}(\sfG)} \rt)\ , \qquad \text{for} \; x \in (0, A)\ .
\end{align}
To be specific, for any $x \in (0, A)$, we use the 
following results (see \cite[pp. 127-128]{Tart14} and \cite[Theorem 1]{Lorden70}), which are crucial to bound $h(x)$ from above. If \eqref{assum:root_cond} holds, then
\begin{align} \label{ineq:bd_of_p}
\frac{e^{-\omega_0 x} - 1}{e^{-\omega_0 x} - \omega_1 e^{\omega_0 (A - x)}} \leq p_x^j &\leq \frac{\omega_2 e^{-\omega_0 x} - 1}{\omega_2 e^{-\omega_0 x} - e^{\omega_0(A - x)}}\ , \qquad \text{for some} \;\; \omega_0 < 0 < \omega_1 < 1 < \omega_2\ .\\ \label{ineq:bd_of_mu}
\mu_x^j &\leq \frac{(A-x) p_x^j - x (1 - p_x^j)}{\tilde{\cJ}_0^j(\sfG)} \; + \; \text{constant}\ . 
\end{align}
Then, for any $x \in (0, A)$, we have 
\begin{align}
h(x) &\leq \frac{(A-x)p_x^j - x (1 - p_x^j)}{\tilde{\cJ}_0^j(\sfG)} + (1 - p_x^j)\frac{A}{\tilde{\cJ}_*^{\kappa(j, 2)}(\sfG)} + \; \text{constant} 
\\ \label{eq:h_bd1}
&= \frac{A-x}{\tilde{\cJ}_0^j(\sfG)} + A (1 - p_x^j) \lt(\frac{1}{\tilde{\cJ}_*^{\kappa(j, 2)}(\sfG)} - \frac{1}{\tilde{\cJ}_0^j(\sfG)}\rt) + \; \text{constant}\ ,
\end{align}
where the first inequality follows by using \eqref{ineq:bd_of_mu}. Now, if $\tilde{\cJ}_0^j(\sfG) \geq \tilde{\cJ}_*^{\kappa(j, 2)}(\sfG)$, then following \eqref{eq:h_bd1} and using the lower bound in \eqref{ineq:bd_of_p} we can further upper bound
\begin{align}
h(x) 
&\leq \frac{A-x}{\tilde{\cJ}_0^j(\sfG)} + A \lt(\frac{1}{\tilde{\cJ}_*^{\kappa(j, 2)}(\sfG)} - \frac{1}{\tilde{\cJ}_0^j(\sfG)}\rt) \frac{1 - \omega_1 e^{\omega_0 (A - x)}}{e^{-\omega_0 x} - \omega_1 e^{\omega_0 (A - x)}} + \; \text{constant}
\\ \label{ineq:to_use_f1}
&\leq \frac{A-x}{\tilde{\cJ}_0^j(\sfG)} + A \lt(\frac{1}{\tilde{\cJ}_*^{\kappa(j, 2)}(\sfG)} - \frac{1}{\tilde{\cJ}_0^j(\sfG)}\rt) e^{\omega_0 x} + \; \text{constant}\ , 
\end{align}
where the last inequality follows by using the inequality that,
\begin{align}
\frac{1-b}{B-b} \leq \frac{1}{B}\ , \quad \text{for any}\;\; 0 \leq b < 1 \leq B\ . 
\end{align}
Therefore, following \eqref{ineq:to_use_f1} and by using Lemma \ref{lem:f1(x)} we have
\begin{align}\label{ineq:bd_of_h_1}
\max_{x \in (0, A)} h(x) \leq \frac{A}{\tilde{\cJ}_*^{\kappa(j, 2)}(\sfG)}\ + \; \text{constant}\ .
\end{align}
On the other hand, if $\tilde{\cJ}_*^{\kappa(j, 2)}(\sfG) > \tilde{\cJ}_0^j(\sfG)$, then following \eqref{eq:h_bd1} and using the upper bound in \eqref{ineq:bd_of_p} we can further bound
\begin{align}
h(x) 
&\leq \frac{A-x}{\tilde{\cJ}_0^j(\sfG)} + A \lt(\frac{1}{\tilde{\cJ}_*^{\kappa(j, 2)}(\sfG)} - \frac{1}{\tilde{\cJ}_0^j(\sfG)}\rt) \frac{1 - e^{\omega_0 (A - x)}}{\omega_2e^{-\omega_0 x} - e^{\omega_0 (A - x)}} + \; \text{constant}\\ 
&\leq \frac{A-x}{\tilde{\cJ}_0^j(\sfG)} + A \lt(\frac{1}{\tilde{\cJ}_*^{\kappa(j, 2)}(\sfG)} - \frac{1}{\tilde{\cJ}_0^j(\sfG)}\rt) \frac{e^{\omega_0 x} - e^{\omega_0 A}}{\omega_2 - e^{\omega_0 A}} + \; \text{constant}\\ \label{ineq:to_use_f2} 
&\leq \frac{A-x}{\tilde{\cJ}_0^j(\sfG)} + A \lt(\frac{1}{\tilde{\cJ}_*^{\kappa(j, 2)}(\sfG)} - \frac{1}{\tilde{\cJ}_0^j(\sfG)}\rt) \frac{e^{\omega_0 x} - e^{\omega_0 A}}{\omega_2} + \; \text{constant}\ . 
\end{align}
Therefore, following \eqref{ineq:to_use_f2} and again by using Lemma \ref{lem:f2(x)} we have
\begin{align}\label{eq:h_bd2}
\begin{split}
\max_{x \in (0, A)} h(x) &\leq \frac{A +\frac{1}{\omega_0}\lt(\log (-A \omega_0) + \log \lt(1 -{\tilde{\cJ}_0^j(\sfG)}/{\tilde{\cJ}_*^{\kappa(j, 2)}(\sfG)}\rt) - \log \omega_2 + 1\rt)}{\tilde{\cJ}_0^j(\sfG)}\\
&\qquad - \frac{Ae^{\omega_0 A}}{\omega_2} \lt(\frac{1}{\tilde{\cJ}_*^{\kappa(j, 2)}(\sfG)} - \frac{1}{\tilde{\cJ}_0^j(\sfG)}\rt) + \; \text{constant}\ .
\end{split}
\end{align}

Next, following \eqref{eq:max_of_g} we also need to upper bound $g(0)$, which is done as follows.

\begin{align}
g(0) &= \mu_0^j + (1 - p^j_0)
\times \sum_{l = 1}^{|\cA(\sfG) \cap \cU|}w_{l}^{\kappa(j, 2)} \; \frac{\tilde{\mu}_A^{\kappa(\kappa(j, 2), l)}(\sfG)}{\tilde{q}_A^{\kappa(\kappa(j, 2), l)}(\sfG)}\\ \label{eq:muandp}
&= \tilde{\mu}_{A}^j(\sfG) + 
\sum_{l = 1}^{|\cA(\sfG) \cap \cU|}(1 - \tilde{q}_A^j(\sfG)) w_{l}^{\kappa(j, 2)} \; \frac{\tilde{\mu}_A^{\kappa(j, l+1)}(\sfG)}{\tilde{q}_A^{\kappa(j, l+1)}(\sfG)}\\ \label{eq:qw_w}
&= \tilde{\mu}_{A}^j(\sfG) + 
\sum_{l = 1}^{|\cA(\sfG) \cap \cU| - 1} w_{l+1}^{j} \; \frac{\tilde{\mu}_A^{\kappa(j, l+1)}(\sfG)}{\tilde{q}_A^{\kappa(j, l+1)}(\sfG)} \; + \; 
\tilde{\mu}_A^{\kappa(j, |\cA(\sfG)\cap\cU|+1)}(\sfG)\frac{\prod_{s= 1}^{|\cA(\sfG) \cap \cU|}(1- \tilde{q}_A^{\kappa(j, s)}(\sfG))}{1 - \prod_{s = 1}^{|\cA(\sfG) \cap \cU|} (1 - \tilde{q}_A^{\kappa(j, s)}(\sfG))}
\\
&= \tilde{\mu}_{A}^j(\sfG) + 
\sum_{l = 1}^{|\cA(\sfG) \cap \cU| - 1} w_{l+1}^{j} \; \frac{\tilde{\mu}_A^{\kappa(j, l+1)}(\sfG)}{\tilde{q}_A^{\kappa(j, l+1)}(\sfG)} \; + \; 
\tilde{\mu}_A^{j}(\sfG)\frac{\prod_{s= 1}^{|\cA(\sfG) \cap \cU|}(1- \tilde{q}_A^{\kappa(j, s)}(\sfG)) - 1 + 1}{1 - \prod_{s = 1}^{|\cA(\sfG) \cap \cU|} (1 - \tilde{q}_A^{\kappa(j, s)}(\sfG))}
\\
\label{eq:kappa_reln}
&= \tilde{\mu}_{A}^j(\sfG) - \tilde{\mu}_{A}^j(\sfG) + 
\sum_{l = 2}^{|\cA(\sfG) \cap \cU|} w_{l}^{j} \; \frac{\tilde{\mu}_A^{\kappa(j, l)}(\sfG)}{\tilde{q}_A^{\kappa(j, l)}(\sfG)} \; + \; 
\frac{\tilde{\mu}_A^{j}(\sfG)}{\tilde{q}_A^{j}(\sfG)} w^j_1 
\\
&=\sum_{l = 1}^{|\cA(\sfG) \cap \cU|} w_{l}^{j} \; \frac{\tilde{\mu}_A^{\kappa(j, l)}(\sfG)}{\tilde{q}_A^{\kappa(j, l)}(\sfG)}\\ \label{ineq:finalAIbound}
&\leq \sum_{l = 1}^{|\cA(\sfG) \cap \cU|} w_{l}^{j} \; \lt( \frac{A}{\tilde{\cJ}^{\kappa(j, l)}_0(\sfG)} + \; \text{constant}\rt)\\ \label{ineq:finalAIbound2}
&= \frac{A}{\tilde{\cJ}_*^{j}(\sfG)} +  \; \text{constant}\ , 
\end{align}
where the equality in \eqref{eq:muandp} follows from the observations that $\mu_{0}^j = \tilde{\mu}_{A}^j(\sfG)$ and $p^j_{0} = \tilde{q}_A^j(\sfG)$, the equality in \eqref{eq:qw_w} follows from the fact that, for every $l \in [|\cA(\sfG) \cap \cU|]$,
\begin{align}
(1 - \tilde{q}_A^j(\sfG)) w_{l}^{\kappa(j, 2)} = \frac{(1 - \tilde{q}_A^j(\sfG)) \tilde{q}_A^{\kappa(j, l + 1)} \prod_{s= 1}^{l-1}(1- \tilde{q}_A^{\kappa(j, s+1)}(\sfG))}{1 - \prod_{s = 1}^{|\cA(\sfG) \cap \cU|} (1 - \tilde{q}_A^{\kappa(j, s+1)}(\sfG))} = 
\frac{\tilde{q}_A^{\kappa(j, l + 1)} \prod_{s= 1}^{l}(1- \tilde{q}_A^{\kappa(j, s)}(\sfG))}{1 - \prod_{s = 1}^{|\cA(\sfG) \cap \cU|} (1 - \tilde{q}_A^{\kappa(j, s)}(\sfG))}\ ,
\end{align}
which coincides with $w_{l+1}^j$ when $l \in [|\cA(\sfG) \cap \cU| - 1]$.
The inequality in \eqref{ineq:finalAIbound} follows from Lemma \ref{lem:mu/p}.

Therefore, from the above results we derive the final upper bound as follows. 
If $\tilde{\cJ}_0^j(\sfG) \geq \tilde{\cJ}_*^{\kappa(j, 2)}(\sfG)$, then following \eqref{eq:max_of_g}, \eqref{ineq:obj_fn}, \eqref{ineq:bd_of_h_1}, and \eqref{ineq:finalAIbound2}, we have
\begin{align}
&\bE_\nu^\sfG\lt[T_1 \med  \tilde{S}_\nu = \e_{i_j(\sfG)}, Y_\nu, \tilde{T} > \nu\rt]
\leq \lt(\frac{A}{\tilde{\cJ}_*^{j}(\sfG)} \bigvee \frac{A}{\tilde{\cJ}_*^{\kappa(j, 2)}(\sfG)}\rt) + \; \text{constant}\ . 
\end{align}
On the other hand, if $\tilde{\cJ}_*^{\kappa(j, 2)}(\sfG) > \tilde{\cJ}_0^j(\sfG)$, then 
we have
\begin{align}
&\bE_\nu^\sfG\lt[T_1 \med  \tilde{S}_\nu = \e_{i_j(\sfG)}, Y_\nu, \tilde{T} > \nu\rt]\\
&\leq \Bigg(\frac{A}{\tilde{\cJ}_*^{j}(\sfG)} \bigvee \Bigg(\frac{A +\frac{1}{\omega_0}\lt(\log (-A \omega_0) + \log \lt(1 -{\tilde{\cJ}_0^j(\sfG)}/{\tilde{\cJ}_*^{\kappa(j, 2)}(\sfG)}\rt) - \log \omega_2 + 1\rt)}{\tilde{\cJ}_0^j(\sfG)}\\
&\qquad \qquad \qquad - \frac{Ae^{\omega_0 A}}{\omega_2} \lt(\frac{1}{\tilde{\cJ}_*^{\kappa(j, 2)}(\sfG)} - \frac{1}{\tilde{\cJ}_0^j(\sfG)}\rt)\Bigg)\Bigg)
+ \; \text{constant}\ .
\end{align}
Thus, the exact forms of the functions $\{\Phi_j : j \in [|\cA(\sfG) \cap \cU|]\}$ can be derived from the above, where the constants are obtained from Lemma \ref{lem:mu/p}. This completes the proof.
\end{IEEEproof}

\bibliographystyle{IEEEtran}
\bibliography{ISIT_2021_AT}

\begin{thebibliography}{10}
\providecommand{\url}[1]{#1}
\csname url@samestyle\endcsname
\providecommand{\newblock}{\relax}
\providecommand{\bibinfo}[2]{#2}
\providecommand{\BIBentrySTDinterwordspacing}{\spaceskip=0pt\relax}
\providecommand{\BIBentryALTinterwordstretchfactor}{4}
\providecommand{\BIBentryALTinterwordspacing}{\spaceskip=\fontdimen2\font plus
\BIBentryALTinterwordstretchfactor\fontdimen3\font minus
  \fontdimen4\font\relax}
\providecommand{\BIBforeignlanguage}[2]{{%
\expandafter\ifx\csname l@#1\endcsname\relax
\typeout{** WARNING: IEEEtran.bst: No hyphenation pattern has been}%
\typeout{** loaded for the language `#1'. Using the pattern for}%
\typeout{** the default language instead.}%
\else
\language=\csname l@#1\endcsname
\fi
#2}}
\providecommand{\BIBdecl}{\relax}
\BIBdecl

\bibitem{ChFeAl21}
A.~Chaudhuri, G.~Fellouris, and A.~Tajer, ``Sequential change detection of a
  correlation structure under a sampling constraint,'' in \emph{Proc. IEEE
  International Symposium on Information Theory}, Melbourne, Australia, Jul.
  2021, pp. 605--610.

\bibitem{CohenZhao2015}
K.~{Cohen} and Q.~{Zhao}, ``Active hypothesis testing for anomaly detection,''
  \emph{IEEE Transactions on Information Theory}, vol.~61, no.~3, pp.
  1432--1450, 2015.

\bibitem{HuangCohenZhao2019}
B.~{Huang}, K.~{Cohen}, and Q.~{Zhao}, ``Active anomaly detection in
  heterogeneous processes,'' \emph{IEEE Transactions on Information Theory},
  vol.~65, no.~4, pp. 2284--2301, 2019.

\bibitem{TsopFell20}
A.~{Tsopelakos} and G.~{Fellouris}, ``Sequential anomaly detection with
  observation control under a generalized error metric,'' in \emph{Proc. IEEE
  International Symposium on Information Theory}, Los Angeles, CA, Jun. 2020,
  pp. 1165--1170.

\bibitem{Draga96}
V.~Dragalin, ``A simple and effective scanning rule for a multi-channel
  system,'' \emph{Metrika}, vol.~43, no.~1, pp. 165--182, 1996.

\bibitem{zigan66}
K.~S. Zigangirov, ``On a problem in optimal scanning,'' \emph{Theory of
  Probability \& Its Applications}, vol.~11, no.~2, pp. 294--298, 1966.

\bibitem{HeyTaj16}
A.~Tajer, J.~Heydari, and H.~V. Poor, ``Active sampling for the quickest
  detection of markov networks,'' \emph{IEEE Transactions on Information
  Theory}, vol.~68, no.~4, pp. 2479--2508, Apr. 2022.

\bibitem{HeyTaj17}
J.~{Heydari} and A.~{Tajer}, ``Quickest search for local structures in random
  graphs,'' \emph{IEEE Transactions on Signal and Information Processing over
  Networks}, vol.~3, no.~3, pp. 526--538, Sep. 2017.

\bibitem{Cher59}
H.~Chernoff, ``Sequential design of experiments,'' \emph{Annals of Mathematical
  Statistics}, vol.~30, no.~3, pp. 755--770, Sep. 1959.

\bibitem{NitAtiaVeer2013}
S.~{Nitinawarat}, G.~K. {Atia}, and V.~V. {Veeravalli}, ``Controlled sensing
  for multihypothesis testing,'' \emph{IEEE Transactions on Automatic Control},
  vol.~58, no.~10, pp. 2451--2464, Oct. 2013.

\bibitem{OlympiaPoorBook}
H.~Poor and O.~Hadjiliadis, \emph{\BIBforeignlanguage{English (US)}{Quickest
  Detection}}.\hskip 1em plus 0.5em minus 0.4em\relax United Kingdom: Cambridge
  University Press, 2008.

\bibitem{Tart14}
A.~Tartakovsky, I.~Nikiforov, and M.~Basseville, \emph{Sequential Analysis:
  Hypothesis Testing and Changepoint Detection}.\hskip 1em plus 0.5em minus
  0.4em\relax CRC Press, 08 2014.

\bibitem{zhang-mei2020banditQCD_journal}
W.~Zhang and Y.~Mei, ``Bandit change-point detection for real-time monitoring
  high-dimensional data under sampling control,'' \emph{Technometrics},
  vol.~65, no.~1, pp. 33--43, 2023.

\bibitem{veeravalli2023quickest}
V.~V. Veeravalli, G.~Fellouris, and G.~V. Moustakides, ``Quickest change
  detection with controlled sensing,'' \emph{IEEE Journal on Selected Areas in
  Information Theory}, pp. 1--1, 2024.

\bibitem{XuMeiMous21}
Q.~Xu, Y.~Mei, and G.~V. Moustakides, ``Optimum multi-stream sequential
  change-point detection with sampling control,'' \emph{IEEE Transactions on
  Information Theory}, vol.~67, no.~11, pp. 7627--7636, 2021.

\bibitem{XuMeiPostUncert}
Q.~Xu and Y.~Mei, ``Multi-stream quickest detection with unknown post-change
  parameters under sampling control,'' in \emph{Proc. IEEE International
  Symposium on Information Theory}, Melbourne, Australia, Jul. 2021, pp.
  112--117.

\bibitem{XuMeiPostUncert_SeqAn}
------, ``Asymptotic optimality theory for active quickest detection with
  unknown postchange parameters,'' \emph{Sequential Analysis}, vol.~42, no.~2,
  pp. 150--181, 2023.

\bibitem{XuMei2streams}
------, ``Active quickest detection when monitoring multi-streams with two
  affected streams,'' in \emph{Proc. IEEE International Symposium on
  Information Theory}, Espoo, Finland, Jul. 2022, pp. 1915--1920.

\bibitem{LiMeiShi15}
K.~Liu, Y.~Mei, and J.~Shi, ``An adaptive sampling strategy for online
  high-dimensional process monitoring,'' \emph{Technometrics}, vol.~57, no.~3,
  pp. 305--319, 2015.

\bibitem{HeyTaj17_2}
J.~{Heydari} and A.~{Tajer}, ``Quickest change detection in structured data
  with incomplete information,'' in \emph{Proc. IEEE International Conference
  on Acoustics, Speech and Signal Processing}, New Orleans, LA, Mar. 2017, pp.
  6434--6438.

\bibitem{ChauFell20}
A.~{Chaudhuri} and G.~{Fellouris}, ``Sequential detection and isolation of a
  correlated pair,'' in \emph{Proc. IEEE International Symposium on Information
  Theory}, Los Angeles, CA, Jun. 2020, pp. 1141--1146.

\bibitem{chaudhuri2022joint}
A.~Chaudhuri and G.~Fellouris, ``Joint sequential detection and isolation for
  dependent data streams,'' \emph{arXiv preprint arXiv:2207.00120}, 2022.

\bibitem{Xie2020}
L.~Xie, Y.~Xie, and G.~V. Moustakides, ``Sequential subspace change point
  detection,'' \emph{Sequential Analysis}, vol.~39, no.~3, pp. 307--335, 2020.

\bibitem{LaiZhang01}
L.~K. Chan and J.~Zhang, ``Cumulative sum control charts for the covariance
  matrix,'' \emph{Statistica Sinica}, vol.~11, no.~3, pp. 767--790, 2001.

\bibitem{Lorden1971}
G.~Lorden, ``Procedures for reacting to a change in distribution,''
  \emph{Annals of Mathematical Statistics}, vol.~42, no.~6, pp. 1897--1908,
  Dec. 1971.

\bibitem{Lai98}
T.~L. Lai, ``Information bounds and quick detection of parameter changes in
  stochastic systems,'' \emph{IEEE Transactions on Information Theory},
  vol.~44, no.~7, pp. 2917--2929, 1998.

\bibitem{BookSeig}
D.~Siegmund, \emph{\BIBforeignlanguage{eng}{Sequential Analysis Tests and
  Confidence Intervals}}, ser. Springer Series in Statistics.\hskip 1em plus
  0.5em minus 0.4em\relax New York, NY: Springer New York, 1985.

\bibitem{Lorden70}
G.~Lorden, ``On excess over the boundary,'' \emph{Annals of Mathematical
  Statistics}, vol.~41, no.~2, pp. 520--527, Apr. 1970.

\end{thebibliography}

\vfill

\end{document}